\PassOptionsToPackage{unicode}{hyperref}
\PassOptionsToPackage{hyphens}{url}
\PassOptionsToPackage{dvipsnames,svgnames*,x11names*}{xcolor}
\documentclass[
  english,
  ,man,floatsintext]{apa6}
\usepackage{lmodern}
\usepackage{amssymb,amsmath}
\usepackage{ifxetex,ifluatex}
\ifnum 0\ifxetex 1\fi\ifluatex 1\fi=0 % if pdftex
  \usepackage[T1]{fontenc}
  \usepackage[utf8]{inputenc}
  \usepackage{textcomp} % provide euro and other symbols
\else % if luatex or xetex
  \usepackage{unicode-math}
  \defaultfontfeatures{Scale=MatchLowercase}
  \defaultfontfeatures[\rmfamily]{Ligatures=TeX,Scale=1}
\fi
\IfFileExists{upquote.sty}{\usepackage{upquote}}{}
\IfFileExists{microtype.sty}{% use microtype if available
  \usepackage[]{microtype}
  \UseMicrotypeSet[protrusion]{basicmath} % disable protrusion for tt fonts
}{}
\makeatletter
\@ifundefined{KOMAClassName}{% if non-KOMA class
  \IfFileExists{parskip.sty}{%
    \usepackage{parskip}
  }{% else
    \setlength{\parindent}{0pt}
    \setlength{\parskip}{6pt plus 2pt minus 1pt}}
}{% if KOMA class
  \KOMAoptions{parskip=half}}
\makeatother
\usepackage{xcolor}
\IfFileExists{xurl.sty}{\usepackage{xurl}}{} % add URL line breaks if available
\IfFileExists{bookmark.sty}{\usepackage{bookmark}}{\usepackage{hyperref}}
\hypersetup{
  pdftitle={Workflow Techniques for the Robust Use of Bayes Factors},
  pdfauthor={Daniel J. Schad1, 2, 3, Bruno Nicenboim2, 3, Paul-Christian Bürkner4, Michael Betancourt5, \& Shravan Vasishth3},
  pdflang={en-EN},
  pdfkeywords={Bayes factors, Bayesian model comparison, Prior, Posterior, Simulation-based calibration},
  colorlinks=true,
  linkcolor=Maroon,
  filecolor=Maroon,
  citecolor=Blue,
  urlcolor=blue,
  pdfcreator={LaTeX via pandoc}}
\usepackage{color}
\usepackage{fancyvrb}

\DefineVerbatimEnvironment{Highlighting}{Verbatim}{commandchars=\\\{\}}
\usepackage{framed}
\definecolor{shadecolor}{RGB}{248,248,248}
\newenvironment{Shaded}{\begin{snugshade}}{\end{snugshade}}

\newcommand{\CommentTok}[1]{\textcolor[rgb]{0.56,0.35,0.01}{\textit{#1}}}

\newcommand{\ControlFlowTok}[1]{\textcolor[rgb]{0.13,0.29,0.53}{\textbf{#1}}}
\newcommand{\DataTypeTok}[1]{\textcolor[rgb]{0.13,0.29,0.53}{#1}}
\newcommand{\DecValTok}[1]{\textcolor[rgb]{0.00,0.00,0.81}{#1}}

\newcommand{\FloatTok}[1]{\textcolor[rgb]{0.00,0.00,0.81}{#1}}

\newcommand{\KeywordTok}[1]{\textcolor[rgb]{0.13,0.29,0.53}{\textbf{#1}}}
\newcommand{\NormalTok}[1]{#1}
\newcommand{\OperatorTok}[1]{\textcolor[rgb]{0.81,0.36,0.00}{\textbf{#1}}}
\newcommand{\OtherTok}[1]{\textcolor[rgb]{0.56,0.35,0.01}{#1}}

\newcommand{\StringTok}[1]{\textcolor[rgb]{0.31,0.60,0.02}{#1}}

\usepackage{longtable,booktabs}
\usepackage{etoolbox}
\makeatletter
\patchcmd\longtable{\par}{\if@noskipsec\mbox{}\fi\par}{}{}
\makeatother
\IfFileExists{footnotehyper.sty}{\usepackage{footnotehyper}}{\usepackage{footnote}}
\makesavenoteenv{longtable}
\usepackage{graphicx}
\makeatletter
\def\maxwidth{\ifdim\Gin@nat@width>\linewidth\linewidth\else\Gin@nat@width\fi}
\def\maxheight{\ifdim\Gin@nat@height>\textheight\textheight\else\Gin@nat@height\fi}
\makeatother
\setkeys{Gin}{width=\maxwidth,height=\maxheight,keepaspectratio}
\makeatletter
\def\fps@figure{htbp}
\makeatother
\providecommand{\tightlist}{%
  \setlength{\itemsep}{0pt}\setlength{\parskip}{0pt}}
\ifx\paragraph\undefined\else
  \let\oldparagraph\paragraph
  \renewcommand{\paragraph}[1]{\oldparagraph{#1}\mbox{}}
\fi
\ifx\subparagraph\undefined\else
  \let\oldsubparagraph\subparagraph
  \renewcommand{\subparagraph}[1]{\oldsubparagraph{#1}\mbox{}}
\fi
\usepackage{upgreek}
\usepackage{longtable}
\usepackage{lscape}
\usepackage{multirow}		% Table styling
\usepackage{tabularx}		% Control Column width
\usepackage[flushleft]{threeparttable}	% Allows for three part tables with a specified notes section
\usepackage{threeparttablex}            % Lets threeparttable work with longtable

\makeatletter
\newcommand\LastLTentrywidth{1em}
\newlength\longtablewidth
\newcommand{\getlongtablewidth}{\begingroup \ifcsname LT@\roman{LT@tables}\endcsname \global\longtablewidth=0pt \renewcommand{\LT@entry}[2]{\global\advance\longtablewidth by ##2\relax\gdef\LastLTentrywidth{##2}}\@nameuse{LT@\roman{LT@tables}} \fi \endgroup}

\makeatletter
\patchcmd{\HyOrg@maketitle}
  {\section{\normalfont\normalsize\abstractname}}
  {\section*{\normalfont\normalsize\abstractname}}
  {}{\typeout{Failed to patch abstract.}}
\patchcmd{\HyOrg@maketitle}
  {\section{\protect\normalfont{\@title}}}
  {\section*{\protect\normalfont{\@title}}}
  {}{\typeout{Failed to patch title.}}
\makeatother
\shorttitle{Bayes factor workflow}
\keywords{Bayes factors, Bayesian model comparison, Prior, Posterior, Simulation-based calibration}
\usepackage{csquotes}
\usepackage{amsmath}
\usepackage[utf8]{inputenc}
\usepackage[T1]{fontenc}
\usepackage[backend=biber,useprefix=true,style=apa,url=false,doi=false,sorting=nyt,eprint=false]{biblatex}
\DeclareLanguageMapping{american}{american-apa}
\usepackage{fancyvrb}
\usepackage{newfloat}
\usepackage{graphicx}
\usepackage{caption}
\usepackage{listings}
\usepackage{glossaries}
\makeglossaries
\DeclareFloatingEnvironment[fileext=los, listname=List of Schemes, name=Listing, placement=!htbp, within=section]{listing}
\usepackage{placeins}
\usepackage{todonotes}
\ifxetex
  \usepackage{polyglossia}
  \setmainlanguage[]{english}
\else
  \usepackage[shorthands=off,main=english]{babel}
\fi
\ifluatex
  \usepackage{selnolig}  % disable illegal ligatures
\fi
\newlength{\cslhangindent}
\newenvironment{cslreferences}%
  {\setlength{\parindent}{0pt}%
  \everypar{\setlength{\hangindent}{\cslhangindent}}\ignorespaces}%
  {\par}

\title{Workflow Techniques for the Robust Use of Bayes Factors}
\author{Daniel J. Schad\textsuperscript{1, 2, 3}, Bruno Nicenboim\textsuperscript{2, 3}, Paul-Christian Bürkner\textsuperscript{4}, Michael Betancourt\textsuperscript{5}, \& Shravan Vasishth\textsuperscript{3}}
\date{}

\authornote{

Correspondence concerning this article should be addressed to Daniel J. Schad, Health and Medical University, Potsdam, Germany. E-mail: \href{mailto:danieljschad@gmail.com}{\nolinkurl{danieljschad@gmail.com}}

}

\affiliation{\vspace{0.5cm}\textsuperscript{1} Health and Medical University Potsdam, Germany\\\textsuperscript{2} Tilburg University, Netherlands\\\textsuperscript{3} University of Potsdam, Germany\\\textsuperscript{4} University of Stuttgart, Germany\\\textsuperscript{5} Symplectomorphic, New York, USA}

\abstract{
Inferences about hypotheses are ubiquitous in the cognitive sciences. Bayes factors provide one general way to compare different hypotheses by their compatibility with the observed data. Those quantifications can then also be used to choose between hypotheses. While Bayes factors provide an immediate approach to hypothesis testing, they are highly sensitive to details of the data/model assumptions. Moreover it's not clear how straightforwardly this approach can be implemented in practice, and in particular how sensitive it is to the details of the computational implementation. Here, we investigate these questions for Bayes factor analyses in the cognitive sciences. We explain the statistics underlying Bayes factors as a tool for Bayesian inferences and discuss that utility functions are needed for principled decisions on hypotheses. Next, we study how Bayes factors misbehave under different conditions. This includes a study of errors in the estimation of Bayes factors. Importantly, it is unknown whether Bayes factor estimates based on bridge sampling are unbiased for complex analyses. We are the first to use simulation-based calibration as a tool to test the accuracy of Bayes factor estimates. Moreover, we study how stable Bayes factors are against different MCMC draws. We moreover study how Bayes factors depend on variation in the data. We also look at variability of decisions based on Bayes factors and how to optimize decisions using a utility function. We outline a Bayes factor workflow that researchers can use to study whether Bayes factors are robust for their individual analysis, and we illustrate this workflow using an example from the cognitive sciences. We hope that this study will provide a workflow to test the strengths and limitations of Bayes factors as a way to quantify evidence in support of scientific hypotheses.
Reproducible code is available from \url{https://osf.io/y354c/}.
}

\begin{document}
\maketitle

{
\hypersetup{linkcolor=}
\setcounter{tocdepth}{3}
\tableofcontents
}
\hypertarget{introduction}{%
\section{Introduction}\label{introduction}}

In the cognitive sciences and related areas, recent years have seen a rise in Bayesian approaches to data analysis. Many cognitive science journals have published special issues on Bayesian data analysis, including methodological journals such as the Journal of Mathematical Psychology (Lee, 2011; Mulder \& Wagenmakers, 2016) and Psychological Methods (Chow \& Hoijtink, 2017; Hoijtink \& Chow, 2017), but also the more experimental journal Psychonomic Bulletin \& Review (Vandekerckhove, Rouder, \& Kruschke, 2018). Further introductory articles have been contributed (see Etz \& Vandekerckhove, 2018; Doorn, Aust, Haaf, Stefan, \& Wagenmakers, 2021; Etz et al., 2018; Nicenboim \& Vasishth, 2016; Sorensen, Hohenstein, \& Vasishth, 2016; Vasishth, Nicenboim, Beckman, Li, \& Kong, 2018). That Bayesian analyses are so prominently discussed and used is an indication that Bayesian approaches are becoming increasingly mainstream (Gelman et al., 2014).

Bayesian approaches provide tools for different aspects of data analysis. Bayesian data analysis plays an important role in cognitive science as it allows us to carry out inference, i.e., a way to quantify the evidence that data provide in support of one hypothesis or another. Such Bayesian hypothesis testing can be implemented using Bayes factors (Gronau et al., 2017a; Heck et al., 2020; Jeffreys, 1939; Kass \& Raftery, 1995; Rouder, Haaf, \& Vandekerckhove, 2018; Schönbrodt \& Wagenmakers, 2018; Wagenmakers, Lodewyckx, Kuriyal, \& Grasman, 2010), which quantify evidence in favor of one model over another, where each model implements one scientific hypothesis about the data (for a critique of Bayes factors see Navarro, 2019).

Bayes factors are increasingly used in the cognitive sciences and other fields of science (Heck et al., 2020). However, while Bayes factors provide an immediate approach to hypothesis testing, it is known that they are highly sensitive to details of the data and model assumptions. Moreover, it is unclear how implementable it is in practice and how sensitive it is to the details of the computational implementation.

First, the results of Bayes factor analyses are highly sensitive to and crucially depend on prior assumptions about model parameters (we illustrate this below) (Aitkin, 1991; Gelman et al., 2013; Grünwald, 2000; Liu \& Aitkin, 2008; Myung \& Pitt, 1997; Vanpaemel, 2010). That is, in Bayesian inference, researchers specify a priori assumptions about which parameter values they consider most likely before seeing the data. These priors can vary between experiments/research problems and even differ subjectively between different researchers, which will change the resulting evidence based on Bayes factors. Note that the dependency of Bayes factors on the prior goes beyond the dependency of the posterior on the prior.

Importantly, for most interesting problems and models, Bayes factors cannot be computed analytically. Instead, approximations are needed. One major approach is to estimate Bayes factors based on posterior MCMC draws (Betancourt, 2020a) via an algorithm termed bridge sampling (Bennett, 1976; Meng \& Wong, 1996), which is implemented in the R package \texttt{bridgesampling} (Gronau, Singmann, \& Wagenmakers, 2020). An alternative algorithm that we will discuss is the Savage--Dickey method (Dickey, Lientz, \& others, 1970). The approximate Bayes factor estimate may be unstable if insufficient MCMC draws are used (for the bridge sampling or the Savage--Dickey method), leading to different Bayes factors each time the analysis is performed (see Gronau et al., 2020). This sensitivity of the estimator to the particular Markov chain realization is also known as the variance of the estimator.

Even if the estimation of Bayes factors via bridge sampling yields stable results, it is still unclear whether the computations are accurate or biased for complex problems, i.e., whether the approximate Bayes factor estimate actually corresponds to the true Bayes factor. This stable error in the estimator is also known as the bias of the estimator. This potential bias is concerning, as - for realistic complex models - there are no guaranties that the Bayes factor estimates we obtain are correct. It is therefore crucial to calibrate Bayes factor estimates, which we do in the present work.

As a further important aspect, any variability that is present in the data will also impact the results from Bayes factor analyses. Any inferences and decisions will always depend on the particular details of observed data and there's no way around that. Accordingly, computing Bayes factors does not mean that we can obtain some abstract and reliable ``truth'' from some observed data, which is still sampled with considerable noise. Bayes factors - just like frequentist p-values or any quantification of evidence - can vary considerably between replications of the same experiment. Excessive variation is a common consequence of poor experimental design, which limits the conclusions that can be drawn from individual data sets (Oelrich, Ding, Magnusson, Vehtari, \& Villani, 2020). To avoid fragile discovery claims we need to ensure that testing based on Bayes factors is relatively stable across possible realizations of the data.

Last, we should not confuse inferences with decisions. Bayes factors provide inference on hypotheses. However, to obtain discrete decisions, such as to claim discovery, from continuous inferences in a principled way requires utility functions. Common decision heuristics (e.g., using Bayes factor larger than 10 as a discovery threshold) do not provide a principled way to perform decisions, but are merely heuristic conventions.
Indeed, simply selecting the hypothesis most compatible with the observed data does not need to result in useful outcomes. Frequentist null hypothesis significance testing, for example, bases testing not on inferences but rather on false discovery rates and true discovery rates, which are examples of \emph{utility functions}. To ensure that Bayes factors inform useful hypothesis tests, we need to define relevant utility functions and investigate the performance of Bayes factors in that context.

\begin{figure}

{\centering \includegraphics{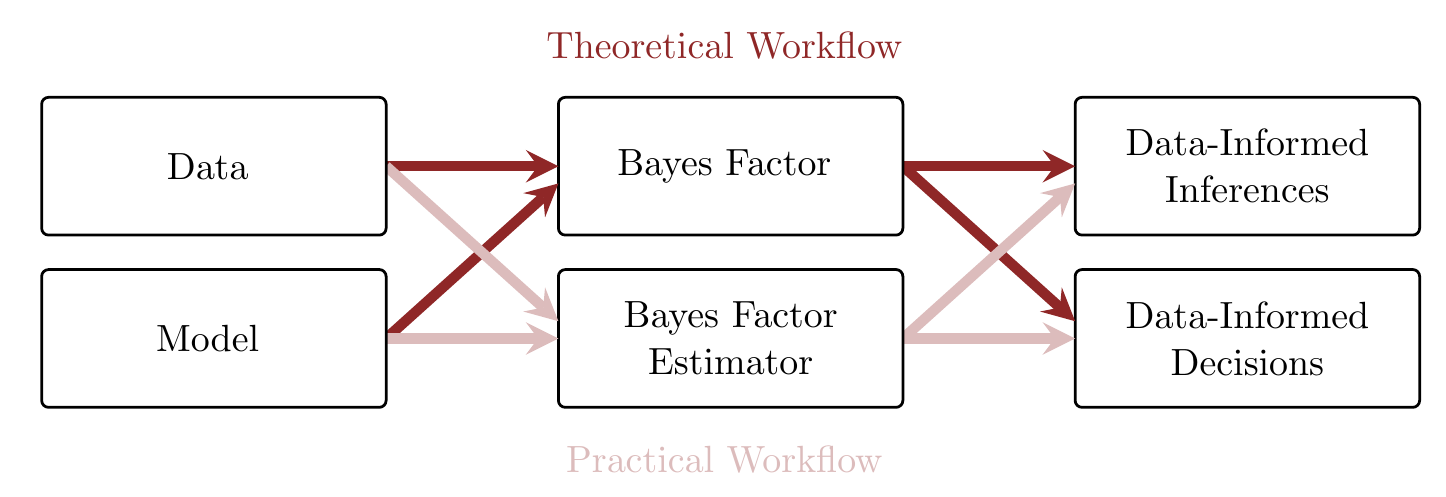} 

}

\caption{Shown are the schematic relations between the data and the model, Bayes factors, and resulting inferences and decisions. The data and the model constitute a true Bayes factor, that can be used for data informed inferences and decisions (dark red arrows). However, the true Bayes factor is unknown for complex models. Therefore, we use the data and the model to obtain an approximate Bayes factor estimator, and we use this for data informed inferences and decisions (light red arrows).}\label{fig:FigureBF}
\end{figure}

In this paper, we investigate these different aspects of the performance of Bayes factors (see Fig.~\ref{fig:FigureBF}). We investigate how Bayes factors are influenced by prior assumptions, we will investigate the stability of Bayes factors, i.e., how many MCMC draws are needed so that Bayes factor estimates won't change in different runs of the bridge sampling algorithm; we will study accuracy, i.e., whether the approximations are biased or correspond to the true Bayes factor; we will look at the variability of Bayes factors with artificial and real replications of empirical data; and we will look at decision-making based on Bayes factors using utility functions.

Note that in Bayesian approaches to data analysis in the cognitive sciences, other approaches than Bayes factors are sometimes used to investigate the viability of some hypotheses. For example, some researchers use the posterior of a fitted model to test whether the e.g., 95\% posterior credible interval for some critical parameter overlaps with zero, and treat this as a Bayesian hypothesis test. Other approaches compute the probability that a parameter is larger than zero. However, importantly, these approaches cannot really answer the question: How much evidence do we have in support for an effect at all (i.e., versus the hypothesis of no effect)? A 95\% credible interval that doesn't overlap with zero or a high probability that the parameter is positive may \emph{hint} that the predictor may be needed to explain the data, but they are not really answering this question how much evidence there is that the parameter is needed to explain the data (versus the null hypothesis that the parameter can be set e.g., to zero) (see Wagenmakers, Lee, Rouder, \& Morey, 2019; Rouder et al., 2018). This is a very important point. Indeed, this is often overlooked in the literature, and many papers use 95\% posterior credibility intervals to argue that there is evidence for or against an effect. This is a mistake that indeed the second and last authors of this paper made in the past (e.g., Nicenboim \& Vasishth, 2016; Jäger, Engelmann, \& Vasishth, 2017). However, the approach of using 95\% posterior credibility intervals to argue there is evidence for an effect is in fact not well defined. In this work, we introduce proper approaches to Bayesian inferences and decision making on hypotheses using Bayes factors, which allow us to explicitly quantify the evidence that the data provide for the hypothesis that a certain model parameter is needed to explain the data.

\hypertarget{a-quick-review-of-bayesian-methodology}{%
\subsection{A quick review of Bayesian methodology}\label{a-quick-review-of-bayesian-methodology}}

Statistical analyses in the cognitive sciences often pursue two goals: to estimate parameters and to test hypotheses. Both of these goals can be achieved using Bayesian data analysis. Bayesian approaches to data analysis focus on a model, which can range from a relatively simple statistical model, such as a linear regression or a multilevel (i.e., linear mixed-effects) model, to a complex non-linear model, such as a computational model of cognition. Indeed, when dealing with Bayes factors, this always implies a set of models, where a Bayes factor comprises a comparison of evidence between two models.
Critical for the model is that it specifies an ``observational'' model \(\mathcal{M}\), which is a mathematical function that specifies the probability density of the data \(y\) given the vector of model parameters \(\Theta\) and the model \(\mathcal{M}\). This is usually written as \(p(y \mid \Theta, \mathcal{M})\), or by dropping the model \(\mathcal{M}\) simply as \(p(y \mid \Theta)\). Since \(y\) is a free variable in the model, it is possible to use the observational model to simulate data, by selecting some model parameters \(\Theta\) and drawing random samples for the data \(\tilde{y}\). We use this approach heavily in our simulated data below. However, the model is also highly useful once we have collected (or simulated) some data and want to estimate parameters and make inferences. When the data is given (fixed), then the observational model turns into a likelihood function: \(p(y \mid \Theta) = L_y(\Theta)\), where the likelihood varies as a function of the model parameters \(\Theta\). This can be used to estimate model parameters or to compute evidence for the model relative to other models.

Let's consider an example, where for each of \(N\) subjects \(n\), we observe one data point \(y_n\) (e.g., the person's IQ). Let's assume in model \(\mathcal{M}_1\) that the data points follow a normal distribution. We can now describe the probability density\footnote{To be precise, note that the likelihood function is technically defined without any terms that don't depend on the parameters. Thus, technically \(\sqrt{2\pi}\) wouldn't be part of the likelihood function even though it's part of the observational model. These technicalities, however, don't affect our inferences and so here we write down the full observational model as the likelihood for simplicity.} for each observed data point \(y_n\) in subject \(n\) based on model parameters for the mean \(\mu\) and the standard deviation \(\sigma\) as:

\begin{equation}
p(y_n \mid \mu, \sigma, \mathcal{M}_1) = \frac{1}{\sigma \sqrt{2\pi}}e^{\frac{(\mu - y_n)^2}{-2\sigma^2}}
\end{equation}

This formula gives the likelihood for the data point \(y_n\) from one subject \(n\). However, we have data from multiple subjects. We assume that the data from the different subjects are conditionally independent from each other (given the parameters). This yields the following formula for the likelihood for the described simple linear model example: \(p(y \mid \mu, \sigma, \mathcal{M}_1) = \prod_n \frac{1}{\sigma \sqrt{2\pi}}e^{\frac{(\mu - y_n)^2}{-2\sigma^2}}\).

Based on this simple model, we can express different hypotheses to explain the data \(y\). For example, we could formulate the general hypothesis that the parameter \(\mu\) can take any possible value \(\mu \neq 0\), i.e., \(p(y \mid \mu \neq 0, \sigma, \mathcal{M}_1)\). However, this model can also be used to specify interval or point hypotheses. An example for a point hypothesis could be to postulate that the parameter \(\mu\) takes the value \(\mu = 100\).
This would yield the probability density: \(p(y \mid \mu=100, \sigma, \mathcal{M}_0) = \prod_n \frac{1}{\sigma \sqrt{2\pi}}e^{\frac{(100 - y_n)^2}{-2\sigma^2}}\). An example for an interval hypothesis could be that we assume that the parameter \(\mu\) is larger than \(100\), which we could specify as

\begin{equation}
p(y \mid \mu > 100, \sigma, \mathcal{M}_0) = \left\{ \begin{array}{ll} \prod_n \frac{1}{\sigma \sqrt{2\pi}}e^{\frac{(\mu - y_n)^2}{-2\sigma^2}} &, \mu > 100 \\
0 &, \mu \le 100 \end{array} \right.
\end{equation}

We will discuss below how Bayes factors can be used to quantify relative evidence for such different hypotheses.

In these models, one key goal is to estimate model parameters from data. In Bayesian data analysis inferences are constructed by complementing the likelihood with the prior model, written \(p(\Theta)\), that defines a probability distribution that encodes whatever domain expertise we want to incorporate into the analysis. From a strict Bayesian perspective the information encoded in the prior model should be independent from the observed data; this can be accomplished, for example, by specifying the prior model before making an observation but this is not always necessary. To inform prior distributions, it is often useful to rely on analyses of previous data sets, meta analyses, or on theoretical models.

Based on the likelihood and the prior, it is possible to compute the posterior distribution of the model parameters. The posterior distribution represents the results of inferences about which values of the model's parameters are most probable given the likelihood and the priors. The posterior is usually written as \(p(\Theta \mid y, \mathcal{M}_1)\) and represents posterior probability distributions specifying how likely each value of a model parameter is a posteriori, that is after seeing the data \(y\) and given the model \(\mathcal{M}_1\). Bayes' rule specifies how the posterior distributions \(p(\Theta \mid y, \mathcal{M}_1)\) can be computed by combining the prior \(p(\Theta \mid \mathcal{M}_1)\) with the likelihood \(p(y \mid \Theta, \mathcal{M}_1)\), reflecting updates of beliefs in the light of data:

\begin{equation}
p(\Theta \mid y, \mathcal{M}_1) = \frac{p(y \mid \Theta, \mathcal{M}_1) p(\Theta \mid \mathcal{M}_1)}{p(y \mid \mathcal{M}_1)} \label{eq:marginall}
\end{equation}

\noindent
Here, \(p(y \mid \mathcal{M}_1)\) is a normalizing constant termed the ``evidence'' or ``marginal likelihood'', which is the likelihood of the data based on the model independent of the parameters \(\Theta\), and is derived as \(p(y \mid \mathcal{M}_1) = \int p(y \mid \Theta, \mathcal{M}_1) p(\Theta \mid \mathcal{M}_1) d \Theta\). This quantity plays a central role in Bayesian model comparison via Bayes factors, as we will describe below.

Note that the marginal likelihood is a single number that tells you the likelihood of the observed data \(y\) given the model \(\mathcal{M}_1\) (and only in the discrete case, it tells you the probability of the observed data \(y\) given the model; in the continuous case, the probability for a specific data point is always zero, and the density for a single data point is evaluated instead). The marginal likelihood is not a function of the model parameters and the marginal likelihood does not depend on the model parameters \(\Theta\) any more; the parameters are ``marginalized'' or integrated out. Instead the marginal likelihood maps entire models to likelihood values. The likelihood is evaluated for all possible parameter values (according to the prior), weighted by the prior plausibility and summed together. For this reason, \emph{the prior here is as important as the likelihood}! The marginal likelihood itself is not particularly interpretable until we consider multiple models: it can only be interpreted relative to another marginal likelihood; we will illustrate this issue below.

Priors play a key role in the performance of Bayesian inference; in particular they can regularize inferences when the data do not inform the likelihood functions sufficiently strongly. We will see below, however, that they will influence marginal likelihoods and thus Bayes factors, and anything informed by Bayes factors, even when the data are strongly informative. Thus, priors are even more crucial for Bayes factors than for posterior distributions (Aitkin, 1991; Gelman et al., 2013; Grünwald, 2000; Liu \& Aitkin, 2008; Myung \& Pitt, 1997; Vanpaemel, 2010).

For very simple models, posterior density functions can be computed analytically, which then allows certain expectation values (e.g., the posterior mean) to be evaluated analytically as well. That is, mathematical formulas can be derived from the likelihood and the prior to obtain a closed form formula for the posterior densities. However, for most interesting models, e.g., for multilevel models, which we will deal with in the current paper, such closed-form analytical solutions are not available and we have to rely on methods that approximate posterior expectation values. An alternative approach to estimating the posterior is to use sampling methods such as Markov Chain Monte Carlo sampling, which is the method behind popular software implementing Bayesian analysis such as Stan (Carpenter et al., 2017), JAGS (Plummer \& others, 2003), WinBUGS (Lunn, Thomas, Best, \& Spiegelhalter, 2000), PYMC3 (Salvatier, Wiecki, \& Fonnesbeck, 2016), Turing (Ge, Xu, \& Ghahramani, 2018), and others. These methods allow us to obtain samples from the posterior distribution, which can be used to obtain approximate estimates for posterior expectations, such as the mean of the posterior distribution or the standard deviation.

\hypertarget{inference-and-discovery}{%
\section{Inference and discovery}\label{inference-and-discovery}}

\hypertarget{hypotheses}{%
\subsection{Hypotheses}\label{hypotheses}}

Three different kinds of hypotheses can be derived from an observational model: general hypotheses (full parameter range), point hypotheses (one specific parameter value), and interval hypotheses (interval of parameter within a model) (also see Betancourt, 2018).

A point hypothesis is defined by restricting one or more of the model parameters to specific values. The other model parameters, however, for example nuisance parameters, will generally be unconstrained. One example of a point hypothesis is that a model parameter is hypothesized to be zero. By contrast, in general hypotheses, all different values for the model parameter are possible. That is, it is hypothesized that the parameter exists, i.e., such as a parameter representing a difference between two experimental conditions, and that it takes some value, which can be estimated from the data. Sometimes, no constraints are put on the possible parameter values by using an improper uniform prior. At other times, some parameter values are considered more likely than others, but still, all values for the model parameter are possible in principle.

By contrast, interval hypotheses specify that a given model parameter is within a given interval or range. For example, an interval could involve the hypothesis that a parameter takes a positive value, and not a negative value. An alternative for an interval hypothesis could be that we specify one parameter to be bounded, e.g., that the parameter lies in the range between 0 and 1. Sometimes, an interval hypothesis can be used to capture the intent of a point hypothesis: i.e.~a parameter might be hypothesized to be very cloze to zero, e.g., between -0.1 and +0.1, such that it can be treated as being zero from a practical perspective (i.e., in a region of practical equivalence; ROPE; Kruschke, 2011; Freedman, Lowe, \& Macaskill, 1984; Spiegelhalter, Freedman, \& Parmar, 1994).

\begin{figure}

{\centering \includegraphics{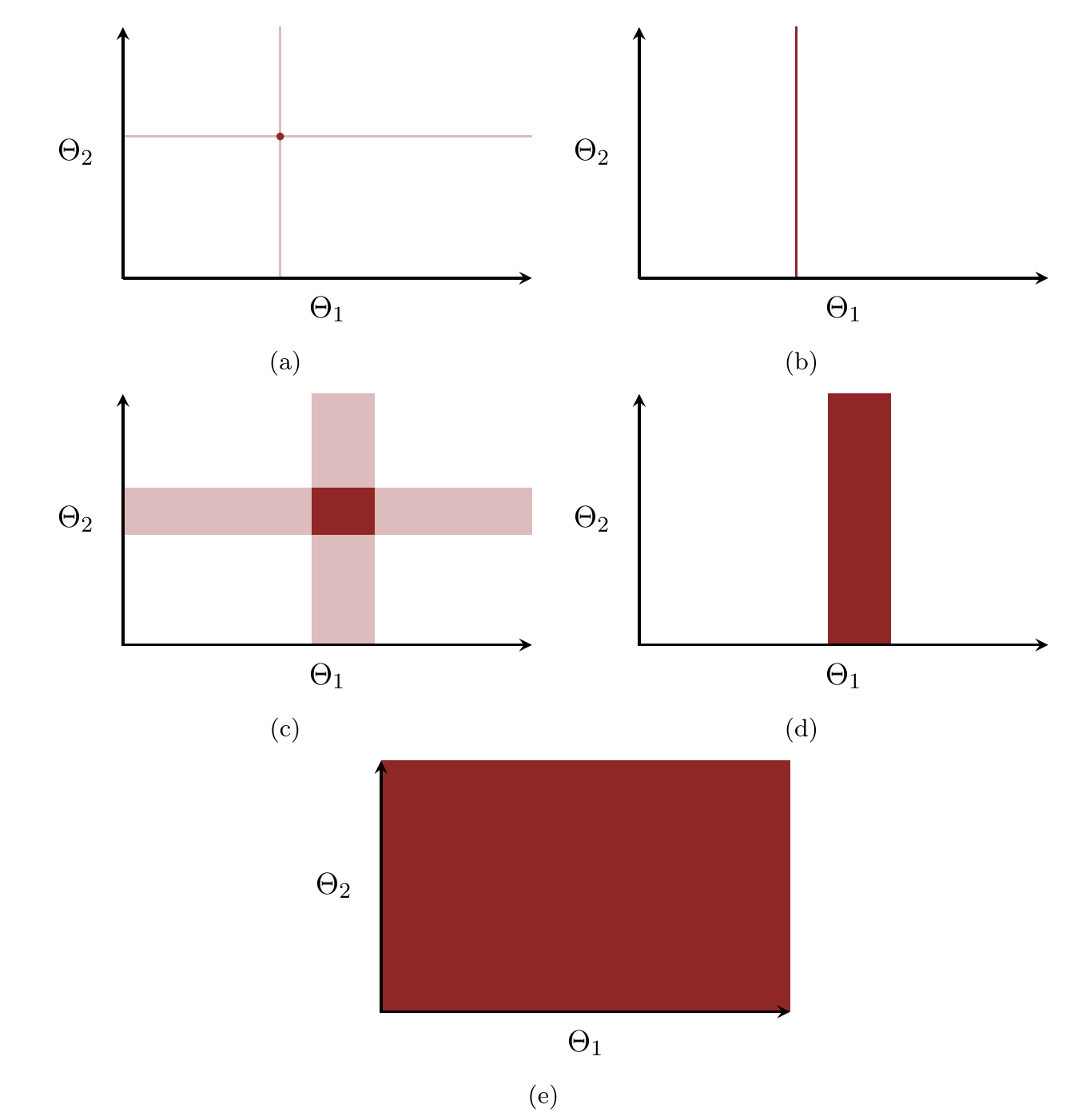} 

}

\caption{Illustration of different types of parameters for two parameters Theta 1 and Theta 2. (a) Point hypothesis in all parameters. (b) Point hypothesis in some parameters. (c) Interval hypothesis in all parameters. (d) Interval hypothesis in some parameters. (e) Full hypothesis.}\label{fig:FigureHyp}
\end{figure}

To illustrate, let's assume an observational model: \(p(y \mid \Theta)\) (e.g., a multilevel/linear mixed effects model). We can partition the model parameters as follows: \(\Theta = \{ \Theta_1, \Theta_2 \}\). That is, we assume the model parameters consist of two blocks, namely \(\Theta_1\) and \(\Theta_2\). For example, in our multilevel models \(\Theta_1\) could contain the fixed effect of interest (e.g., the regression coefficient associated with some predictor variable, e.g., cloze predictability), whereas \(\Theta_2\) may capture all other parameters (e.g., the intercept, random effects, and the residual variance). Based on this partition, we can distinguish a point hypothesis, an interval hypothesis, and a general hypothesis for \(\Theta_1\) (see Fig.~\ref{fig:FigureHyp}). In the point hypothesis, we assume that \(\Theta_1\) takes exactly one specific value, in our example zero: \(\Theta_1 = 0\), leading to the observational model \(p(y \mid \Theta_1 = 0, \Theta_2, \mathcal{M}_0)\). In the interval hypothesis, we assume that \(\Theta_1\) is not zero but takes some range of values, e.g., \(\Theta_1 > 0\), leading to the observational model \(p(y \mid \Theta_1 > 0, \Theta_2, \mathcal{M}_1)\). In the general hypothesis, we assume that \(\Theta_1\) can take any possible value, e.g., \(\Theta_1 \neq 0\), which leads to the observational model \(p(y \mid \Theta_1 \neq 0, \Theta_2, \mathcal{M}_2)\).

\hypertarget{inference-over-hypotheses}{%
\subsection{Inference over hypotheses}\label{inference-over-hypotheses}}

\hypertarget{comparing-two-point-hypotheses}{%
\subsubsection{Comparing two point hypotheses}\label{comparing-two-point-hypotheses}}

Point hypothesis tests are widely used in frequentist statistics. Specifically, frequentist statistics can be used to test the alternative hypothesis (i.e., model \(\mathcal{M}_1\)) that a true parameter value is different from zero by considering a point estimate for the parameter value. It chooses for all parameters the value that exhibits the largest value for the likelihood, that is, the maximum likelihood estimate (MLE) \(\{ \Theta_1=\hat{\Theta}_1, \Theta_2=\hat{\Theta}_2 \}\). Thus, note that while frequentist statistics aims to test a point null hypothesis against a general hypothesis (i.e., that the parameter is different from zero), in fact it reduces this to a comparison between two point hypotheses, by using the MLE for model comparison! Based on the MLE parameters, it is considered how compatible these parameters are with the data, i.e., \(p(y \mid \mathcal{M}_1) = p(y \mid \Theta_1=\hat{\Theta}_1, \Theta_2=\hat{\Theta}_2)\). In the likelihood ratio test, the MLE is compared to a second point hypothesis (\(\mathcal{M}_0\)), namely, that the critical parameter \(\Theta_1\) (e.g., a fixed effect) is zero: \(\Theta_1 = 0\). The other parameters (e.g., intercept or residual variance) are still assumed to be the MLE: \(\Theta_2=\hat{\Theta}_2\). From this, it is again possible to compute how likely the data are under this point parameter value, yielding a second likelihood value: \(p(y \mid \mathcal{M}_0) = p(y \mid \Theta_1=0, \Theta_2=\hat{\Theta}_2)\). In frequentist statistics, evidence for the alternative hypothesis (\(\mathcal{M}_1\)) over the null hypothesis (\(\mathcal{M}_0\)) is computed as the ratio in likelihoods:

\begin{align}
\mathcal{M}_0 &= \{\Theta_1 = 0, \Theta_2 = \hat{\Theta}_2\} \\
\mathcal{M}_1 &= \{\Theta_1 = \hat{\Theta}_1, \Theta_2 = \hat{\Theta}_2\} \\
LR &= \frac{p(y\mid \mathcal{M}_1)}{p(y\mid \mathcal{M}_0)} = \frac{p(y \mid \Theta_1=\hat{\Theta}_1, \Theta_2=\hat{\Theta}_2)}{p(y \mid \Theta_1=0, \Theta_2=\hat{\Theta}_2)}
\end{align}

Thus, the likelihood ratio test depends on the ``best'' estimate for the model parameter(s), that is, the model parameter \(\Theta\) occurs on the right side of the equation for each likelihood.
That means that in the likelihood ratio test, each model is tested on its ability to explain the data conditional on the ``best'' estimate for the model parameter (i.e., the MLE \(\hat{\Theta} = \{ \hat{\Theta}_1, \hat{\Theta}_2 \}\)). Thus, the likelihood ratio reduces interval hypotheses to point hypotheses. A likelihood ratio reduces an entire interval hypothesis/model to a single point hypothesis. Note that this reduction can be problematic.

Importantly, the comparison of point hypotheses completely depends on whether the point estimate for the model parameter(s) is representative of the possible values for the model parameter(s). If the point estimate is not representative, which is often the case in practical data analysis, where there is uncertainty about the precise parameter value, then comparing point hypotheses can be problematic.

Another related issue worth mentioning is that the likelihood ratio test introduces \emph{data-dependent hypotheses}. It is thus not comparing scientific hypotheses any more, but rather algorithmic hypotheses derived from the data. If the likelihood function is sufficiently narrow this might \emph{approximate} a well-defined hypothesis, but in general the difference can be large.

Bayesian analyses can also quantify relative evidence for two point hypotheses. In Bayesian analyses, this relative evidence can be obtained from within one single Bayesian model (\(\mathcal{M}_1\)). In this case, point hypothesis tests can be performed based on the ratio of posterior densities at the point parameter values. For example, one might compare evidence for the hypothesis that a critical parameter \(\Theta_1\) takes a value of e.g., \(\Theta_1 = 100\). Such a point hypothesis could be compared to the assumption that the parameter takes a value of zero (\(\Theta_1 = 0\)). Thus, to compute relative Bayesian evidence, one would take the estimated posterior density at the value of \(\Theta_1 = 100\) (\(p(\Theta_1=100 \mid y)\)) and the posterior density at a parameter value of zero (\(p(\Theta_1=0 \mid y)\)). Taking a ratio between these two posterior densities yields the relative evidence on the comparison of these two point hypotheses, i.e., the posterior evidence in favor of \(\Theta_1 = 100\) over \(\Theta_1 = 0\):

\begin{align}
\text{Posterior density ratio}&=\frac{\int d(\Theta_2) p(\Theta_1=100, \Theta_2 \mid y)}{\int d(\Theta_2) p(\Theta_1=0, \Theta_2 \mid y)}\\&= \frac{\int d(\Theta_2) p(y \mid \Theta_1=100, \Theta_2) \times p(\Theta_1=100, \Theta_2)}{\int d(\Theta_2) p(y \mid \Theta_1=0, \Theta_2) \times p(\Theta_1=0, \Theta_2)}
\end{align}

\begin{equation}
\text{Posterior density ratio} = \mathrm{likelihood\;ratio} \times \mathrm{ratio\;of\;prior\;densities}
\end{equation}

As we can see, the resulting ratio of posterior densities can be rewritten as a product of the ratio of likelihood functions and the ratio of prior densities. In other words the Bayesian comparison of point hypotheses reduces to the frequentist comparison, with a correction that takes into account the information in the prior model.

\hypertarget{comparing-two-interval-hypotheses}{%
\subsubsection{Comparing two interval hypotheses}\label{comparing-two-interval-hypotheses}}

An alternative type of hypotheses refers to intervals or ranges of parameters. I.e., these are cases where the hypothesis simply states that a free model parameter has a certain range of values, but where the precise parameter value is unknown.
As one example, let's assume the hypothesis that a parameter \(\Theta_1\) takes a positive value, \(H_1: \Theta_1 > 0\), is compared to a ROPE: \(H_2: -0.1 < \Theta_1 < 0.1\). In this case, the result is again a ratio of posterior probabilities:

\begin{equation}
Posterior\;ratio = \frac{p(\Theta_1>0 \mid y)}{p(-0.1 < \Theta_1 < 0.1 \mid y)}
\end{equation}

However, let's also look at a more specific example case, which is often of relevance in the cognitive sciences. Specifically, one could specify the hypothesis that a critical model parameter \(\Theta_1\) takes a positive value: \(H_1: \Theta_1 > 0\) and compare this to the hypothesis that the parameter value is zero or smaller: \(H_2: \Theta_1 \leq 0\). In this specific case, where both hypotheses together span the full range of possible parameter values, evidence for hypothesis \(H_1\) can be obtained by computing the posterior probability that the parameter is positive, i.e., \(p(\Theta_1>0 \mid y) = \int p(\Theta_1>0, \Theta_2 \mid y) d \Theta_2\).\footnote{Note that in certain special cases (e.g., with a symmetric prior centered around a point null hypothesis using Savage Dickey estimation), posterior probabilities are in fact Bayes factors.}
When using MCMC sampling to estimate the posterior, one can compute the posterior probability for the hypothesis by taking the proportion of samples that is larger than zero.

\hypertarget{bayes-factors-comparing-two-arbitrary-hypotheses}{%
\subsubsection{Bayes factors: Comparing two arbitrary hypotheses}\label{bayes-factors-comparing-two-arbitrary-hypotheses}}

Comparing more general hypotheses is hard: We can't compare densities to probabilities so we can't compare \emph{different kinds of hypotheses} with simple ratios as we did above. Instead, we need to reduce the posteriors with different parameter spaces to something compatible that can be compared; because all models share the same observational space this has to be the marginal likelihood, which is the basis for computing Bayes factors.

Bayes factors thus provide a way to compare any two model hypotheses against each other. This can e.g., involve comparison between two general hypotheses, or comparison between a general hypothesis and a point hypothesis, or any other comparison.

The Bayes factor tells us, given the data and the model priors, how much we need to update our relative belief between the two models. The Bayes factor is thus the ratio between posterior to prior odds.

To derive Bayes factors, we first compute the model posterior, i.e., the posterior probability for a model \(\mathcal{M}_i\) given the data: \(p(\mathcal{M}_i \mid y) = p(y \mid \mathcal{M}_i) \times P(\mathcal{M}_i)\). This involves the marginal likelihood for each model, that is the average probability density of the data given the model \(p(y \mid \mathcal{M}_i)\). This can be computed by taking integrals over the model parameters; that is, marginal likelihoods are averaged across all possible posterior values of the model parameter(s): \(p(y \mid \mathcal{M}_i) = \int p(y, \Theta \mid \mathcal{M}_i) d \Theta = \int p(y \mid \Theta, \mathcal{M}_i) p(\Theta \mid \mathcal{M}_i) d \Theta\).

Based on this posterior model probability \(p(\mathcal{M}_i \mid y)\), we can compute the model odds for one model over another as:

\begin{equation}
\frac{p(\mathcal{M}_1 \mid y)}{p(\mathcal{M}_2 \mid y)} = \frac{p(y \mid \mathcal{M}_1) \times p(\mathcal{M}_1)}{p(y \mid \mathcal{M}_2) \times p(\mathcal{M}_2)} = \frac{p(y \mid \mathcal{M}_1)}{p(y \mid \mathcal{M}_2)} \times \frac{p(\mathcal{M}_1)}{p(\mathcal{M}_2)} \label{eq:PostRatio}
\end{equation}

\begin{equation}
Posterior\;ratio = Bayes\;factor \times prior\;odds
\end{equation}

The Bayes factor is thus a measure of relative evidence, the comparison of the predictive performance of one model (\(\mathcal{M}_1\)) against another one (\(\mathcal{M}_2\)). This comparison (\(BF_{12}\)) is a ratio of marginal likelihoods:

\begin{equation}
BF_{12} = \frac{P(y \mid \mathcal{M}_1)}{P(y \mid \mathcal{M}_2)}
\end{equation}

\(BF_{12}\) indicates the evidence that the data provide for \(\mathcal{M}_1\) over \(\mathcal{M}_2\), or in other words, which of the two models is more likely to have generated the data, or the relative evidence that we have for \(\mathcal{M}_1\) over \(\mathcal{M}_2\). Under the assumption that all models are equally likely a priori, Bayes factor values larger than one indicate that \(\mathcal{M}_1\) is more compatible with the data, smaller than one indicate \(\mathcal{M}_2\) is more compatible with the data, and values close to one indicate that both models are equally compatible with the data.
Note that this model comparison does not depend on a specific parameter value. Instead, all possible prior parameter values are taken into account simultaneously.

Importantly, Bayes factors are a general way to compare models. When computing the Bayes factor between two point hypotheses, then the Bayes factor reduces to the ratio of posterior densities (after marginalizing out all other parameters not involved in the point hypothesis). When computing the Bayes factor for comparing two interval hypotheses, then the Bayes factor reduces to the ratio of posterior probabilities. Thus, Bayes factors are the general way of providing evidence for any hypothesis over another one in Bayesian data analysis.

Note that the marginal likelihood shares similarities to a quantity termed the prior predictive distribution. This addresses the important question how it is possible to make predictions and sample artificial data \(\tilde{y}\) from a Bayesian model \(\mathcal{M}\). This can be done based on the prior predictive distribution:

\begin{equation}
p(\tilde{y} \mid \mathcal{M}) = \int p(\tilde{y} \mid \Theta, \mathcal{M}) p(\Theta \mid \mathcal{M}) d \Theta
\end{equation}

or written differently:

\begin{align*}
\tilde{\Theta} &\sim \pi(\Theta)
\\
\tilde{y} &\sim \pi(y | \tilde{\Theta})
\end{align*}

Note that this prior predictive distribution averages predictions across the observational model \(p(\tilde{y} \mid \Theta, \mathcal{M})\) weighted by the prior \(p(\Theta \mid \mathcal{M})\). It is visible that the prior predictive distribution looks very similar to the marginal likelihoods.
Conceptually, in Bayes factor analyses, the model is specified with the priors, before seeing the data to be analyzed. Based on these priors and the observational model, it is possible to compute prior predictions (i.e., predictive densities) for observed data. These prior model predictions are then evaluated using the observed data to yield the support that the data give to the model.
In other words, the marginal likelihoods quantify how compatible the observations are with the \emph{prior} predictions.
The prior predictive distribution is highly sensitive to the priors because it evaluates the likelihood of the observed data under prior assumptions. Note that Bayes factor analyses always investigate prior predictions. This stands in contrast to posterior predictions usually evaluated using some kind of cross-validation. Both approaches are ``out-of-sample'', and are therefore valid approaches to investigating predictions.

Importantly, Bayes factors are even \emph{more} sensitive to prior assumptions than intra-model posterior distributions of the model parameters. The issue is that even if the posterior density of a model is hardly influenced by the prior assumptions (e.g., because there's enough data and a good experimental design), the marginal likelihoods and the Bayes factors can still be strongly influenced by the prior, because the models are compared under prior assumptions. Thus, defining priors is a central issue when using Bayes factors. Conceptually, the priors will determine how models will be compared.

In the present work, we will consider the case of nested model comparison, where a null model hypothesizes that a model parameter is zero or absent (a point hypothesis: \(p(\Theta_1 = 0 \mid y)\))\footnote{Note that the fact that we investigate Bayes factors for point null hypotheses doesn't mean we are advocating for point null hypotheses.}, whereas an alternative model hypothesizes that the model parameter is present and has some value different from zero that needs to be estimated from the data (a general hypothesis: \(p(\Theta_1 \neq 0 \mid y)\)). Bayes factors provide one way to generalize the likelihood ratio test beyond true point hypotheses. Note that Bayes factor analyses thus have the advantage (over frequentist analyses) that nuisance parameters (\(\Theta_2\)) can be integrated out.

\begin{equation}
BF_{10} = \frac{p(\Theta_1 \neq 0 \mid y)}{p(\Theta_1 = 0 \mid y)} = \frac{\int p(\Theta_1 \neq 0, \Theta_2 \mid y) d \Theta_2}{\int p(\Theta_1 = 0, \Theta_2 \mid y) d \Theta_2}
\end{equation}

Note, however, that Bayes factors do not only work for such nested hypotheses, but also extend to non-nested models.

For general hypotheses, Bayes factors provide the Bayesian way of quantifying evidence in favor of one model over another, where evidence can be written as \(p(y \mid \mathcal{M})\). Prior model probabilities \(p(\mathcal{M})\) reflect the probabilities of each of the models before seeing the data. Bayes factors allow us to compute the posterior probabilities of the models, i.e., \(p(\mathcal{M} \mid y)\), which reflect the probability of the model given the prior probabilities of the models and the data.
The interpretation of posterior probabilities relies on the assumption that the true model is contained within the observational model (this is often called the \(\mathcal{M}\)-closed assumption). Likewise the interpretation of posterior model probabilities assumes that the true model is one of the observational models being compared. If the true model is not any of the investigated models, the posterior cannot be interpreted as ``probability of truth''. Instead, Bayes factors quantify only how compatible each prior predictive distribution is with the observed data.

Bayes factors have important advantages over frequentist analyses. Bayes factors are immediately applicable to the comparison of any set of well-defined hypotheses, whereas frequentist comparisons often have to be developed bespoke for each particular comparison, and common frequentist methods limit one to only a few possible comparisons.
As we saw above, the common frequentist approach of using likelihood ratio tests to quantify evidence for competing hypotheses depends on the best parameter estimate (i.e., the MLE). If the best estimate for the model parameter(s) is not very representative of the possible values for the model parameter(s), then Bayes factors will be superior to the likelihood ratio test. Indeed, we can also reduce a Bayesian hypothesis test to just test single point values against each other; however, what is much better is to integrate over the parameter space before taking the ratio using Bayes factors.

Note that Bayes factors quantify Bayesian evidence when comparing two models with each other. However, posterior model probabilities can also be computed for the more general case, where two models or more than two models are considered:

\begin{equation}
p(\mathcal{M}_1 \mid y) = \frac{p(y \mid \mathcal{M}_1) p(\mathcal{M}_1)}{\sum_i p(y \mid \mathcal{M}_i) p(\mathcal{M}_i)}
\end{equation}

For simplicity, we here mostly constrain ourselves to two models. (Note that the prior sensitivity analyses we study below are comparing evidence between many models.)

\hypertarget{occams-razor}{%
\subsubsection{Occam's razor}\label{occams-razor}}

The marginal likelihoods can only be interpreted relative to another marginal likelihood (evaluated at the same \(y\)). Thus, we can only obtain \emph{relative} evidence for one model over another model, which is what the Bayes factor does, or over a set of other models. Thus, Bayes factors imply relative evidence.

\begin{figure}

{\centering \includegraphics{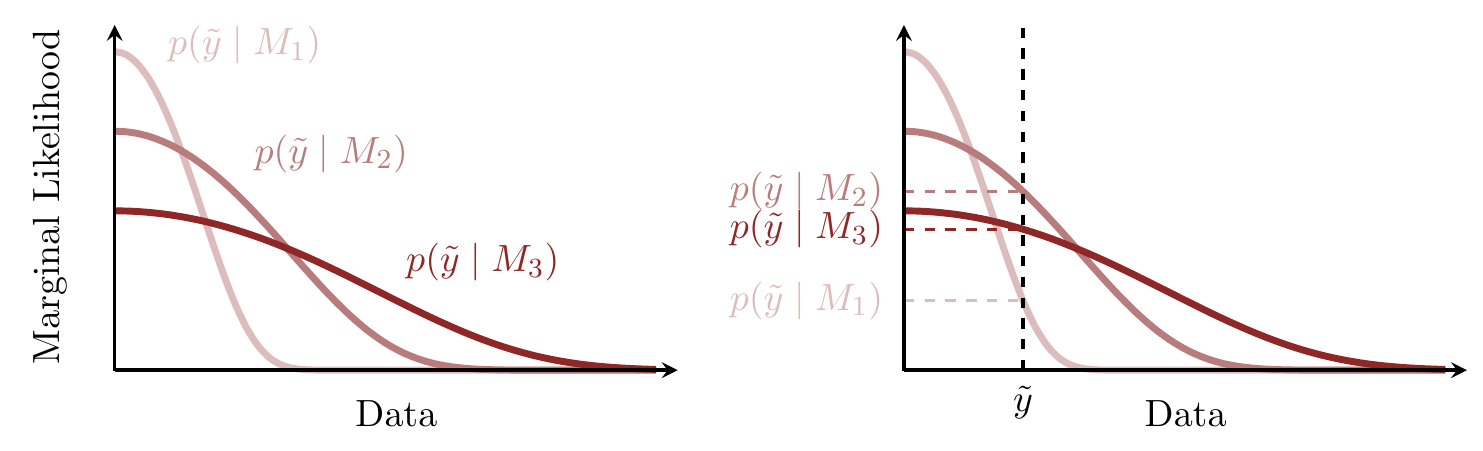} 

}

\caption{Shown are the schematic marginal likelihood functions, p(Data|Model), that each of three models assigns to different possible data (left panel) and evaluated marginal likelihoods at the data y (right panel). The total probability each model assigns to the data is equal to one, i.e., the areas under the curves of all three models are the same. Model 1 assigns all probability to a narrow range of data, and can predict this data with high probability density (low complexity model). Model 3 assigns its probability to a large range of different possible outcomes, but predicts each individual data with low probability density (high complexity model). Model 2 takes an intermediate position (intermediate complexity). Left panel: The vertical dashed line illustrates where the empirically observed data fall. The data most support model 2, since this model predicts the data with highest likelihood. In other words, Model 2 has enough complexity to capture the structure of the observed data. The figure follows Figure 3.13 in Bishop (2006).}\label{fig:OccamFactor}
\end{figure}

Importantly, one would prefer a model that gives a higher marginal likelihood, i.e., a higher likelihood of observing the data after integrating out the influence of the model parameter(s) (here: \(\Theta\)). A model will yield a high marginal likelihood if it makes a high proportion of good prior predictions (i.e., model 2 in Fig.~\ref{fig:OccamFactor}; Figure adapted from Bishop, 2006).
Models that are too flexible (Fig.~\ref{fig:OccamFactor}, model 3) will divide their prior predictive probability density across all of their predictions. They can predict many different outcomes. Thus, they likely can also predict the actually observed outcome. However, due to the normalization, they cannot predict it with high probability, because they also predict all kinds of other outcomes. This is true for both models with priors that are too wide or for models with too many parameters. Bayesian model comparison automatically penalizes such complex models, which is called ``Occam's razor''.

By contrast, good models (Fig.~\ref{fig:OccamFactor}, model 2) will make very specific predictions, where the specific predictions are consistent with the observed data. Here, all the predictive probability density is at the ``location'' where the observed data fall, and little probability density is located at other places, providing good support for the model. Of course, specific predictions can also be wrong, when expectations differ from what the observed data actually look like (Fig.~\ref{fig:OccamFactor}, model 1).

Note that having a natural Occam's razor is good for posterior inference, i.e., for assessing how much (continuous) evidence there is for one model or another. However, it doesn't necessarily imply good decision making or hypothesis testing, i.e., to make discrete decisions about which model explains the data best, or on which model to base further actions. We will discuss such discrete decisions further below (see section ``Selecting between hypotheses'').

\hypertarget{bayes-factor-scale}{%
\subsubsection{Bayes factor scale}\label{bayes-factor-scale}}

For the Bayes factor, a scale (see Table~\ref{tab:BFs}) has been proposed to interpret Bayes factors according to the strength of change of evidence in favor of one model (corresponding to some hypothesis) over another (Jeffreys, 1939); but this scale should not be regarded as a hard and fast rule with clear boundaries.

\begin{longtable}[]{@{}rl@{}}
\caption{\label{tab:BFs} Bayes factor scale as proposed by Jeffreys (1939)}\tabularnewline
\toprule
\begin{minipage}[b]{0.47\columnwidth}\raggedleft
\(BF_{12}\)\strut
\end{minipage} & \begin{minipage}[b]{0.47\columnwidth}\raggedright
Interpretation\strut
\end{minipage}\tabularnewline
\midrule
\endfirsthead
\toprule
\begin{minipage}[b]{0.47\columnwidth}\raggedleft
\(BF_{12}\)\strut
\end{minipage} & \begin{minipage}[b]{0.47\columnwidth}\raggedright
Interpretation\strut
\end{minipage}\tabularnewline
\midrule
\endhead
\begin{minipage}[t]{0.47\columnwidth}\raggedleft
\(>100\)\strut
\end{minipage} & \begin{minipage}[t]{0.47\columnwidth}\raggedright
Extreme change in evidence towards \(\mathcal{M}_1\).\strut
\end{minipage}\tabularnewline
\begin{minipage}[t]{0.47\columnwidth}\raggedleft
\(30-100\)\strut
\end{minipage} & \begin{minipage}[t]{0.47\columnwidth}\raggedright
Very strong change in evidence towards \(\mathcal{M}_1\).\strut
\end{minipage}\tabularnewline
\begin{minipage}[t]{0.47\columnwidth}\raggedleft
\(10-30\)\strut
\end{minipage} & \begin{minipage}[t]{0.47\columnwidth}\raggedright
Strong change in evidence towards \(\mathcal{M}_1\).\strut
\end{minipage}\tabularnewline
\begin{minipage}[t]{0.47\columnwidth}\raggedleft
\(3-10\)\strut
\end{minipage} & \begin{minipage}[t]{0.47\columnwidth}\raggedright
Moderate change in evidence towards \(\mathcal{M}_1\).\strut
\end{minipage}\tabularnewline
\begin{minipage}[t]{0.47\columnwidth}\raggedleft
\(1-3\)\strut
\end{minipage} & \begin{minipage}[t]{0.47\columnwidth}\raggedright
Anecdotal change in evidence towards \(\mathcal{M}_1\).\strut
\end{minipage}\tabularnewline
\begin{minipage}[t]{0.47\columnwidth}\raggedleft
\(1\)\strut
\end{minipage} & \begin{minipage}[t]{0.47\columnwidth}\raggedright
No change in evidence.\strut
\end{minipage}\tabularnewline
\begin{minipage}[t]{0.47\columnwidth}\raggedleft
\(\frac{1}{1}-\frac{1}{3}\)\strut
\end{minipage} & \begin{minipage}[t]{0.47\columnwidth}\raggedright
Anecdotal change in evidence towards \(\mathcal{M}_2\).\strut
\end{minipage}\tabularnewline
\begin{minipage}[t]{0.47\columnwidth}\raggedleft
\(\frac{1}{3}-\frac{1}{10}\)\strut
\end{minipage} & \begin{minipage}[t]{0.47\columnwidth}\raggedright
Moderate change in evidence towards \(\mathcal{M}_2\).\strut
\end{minipage}\tabularnewline
\begin{minipage}[t]{0.47\columnwidth}\raggedleft
\(\frac{1}{10}-\frac{1}{30}\)\strut
\end{minipage} & \begin{minipage}[t]{0.47\columnwidth}\raggedright
Strong change in evidence towards \(\mathcal{M}_2\).\strut
\end{minipage}\tabularnewline
\begin{minipage}[t]{0.47\columnwidth}\raggedleft
\(\frac{1}{30}-\frac{1}{100}\)\strut
\end{minipage} & \begin{minipage}[t]{0.47\columnwidth}\raggedright
Very strong change in evidence towards \(\mathcal{M}_2\).\strut
\end{minipage}\tabularnewline
\begin{minipage}[t]{0.47\columnwidth}\raggedleft
\(<\frac{1}{100}\)\strut
\end{minipage} & \begin{minipage}[t]{0.47\columnwidth}\raggedright
Extreme change in evidence towards \(\mathcal{M}_2\).\strut
\end{minipage}\tabularnewline
\bottomrule
\end{longtable}

\hypertarget{implementation-of-bayes-factors}{%
\subsubsection{Implementation of Bayes factors}\label{implementation-of-bayes-factors}}

One question now is how do we apply the Bayes Factor method to models that we care about, i.e., that represent more realistic data analysis situations that frequently occur in psycholinguistics, cognitive science, and other fields of research. In psycholinguistics and psychology, we typically fit fairly complex hierarchical models with many variance components. The major problem is that we won't be able to calculate the marginal likelihood for hierarchical models (or any other complex model) analytically. There are two very common methods for calculating the Bayes factor for complex models: the Savage--Dickey density ratio method (Dickey et al., 1970) and bridge sampling (Bennett, 1976; Meng \& Wong, 1996). The Savage--Dickey density ratio method is a straightforward way to compute a Bayes factor estimator, but it is limited to nested models. See Wagenmakers et al. (2010) for a complete tutorial. Note that the Savage--Dickey method can be unstable, especially in cases where the posterior is far away from zero. We will revisit this instability later.

Bridge sampling is a much more powerful method. This approach involves approximations of the marginal likelihoods. However, Bayes factor estimates based on bridge sampling can be unstable when based on models with too low effective sample size.\footnote{Posterior MCMC draws are correlated, and depending on the correlation a sample of a given size might contain more or less information. Therefore, ``effective sample size'' is corrected for the autocorrelation and provides an estimate of how much information is contained within the Markov chain relative to the number of independent samples (Vehtari et al., 2020).}
However, estimates of effective sample size are quantity specific (Betancourt, 2020a) and an effective sample size estimate for the posterior mean may not say anything about a potential effective sample size estimate for the bridge sampling estimate. So even high effective sample size for the (unnormalized) posterior density may not yield stable bridge sampling estimators. Instead, effective samples size may still be low for the (unnormalized) likelihood function.
Indeed, bridge sampling relies on posterior densities and requires many more (effective) posterior samples than what is normally required for parameter estimation; see Gronau et al. (2017b) for a general tutorial, and Gronau et al. (2020) for a tutorial using the R package \texttt{bridgesampling}.

Importantly, even when Bayes factor estimates based on bridge sampling are computed in a stable way (i.e., stability over different sets of MCMC draws), it is unclear, whether the estimates are unbiased for the kinds of (multilevel) models that we care about. Bridge sampling doesn't only have a problem with low effective sample size. To understand these problems, it is useful to discuss the typical set, which is the ``set containing the bulk of the posterior probability mass'' (Gabry, Simpson, Vehtari, Betancourt, \& Gelman, 2019, pp. 394--395).
MCMC explores the typical set and uses that exploration to estimate expectation values of functions of the parameters. When the algorithm enjoys a \emph{central limit theorem} that exploration is effective and the error in an estimator is determined by how much the variation of the corresponding function is contained within the typical set. Bayes factors, however, are given by the posterior expectation of the reciprocal likelihood function with usually varies most at extreme values far away from the typical set and even under ideal conditions the MCMC estimators for these expectations can suffer from large errors.
Therefore, calibrations (Betancourt, 2019) are needed to test whether Bayes factor estimates correspond to the true Bayes factor in a given application. We will discuss this issue and perform such calibrations below.

\hypertarget{selecting-between-hypotheses}{%
\subsection{Selecting between hypotheses}\label{selecting-between-hypotheses}}

Importantly, Bayes factors (and posterior model probabilities) tell how much evidence the data provide in favor of one model or another. That is, they allow us to perform inferences on the model space, i.e., to determine how much each hypothesis is consistent with the data.

Based on this evidence, it is also possible to perform decisions about selecting one hypothesis or the other, e.g., to declare discovery based on a Bayes factor analysis. Note however, that such discrete decision making is a completely different issue. Several heuristics have been proposed on how such decisions can be made. For example, Table~\ref{tab:BFs} shows how to put continuous evidence into discrete categories, and these categories could be used for decision-making. One common heuristic sometimes used in basic research is to treat Bayes factors that are larger than 10 (or smaller than 1/10) as a ground to declare discovery. Another heuristic that is often used in machine learning is to select the model with the highest posterior probability.

Importantly, these are just heuristics for deriving decisions, and they are not principled ways of how to derive decisions from evidence. A principled way to obtain decisions from evidence is to explicitly define utility functions. Utilities specify the values of possible actions (i.e., consequences of decisions) if certain hypotheses are in fact true. Thus, one could ask: what is the value of declaring discovery correctly or incorrectly? And what is the value of not declaring discovery correctly or incorrectly? Based on such reasoning about utilities, one can ask the question: which hypothesis should one choose to maximize utility?
For example frequentist null hypothesis significance testing (NHST) considers utilities in the form of the cost of false discoveries and the benefit of true discoveries, and then constructs a decision making process that bounds the worst case utility, at least when the assumptions hold.
Thus, while Bayes factors have a clear rationale and justification in terms of the (continuous) evidence they provide, utility functions are needed to map such evidence to actions, i.e., to perform decisions based on them.

\hypertarget{bayesian-decision-making-processes}{%
\subsubsection{Bayesian Decision Making Processes}\label{bayesian-decision-making-processes}}

To perform decisions in Bayesian analyses, the implementation of Bayesian decision making processes (Gelman et al., 2013; Robert, 2007) is necessary, which convert inferential information, such as the continuous Bayes factor or continuous posterior model probabilities, into discrete decisions. Bayesian inferences are continuous in nature and do not provide such discrete results.

However, there are two important caveats associated with discrete decisions: first, in practice, we often work with estimators of Bayes factors rather than with true Bayes factors (see Fig.~\ref{fig:FigureBF}). Such estimators can be noisy (we will illustrate this below). If the estimation error is not zero then the estimator will influence the decision making process in addition to the posterior distribution. This highlights that it is crucial to calibrate the Bayes factor estimator (we discuss this below) to make sure that the practical implementation of the Bayes factor estimator works appropriately.

As the second caveat, because the inferential information varies with observations (we will discuss this in detail below), so too will the decisions. Thus, random noise in the data can lead to very different inferences, and thus to very different decisions, simply based on chance.

\begin{figure}

{\centering \includegraphics{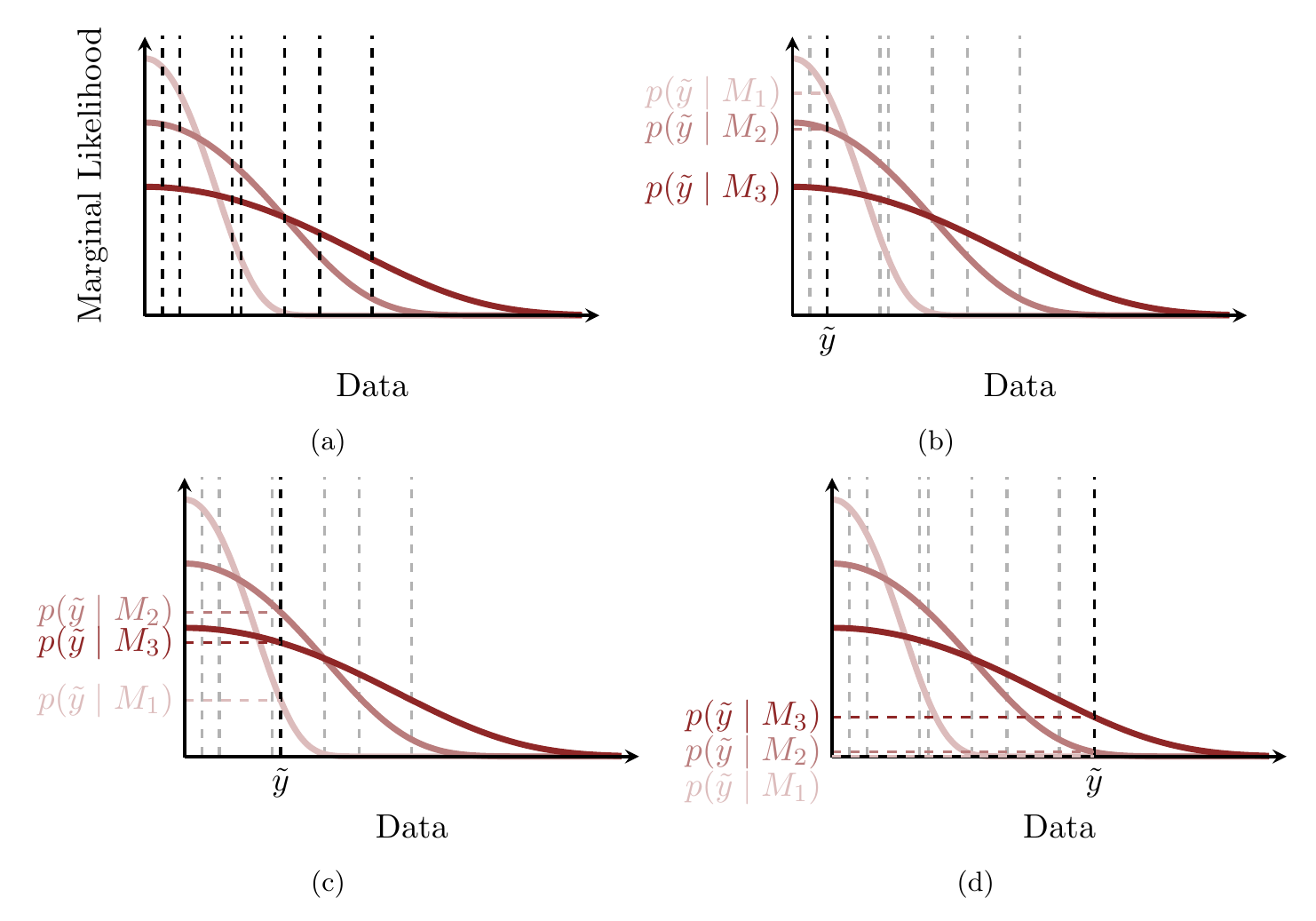} 

}

\caption{Shown are the schematic marginal likelihood functions, p(Data|Model), that each of three models assigns to different possible data. (a) The vertical dashed lines indicate artificial data sampled from the posterior of model 2. (b-c) The three panels illustrate three sampled data points. Although the data are simulated from model 2, some data sets still support model 1, model 2, and model 3.}\label{fig:OccamFactor2}
\end{figure}

Such random fluctuation in the data is illustrated in Figure~\ref{fig:OccamFactor2}. The Figure shows long vertical dashed lines (one in each panel), which illustrate data that is simulated based on model 2. It is clear that some of these simulated data points again support model 2. However, some simulated data points fall at higher or lower locations, and end up providing support for a different model, i.e., models 1 and 3. This illustrates that taking decisions on hypotheses based on observed data can be premature, and can lead to the kind of errors we have discussed in the previous section (e.g., deciding for a false model, be it too simple, or too complex, or simply different, that in fact did not generate the data). Here, we quantify the variability that is inherent in artificial and actual replications of the same data sets, and find that a single data set of conventional size from a fairly standard cognitive experimental design might not contain sufficient information to support clear inferences or even decisions on hypotheses.

Robust decision making requires sufficiently good experimental design to reduce the variation of the inferences, and hence the decisions, as much as possible. At the very least we have to quantify that variation to understand how stable a decision making process will actually be.
Indeed, because of the variation inherent in decisions, often making \emph{no} decision may actually be the best approach! If one just reports the inferences (i.e., Bayes factors), others can make their own decisions using their own utility functions in combination with the full information in the reported inference.

\hypertarget{utility-functions}{%
\subsubsection{Utility functions}\label{utility-functions}}

To perform decisions based on Bayesian analyses, utility functions are needed. The utility of different possible actions, that is, the value of the consequences when accepting and acting based on one hypothesis or another, can differ quite dramatically in different situations. For example, for a researcher trying to implement a life-saving therapy, falsely rejecting this new therapy could have high negative utility (negative utility is loss), whereas falsely adopting the new therapy may have little negative consequences. By contrast, falsely claiming a new discovery in fundamental research may have bad consequences (high loss), whereas falsely missing a new discovery claim may be less problematic if further evidence can be accumulated. Thus the performance of decision making procedures can be determined only in the context of utility functions appropriate to a given analysis.

A decision process has different possible outcomes, for which it is possible to assign different utilities. For example, in the cognitive sciences, when deciding to claim a discovery or not, different situations with different utilities can occur. First, if one claims a discovery based on a (Bayesian) decision-process, this can yield a true discovery (TD), which would have positive value, e.g., a utility of \(U_{TD} = 1\). However, a discovery claim can also be false (FD), yielding a possibly negative utility of \(U_{FD} = -1.5\). Second, an alternative outcome of a decision-making process is to not claim discovery, but to reject it. Again, this can be a true rejection (TR) of a discovery, which may have positive utility (e.g., \(U_{TR} = 0.5\)). However, the rejection of a discovery can also be false (i.e., missing a true new discovery), which might have a negative utility (e.g., of \(U_{FR} = -0.5\)). Note that the utilities that we chose here are arbitrary, and other values could be chosen as well.
In the cognitive sciences, decision making might, in general, be premature. If we can't construct useful utilities then we probably shouldn't be trying to make decisions. Reporting inferences directly and avoiding discovery claims avoids having to worry about utility functions.

One research goal in the cognitive science would be to develop a procedure of how such utilities can be conceived in a way that is not arbitrary, but theoretically motivated. If such utilities are available, this can support Bayesian decision making. We will illustrate this in an example below.

\hypertarget{calibration-methods}{%
\subsubsection{Calibration methods}\label{calibration-methods}}

Decisions based on Bayes factors, and the estimation error of Bayes factors itself, can vary with the observed data. Therefore, we need to quantify that data variation, or calibrate the Bayes factor method relative to the assumed model, if we want to use this method responsibly. Conveniently we can implement this calibration by observing how the Bayes factor outcomes vary across prior predictive simulations.

Calibrations over different data sets (Betancourt, 2019) are thus needed to investigate the properties of Bayes factor estimates (marginal likelihoods), i.e., to test whether Bayes factor estimates correspond to the true Bayes factor for a given study. They are also needed to understand the properties of Bayesian decision-making procedures.

To investigate Bayes factor estimates, in simulation-based calibrations (SBC), one can simulate data based on a data generating process by sampling artificial data from several observational models.
What we do here is that in each simulation run, we simulate data from one of several different models, where the probability for each model is specified as a prior across model space.
Then, it is possible to estimate marginal likelihoods based on the simulated data, which can then be used to estimate Bayes factors and posterior model probabilities. When this is done many times, then it is possible to test whether the posterior model probabilities on average correspond to the true data generating process. Moreover, it is possible to check whether on average the inference that Bayes factors support is correct.

In the previous section, we have introduced utility functions to quantify the values of actions taken based on decision-making processes. However, decision making procedures will vary depending on the data, and may perform well or badly (in terms of utilities) depending on what the data look like. The problem of course is that before running a study we do not know what the data looks like and what the possible outcomes of a study will be. Therefore, we need to quantify how those utilities can vary across different possible data sets. To determine this is the goal of calibration studies (Betancourt, 2018). These can be implemented using artificial data simulations, where we simulate data based on some priors and models, where we thus know which model or hypothesis was true in the data simulation. Then we can run a Bayesian decision-procedure on the simulated data and summarize the results in terms of their average utilities. For example, we can summarize false positives with false positive \emph{rates} that quantify how often an observation informs a false positive decision, and we can compute the utilities associated with such false positive rates.

For example, we can look at simulated data sets where the true model (i.e., the model sampled from the model prior) corresponds to a null hypothesis (H0), perform decisions based on the Bayesian evidence, and obtain the false discovery rate (FDR), i.e., how often the Bayes factor supports an alternative model (H1) when in fact the null hypothesis (H0) is true. This is a Bayesian equivalent of frequentist type I (\(\alpha\)) errors. Likewise, we can look at the simulated data sets where the true model (i.e., the model sampled from the prior) corresponds to an alternative hypothesis (H1), compute Bayesian evidence, perform decisions, and obtain the true discovery rate (TDR), i.e., the probability to choose the alternative hypothesis (H1) when it is actually true. This is a Bayesian equivalent to frequentist power analyses. When combining decisions with utilities, we can then obtain the average utility of a given decision-rule.

We will perform calibration studies here to investigate the accuracy of Bayes factor estimates and to investigate utilities of Bayesian decision-making procedures.

\hypertarget{simulation-based-calibration-and-calibrating-decisions}{%
\subsubsection{Simulation-based calibration and calibrating decisions}\label{simulation-based-calibration-and-calibrating-decisions}}

An important point about approximate computations of Bayes factor estimates using bridge sampling is that there are no strong guarantees for their accuracy. That is, even if we can show that the approximated Bayes factor estimate using bridge sampling is stable across different MCMC draws and across different starting values for the bridge sampling, even then it remains unclear how close the approximated Bayes factor is to the true Bayes factor. Bridge sampling is a form of density estimation. Technically, bridge sampling estimators can be written as a product of expectation values, although those expectation values are particularly hard to estimate with MCMC. In principle, it could very well be that the stably estimated Bayes factors based on bridge sampling are in fact biased, i.e., that they do not yield the correct (true) Bayes factor, but some biased approximation to it. The technique of simulation-based calibration (SBC; Talts, Betancourt, Simpson, Vehtari, \& Gelman, 2018; Betancourt, 2020b; Schad, Betancourt, \& Vasishth, 2021) can be used to investigate this question.

In SBC, the priors are used to simulate artificial data. Then, posterior inference is done on the artificial, simulated data, and the data-averaged posterior can be compared to the prior.
Any differences between the average posterior and the prior are due to errors in the computation and thus indicate a problem with inference.
By contrast, if the data-averaged posterior is equal to the prior, then this is consistent with accurate computations (caution: this consistency condition holds only for the average posterior over prior predictive simulations; we have no guarantees on how any individual posterior distribution will behave in these simulations, let alone for observed data; thus, this statement does not apply to Bayesian inference on a single data set, where a prior is used to infer a posterior distribution, but is specific to SBC). We can formulate SBC for model inference, where \(\mathcal{M}\) is a true model used to simulate artificial data \(y\), and \(\mathcal{M}'\) is a model inferred from the simulated data.

\begin{equation}
p(\mathcal{M}') = \int \int p(\mathcal{M}' \mid y) p(y \mid \mathcal{M}) p(\mathcal{M}) \; \mathrm{d} y \mathrm{d} \mathcal{M} \label{eq:SBC}
\end{equation}

Critically if SBC does not show a difference between the average posterior (i.e., the left-hand side of equation~\eqref{eq:SBC}) and the prior, then this doesn't guarantee that the computation for every posterior will necessarily be good; it is a necessary condition but not a sufficient one.

Applied to Bayes factor analyses, we define a prior on the model space, e.g., we can define the prior probabilities for a null and an alternative model, specifying how likely each model is a priori. From these priors, we can randomly draw one hypothesis (model), e.g., \(n_{sim} = 500\) times. Thus, in each of \(500\) draws we randomly choose one model (either H0 or H1), with the probabilities given by the model priors. For each draw, we first sample model parameters from their prior distributions, and then use these sampled model parameters to simulate artificial data. For each simulated artificial data set, we can then compute marginal likelihoods and Bayes factors (between the models H1 and H0) using bridge sampling, and we can then compute the posterior probabilities for each hypothesis using the true prior model probabilities (i.e., how likely each model is a posteriori). As the last, and critical step in SBC, we can then compare the posterior model probabilities to the prior model probabilities. A key result in SBC is that if the computation of marginal likelihoods and model posteriors is performed accurately by the bridge sampling procedure, i.e., without bias, that is, if the approximate Bayes factor estimate corresponds to the true Bayes factor, then the data-averaged posterior model probabilities should be the same as the prior model probabilities. We show the concrete steps of simulation-based calibration in an example R analysis below.

Conveniently those same simulations can also be used to calibrate inferences, such as how variable the Bayes factor is, or decision making processes, such as selection between models with the Bayes factor.
Thus, one can ask how sensitive the Bayes factor is in detecting the more appropriate model given the data. Moreover, the same simulations from the SBC can also be used to determine true discovery rates (TDR) and false discovery rates (FDR).
Based on these same simulations, it is also possible to calibrate any decision making process based on Bayes factors. That is, for the specific set of simulations from the SBC, one can specify utilities for different actions.
The immediate heuristic for turning Bayes factors into decisions is a threshold (e.g., of \(BF_{10} = 10\)). However, the specific threshold value has no canonical value. By calibrating the consequences of different threshold values, however, one can identify the threshold value best suited to a particular analysis.
Thus, the calibrations allow to determine the overall utility as a function of the threshold value used to determine the decisions. Then, it is possible to optimize the threshold value (or cut-off criteria) to yield optimal total utility. This procedure may even compensate for experimental design limitations. We will illustrate an example analysis for this below.

\hypertarget{bayes-factor-workflow}{%
\section{Bayes factor workflow}\label{bayes-factor-workflow}}

In the coming section on ``Misbehaving Bayes factors'', we discuss various potential problems associated with Bayes factor analyses. We outline a Bayes factor workflow to investigate these potential problems for a concrete analysis. These problems can largely be investigated using one set of artificial data simulations in the context of simulation-based calibration (SBC; Talts et al., 2018; Betancourt, 2020b; Schad et al., 2021) - we will discuss what this is below. Consequently we can integrate a set of analyses (that we illustrate below) into a coherent workflow for determining when Bayes factors are robust in any given application. For this workflow, we define the following steps:

\begin{enumerate}
  \item Define the observational model
  \item Define the prior (prior model probabilities and prior parameter distributions), ideally verified with prior pushforward and prior predictive checks
  \item Fit the model and estimate Bayes factors using bridge sampling on the same empirical data set multiple times (at least twice) to investigate whether the number of MCMC draws are sufficient to obtain stable Bayes factor estimates
  \item Run SBC to check whether Bayes factors are computed accurately
  \item Use simulations to investigate data variability of Bayesian inferences to support realistic expectations concerning their reliability
  \item If SBC supports accurate and reliable Bayes factor estimation, then one can use the Bayes factors obtained for the empirical data to support Bayesian inferences, otherwise, improve experimental design or acknowledge limitations
\end{enumerate}

In many cases, having valid Bayes factor estimates will be sufficient, since an important goal in cognitive science is to provide evidence in support of scientific hypotheses. This evidence is continuous in nature and thus reporting continuous evidence in scientific papers, without making discrete decisions, would be a natural approach. This is especially important given the large data variability inherent in evidence quantification (see below for illustrations), which often makes discrete decisions seem premature based on individual data sets. However, if discrete decisions are needed, for example, in order to make a discrete discovery claim, then the workflow can be expanded with the following steps:

\begin{enumerate}
  \item If one wants to make a decision, e.g., on a discovery claim, then one can define utility functions, i.e., the utility for each action given each truth
  \item Use simulations to optimize the discovery threshold
  \item Use simulations to investigate data variability of decisions (false and true discovery rates)
  \item Make a decision on discovery using an optimized discovery threshold
\end{enumerate}

We consider this workflow, in particular conducting SBC, to be the ideal way to approach Bayes factor analyses. However, we acknowledge that it takes a lot of time and computational resources to run this workflow for realistic research problems. It may therefore be difficult in research practice to implement this ideal workflow in every single analysis that one runs. We therefore suggest to implement this workflow once for a given research program, were different models and experimental designs may be similar to each other.

Based on this definition of the Bayes factor workflow, we now discuss in detail the problems and questions that motivate the workflow.

\hypertarget{misbehaving-bayes-factors}{%
\section{Misbehaving Bayes Factors}\label{misbehaving-bayes-factors}}

Bayes factors are a useful tool for quantifying relative Bayesian support for different models of the data, and they can be used to derive decisions based on the Bayesian evidence. However, there are several problems associated with Bayes factor analyses. First, Bayes factor estimates can exhibit estimation error because they are unstable against MCMC draws and because the estimation is not accurate (for other reasons not related to imprecision caused by finite number of MCMC draws) and does not correspond to the true value. Second, Bayes factor estimates - as any other form of evidence quantification - strongly depends on the particular data set, and can thus strongly vary with noise in the data. Third, Bayes factors can support poor decision-making, either because simple decision heuristics perform badly with respect to relevant utility functions, or because the data variability of Bayes factors leads to highly variable decision-outcomes. In the following, we will discuss these issues in detail. Moreover, we use the analysis of these difficulties to formulate a Bayes factor workflow that can be used to validate robust inference for specific data sets.

\hypertarget{EstimationError}{%
\subsection{Estimation error}\label{EstimationError}}

Two questions that we investigate here are how stable estimates of Bayes factors are when they are computed from different MCMC chains and with different starting values for the bridge sampler, and how accurate the estimates of Bayes factors are relative to the true Bayes factor.

\hypertarget{SBC1}{%
\subsubsection{Simulation-based calibration: Recovering the prior from the data}\label{SBC1}}

An important point about approximate computations of Bayes factor estimates (using bridge sampling) is that we do not know whether Bayes factor estimates are unbiased, i.e., whether the estimates correspond to the true Bayes factor. Here, we use the technique of simulation-based calibration (SBC; Talts et al., 2018; Betancourt, 2020b; Schad et al., 2021) to investigate this question, and we perform one example analysis in R.

First, we create an (artificial) experimental design. We use the R package \texttt{designr} (Rabe, Kliegl, \& Schad, 2021) to create the experimental design with a within-subject factor \texttt{x} with two levels (using sum coding with -1 and +1) and 15 subjects. Each condition (-1/+1) is measured twice per subject (this is what the \texttt{replications=2} argument does).

\begin{Shaded}
\begin{Highlighting}[]
\NormalTok{design \textless{}{-}}\StringTok{ }\KeywordTok{fixed.factor}\NormalTok{(}\StringTok{"x"}\NormalTok{, }\DataTypeTok{levels=}\KeywordTok{c}\NormalTok{(}\StringTok{"{-}1"}\NormalTok{, }\StringTok{"1"}\NormalTok{), }\DataTypeTok{replications=}\DecValTok{2}\NormalTok{) }\OperatorTok{+}
\StringTok{          }\KeywordTok{random.factor}\NormalTok{(}\StringTok{"subj"}\NormalTok{, }\DataTypeTok{instances=}\DecValTok{15}\NormalTok{)}
\NormalTok{simdata \textless{}{-}}\StringTok{ }\KeywordTok{design.codes}\NormalTok{(design)}
\NormalTok{simdata}\OperatorTok{$}\NormalTok{x \textless{}{-}}\StringTok{ }\KeywordTok{as.numeric}\NormalTok{(}\KeywordTok{as.character}\NormalTok{(simdata}\OperatorTok{$}\NormalTok{x))}
\end{Highlighting}
\end{Shaded}

We assume that our dependent variable are response times in milliseconds, and we assume that response times are log-normally distributed.

To explain response times in this experimental design, we aim to test two distinct hypotheses, which are implemented in two different hierarchical (linear mixed-effects) models. The alternative hypothesis (H1) assumes that factor \texttt{x} influences the dependent variable, i.e., that the fixed effects estimate associated with factor \texttt{x}, \(\beta_1\), takes some value that is different from zero \(H_1: \beta_1 \neq 0\). In R, the corresponding model formula can be written as: \texttt{log(rt)\ \textasciitilde{}\ 1\ +\ x\ +\ (1\ +\ x\ \textbar{}\ subj)}. By contrast, the null hypothesis (H0) assumes that factor \texttt{x} does not influence the dependent variable response times, i.e., \(H_0: \beta_1 = 0\). In R, the corresponding model formula can be written as: \texttt{log(rt)\ \textasciitilde{}\ 1\ +\ (1\ +\ x\ \textbar{}\ subj)}. To compare this general hypothesis H1 to the point hypothesis H0, we will use Bayes factors.

The next step in SBC is to define the prior model probabilities. For simplicity, we assume that both hypotheses (H0 and H1) are both equally likely a priori, which also has the advantage that both hypotheses are equally frequently sampled in the SBC. (However, see Schad \& Vasishth, 2019, for a different prior with higher probability for the null.)

\begin{Shaded}
\begin{Highlighting}[]
\NormalTok{priorsHypothesis \textless{}{-}}\StringTok{ }\KeywordTok{c}\NormalTok{(}\DataTypeTok{H0 =} \FloatTok{0.5}\NormalTok{, }\DataTypeTok{H1 =} \FloatTok{0.5}\NormalTok{)}
\end{Highlighting}
\end{Shaded}

Moreover, we define hypothetical priors for the model parameters. Note that we assume the dependent variable response times to be log-normally distributed; the priors are thus defined in this log-normal distribution model. They can be interpreted as the priors for a linear mixed-effects model on log-transformed response times. Specifically, for the intercept we assume a normal distribution with mean \(6\) and standard deviation \(0.5\). Note that a prior mean for the intercept of \(6\) reflects the a priori expectation that response times are an average of \texttt{exp(6)\ =\ 403} ms. For the fixed effect estimate for factor \texttt{x} (i.e., \texttt{b}), we assume a normal distribution with mean \(0\) and standard deviation of \(1.0\). For the random effects standard deviations, we assume a half normal distribution with mean \(0\) and standard deviation of \(1.5\), which is truncated to take only positive values. For the residual noise term, we assume a normal distribution with mean \(0\) and standard deviation of \(0.5\), which is again truncated to take only positive values. For the random effects correlation between the intercept and the estimate for \texttt{x}, we assume an LKJ prior (Lewandowski, Kurowicka, \& Joe, 2009) with parameter value \(2\). We write these priors in \texttt{brms} (Bürkner, 2017, 2018):

\begin{Shaded}
\begin{Highlighting}[]
\NormalTok{priors \textless{}{-}}\StringTok{ }\KeywordTok{c}\NormalTok{(}\KeywordTok{set\_prior}\NormalTok{(}\StringTok{"normal(6, 0.5)"}\NormalTok{, }\DataTypeTok{class =} \StringTok{"Intercept"}\NormalTok{),}
            \KeywordTok{set\_prior}\NormalTok{(}\StringTok{"normal(0, 1.0)"}\NormalTok{, }\DataTypeTok{class =} \StringTok{"b"}\NormalTok{),}
            \KeywordTok{set\_prior}\NormalTok{(}\StringTok{"normal(0, 1.5)"}\NormalTok{, }\DataTypeTok{class =} \StringTok{"sd"}\NormalTok{),}
            \KeywordTok{set\_prior}\NormalTok{(}\StringTok{"normal(0, 0.5)"}\NormalTok{, }\DataTypeTok{class =} \StringTok{"sigma"}\NormalTok{),}
            \KeywordTok{set\_prior}\NormalTok{(}\StringTok{"lkj(2)"}\NormalTok{, }\DataTypeTok{class =} \StringTok{"cor"}\NormalTok{))}
\end{Highlighting}
\end{Shaded}

Based on these priors, it is now possible to simulate a priori data for the artificial experimental design. First, we use the prior probabilities for the hypotheses to sample a hypothesis from the prior. We do so 500 times (i.e., 500 runs of SBC).

\begin{Shaded}
\begin{Highlighting}[]
\NormalTok{nsim \textless{}{-}}\StringTok{ }\DecValTok{500}
\NormalTok{u \textless{}{-}}\StringTok{ }\KeywordTok{runif}\NormalTok{(nsim)}
\NormalTok{hypothesis\_samples \textless{}{-}}\StringTok{ }\NormalTok{(u }\OperatorTok{\textgreater{}}\StringTok{ }\NormalTok{priorsHypothesis[}\DecValTok{1}\NormalTok{])}\OperatorTok{/}\KeywordTok{sum}\NormalTok{(priorsHypothesis)}
\end{Highlighting}
\end{Shaded}

\begin{Shaded}
\begin{Highlighting}[]
\KeywordTok{table}\NormalTok{(hypothesis\_samples)}
\end{Highlighting}
\end{Shaded}

\begin{verbatim}
## hypothesis_samples
##   0   1 
## 245 255
\end{verbatim}

We see that the H0 and the H1 are each sampled approximately 250 times. We will perform a formal SBC analysis below.

Next, we sample model parameters from the priors based on the model that was sampled in each run. For this, we use the custom R function \texttt{SimFromPrior()} {[}taken from Schad et al. (2021); \url{https://osf.io/b2vx9/}{]}. First, we choose the alternative hypothesis (H1) to sample values for the model parameters, i.e., to sample parameters from their prior distributions.

\begin{Shaded}
\begin{Highlighting}[]
\NormalTok{beta0 \textless{}{-}}\StringTok{ }\NormalTok{beta1 \textless{}{-}}\StringTok{ }\NormalTok{sigma\_u0 \textless{}{-}}\StringTok{ }\NormalTok{sigma\_u1 \textless{}{-}}\StringTok{ }\NormalTok{rho\_u \textless{}{-}}\StringTok{ }\NormalTok{sigma \textless{}{-}}\StringTok{ }\OtherTok{NA}
\KeywordTok{set.seed}\NormalTok{(}\DecValTok{123}\NormalTok{)}
\ControlFlowTok{for}\NormalTok{ (i }\ControlFlowTok{in} \DecValTok{1}\OperatorTok{:}\NormalTok{nsim) \{}
\NormalTok{  tmp \textless{}{-}}\StringTok{ }\DecValTok{{-}1}\NormalTok{; }\ControlFlowTok{while}\NormalTok{ (tmp}\OperatorTok{\textless{}}\DecValTok{0}\NormalTok{) }\CommentTok{\# sample from a half{-}normal distribution}
\NormalTok{    tmp \textless{}{-}}\StringTok{ }\KeywordTok{SimFromPrior}\NormalTok{(priors,}\DataTypeTok{class=}\StringTok{"Intercept"}\NormalTok{,}\DataTypeTok{coef=}\StringTok{""}\NormalTok{)}
\NormalTok{  beta0[i]    \textless{}{-}}\StringTok{ }\NormalTok{tmp}
\NormalTok{  beta1[i]    \textless{}{-}}\StringTok{ }\KeywordTok{SimFromPrior}\NormalTok{(priors,}\DataTypeTok{class=}\StringTok{"b"}\NormalTok{)}
\NormalTok{  sigma\_u0[i] \textless{}{-}}\StringTok{ }\KeywordTok{SimFromPrior}\NormalTok{(priors,}\DataTypeTok{class=}\StringTok{"sd"}\NormalTok{)}
\NormalTok{  sigma\_u1[i] \textless{}{-}}\StringTok{ }\KeywordTok{SimFromPrior}\NormalTok{(priors,}\DataTypeTok{class=}\StringTok{"sd"}\NormalTok{)}
\NormalTok{  rho\_u[i]    \textless{}{-}}\StringTok{ }\KeywordTok{SimFromPrior}\NormalTok{(priors,}\DataTypeTok{class=}\StringTok{"cor"}\NormalTok{)}
\NormalTok{  sigma[i]    \textless{}{-}}\StringTok{ }\KeywordTok{SimFromPrior}\NormalTok{(priors,}\DataTypeTok{class=}\StringTok{"sigma"}\NormalTok{)}
\NormalTok{\}}
\end{Highlighting}
\end{Shaded}

Then we set the \texttt{beta1} parameter to zero in all runs where the null hypothesis was drawn.

\begin{Shaded}
\begin{Highlighting}[]
\NormalTok{beta1[ hypothesis\_samples}\OperatorTok{==}\DecValTok{0}\NormalTok{ ] \textless{}{-}}\StringTok{ }\DecValTok{0}
\end{Highlighting}
\end{Shaded}

Now that we have simulated the model parameters, we can simulate data based on the sampled hypothesis. For the fake data simulation from a generalized linear mixed-effects model, we use the R function \texttt{simLMM()} from the \texttt{designr} package.

\begin{Shaded}
\begin{Highlighting}[]
\NormalTok{rtsimmat \textless{}{-}}\StringTok{ }\KeywordTok{matrix}\NormalTok{(}\OtherTok{NA}\NormalTok{,}\KeywordTok{nrow}\NormalTok{(fakedata),nsim)}
\CommentTok{\# We take exp() since we assume response times are log{-}normally distributed}
\ControlFlowTok{for}\NormalTok{ (i }\ControlFlowTok{in} \DecValTok{1}\OperatorTok{:}\NormalTok{nsim) }
\NormalTok{  rtsimmat[,i] \textless{}{-}}\StringTok{ }\KeywordTok{exp}\NormalTok{(}\KeywordTok{simLMM}\NormalTok{(}\DataTypeTok{formula=}\OperatorTok{\textasciitilde{}}\StringTok{ }\NormalTok{x }\OperatorTok{+}\StringTok{ }\NormalTok{(x }\OperatorTok{|}\StringTok{ }\NormalTok{subj), }
                              \DataTypeTok{dat=}\NormalTok{simdata, }
                              \DataTypeTok{Fixef=}\KeywordTok{c}\NormalTok{(beta0[i], beta1[i]), }
                              \DataTypeTok{VC\_sd=}\KeywordTok{list}\NormalTok{(}\KeywordTok{c}\NormalTok{(sigma\_u0[i], sigma\_u1[i]), sigma[i]), }
                              \DataTypeTok{CP=}\NormalTok{rho\_u[i], }\DataTypeTok{empirical=}\OtherTok{FALSE}\NormalTok{))}
\end{Highlighting}
\end{Shaded}

The next step is to estimate the Bayesian (brms) models on the simulated data. For each simulated data set, we estimate the posterior of the H0 and the H1, then we perform bridge sampling, and then we use this to compute a Bayes factor for each of the \(500\) simulated data sets.

For the hierarchical modeling, we use the R-package \texttt{brms} (Bürkner, 2017, 2018). We specify a large number of sampling iterations for each of four chains (\texttt{s\ =\ 10,000}, warmup samples: \texttt{s\ =\ 2,000}). This large number is required to obtain stable Bayes factor estimates. Note that it is a much larger number than the default number of iterations (\texttt{s\ =\ 2,000}), which was not set to estimate Bayes factors, but instead to estimate posterior expectations.
Moreover, \texttt{adapt\_delta}, which is set to \texttt{adapt\_delta\ =\ 0.9}, and \texttt{max\_treedepth}, which is set to \texttt{max\_treedepth\ =\ 15} are control parameters for ensuring the posterior sampler is working correctly (Betancourt, 2016, 2017; Gabry et al., 2019). Importantly, it's necessary to set the argument \texttt{save\_pars\ =\ save\_pars(all\ =\ TRUE)}. This setting is a precondition for later performing bridge sampling for computing the Bayes factor analysis.

For each model (H0 and H1), we use the function \texttt{bridge\_sampler()} to compute marginal likelihoods, and we compute the Bayes factor by comparing marginal likelihoods using the function \texttt{bayes\_factor(lml\_Full,\ lml\_Null)}.

\begin{Shaded}
\begin{Highlighting}[]
\NormalTok{BF10\_SBC \textless{}{-}}\StringTok{ }\KeywordTok{rep}\NormalTok{(}\OtherTok{NA}\NormalTok{,nsim)}
\ControlFlowTok{for}\NormalTok{ (i }\ControlFlowTok{in} \DecValTok{1}\OperatorTok{:}\NormalTok{nsim) \{}
\NormalTok{    simdata}\OperatorTok{$}\NormalTok{simrt \textless{}{-}}\StringTok{ }\NormalTok{rtsimmat[,i]}
    \CommentTok{\# estimate model for alternative hypothesis}
\NormalTok{    brm1   \textless{}{-}}\StringTok{ }\KeywordTok{brm}\NormalTok{(simrt }\OperatorTok{\textasciitilde{}}\StringTok{ }\NormalTok{x }\OperatorTok{+}\StringTok{ }\NormalTok{(}\DecValTok{1}\OperatorTok{+}\NormalTok{x}\OperatorTok{|}\NormalTok{subj), simdata, }
            \DataTypeTok{family=}\KeywordTok{lognormal}\NormalTok{(), }\DataTypeTok{prior=}\NormalTok{priors, }\DataTypeTok{cores=}\DecValTok{4}\NormalTok{,}
            \DataTypeTok{save\_pars =} \KeywordTok{save\_pars}\NormalTok{(}\DataTypeTok{all =} \OtherTok{TRUE}\NormalTok{),}
            \DataTypeTok{warmup=}\DecValTok{2000}\NormalTok{, }\DataTypeTok{iter=}\DecValTok{10000}\NormalTok{, }
            \DataTypeTok{control=}\KeywordTok{list}\NormalTok{(}\DataTypeTok{adapt\_delta=}\FloatTok{0.99}\NormalTok{, }\DataTypeTok{max\_treedepth=}\DecValTok{15}\NormalTok{))}
\NormalTok{    lml\_Full \textless{}{-}}\StringTok{ }\KeywordTok{bridge\_sampler}\NormalTok{(brm1, }\DataTypeTok{silent=}\OtherTok{TRUE}\NormalTok{)}
    \KeywordTok{rm}\NormalTok{(brm1)}
    \CommentTok{\# estimate model for null hypothesis}
\NormalTok{    brm0   \textless{}{-}}\StringTok{ }\KeywordTok{brm}\NormalTok{(simrt }\OperatorTok{\textasciitilde{}}\StringTok{ }\DecValTok{1} \OperatorTok{+}\StringTok{ }\NormalTok{(}\DecValTok{1}\OperatorTok{+}\NormalTok{x}\OperatorTok{|}\NormalTok{subj), simdata, }
            \DataTypeTok{family=}\KeywordTok{lognormal}\NormalTok{(), }\DataTypeTok{prior=}\NormalTok{priors[}\OperatorTok{{-}}\DecValTok{2}\NormalTok{,], }\DataTypeTok{cores=}\DecValTok{4}\NormalTok{,}
            \DataTypeTok{save\_pars =} \KeywordTok{save\_pars}\NormalTok{(}\DataTypeTok{all =} \OtherTok{TRUE}\NormalTok{),}
            \DataTypeTok{warmup=}\DecValTok{2000}\NormalTok{, }\DataTypeTok{iter=}\DecValTok{10000}\NormalTok{,}
            \DataTypeTok{control=}\KeywordTok{list}\NormalTok{(}\DataTypeTok{adapt\_delta=}\FloatTok{0.99}\NormalTok{, }\DataTypeTok{max\_treedepth=}\DecValTok{15}\NormalTok{))}
\NormalTok{    lml\_Null \textless{}{-}}\StringTok{ }\KeywordTok{bridge\_sampler}\NormalTok{(brm0, }\DataTypeTok{silent=}\OtherTok{TRUE}\NormalTok{)}
    \KeywordTok{rm}\NormalTok{(brm0)}
\NormalTok{    BF10\_SBC[i] \textless{}{-}}\StringTok{ }\KeywordTok{bayes\_factor}\NormalTok{(lml\_Full, lml\_Null)}\OperatorTok{$}\NormalTok{bf}
\NormalTok{\}}
\end{Highlighting}
\end{Shaded}

Note that in the null-model, we do keep the random effects of factor \texttt{x} varying across subjects and across items, i.e., \texttt{simrt\ \textasciitilde{}\ 1\ +\ (1+x\textbar{}subj)}. That is, we do assume that effects of factor \texttt{x} could be present for individual subjects, but importantly, by removing the fixed-effect of \texttt{x} we assume a priori that the overall mean effect across all subjects is zero. The model comparison therefore targets only this fixed effect of factor \texttt{x}, but not the random effects.

\begin{figure}

{\centering \includegraphics{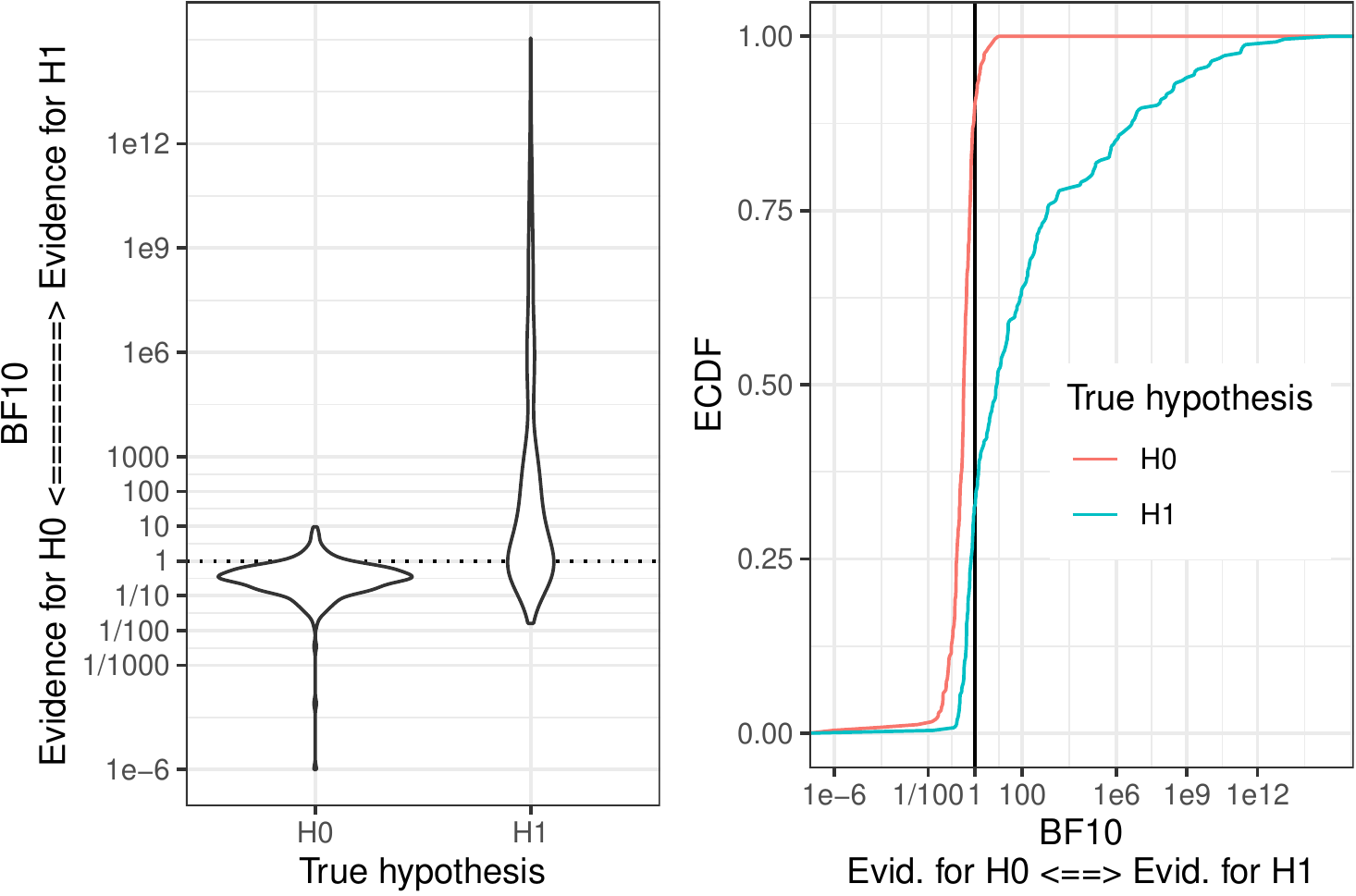} 

}

\caption{Distribution of Bayes factors (BF10) as a function of which hypothesis was true in the simulations from the SBC. Left panel: Distributions are shown as violin plots. Right panel: Empirical cumulative distribution functions (ECDFs). One outlier data point for the H0 with a BF10 of -3e-86 resulted from an unstable marginal likelihood (i.e., the bridge sampling did not converge) and was removed for visualization.}\label{fig:ViolinPlot}
\end{figure}

While this is not required in and part of SBC, we show here the distributions of Bayes factors given the true hypotheses (see Fig.~\ref{fig:ViolinPlot}). The results show that the Bayes factor estimates exhibit wide distributions when either the H0 or the H1 are true. It is clear that when the H1 is the true hypothesis in the data simulation, then the Bayes factors provides more evidence to the H1 on average. By contrast, when the H0 is the true hypothesis in the data simulation, then the distribution of Bayes factors is clearly shifted towards evidence for the H0. Interestingly, these distributions are quite asymmetric such that strong evidence for the correct hypothesis is rather rare, and weaker evidence is more frequent.

Note that there was one outlier data point for the H0 with a \(BF_{10} = -3e-86\). This resulted from an unstable marginal likelihood, since the bridge sampling did not converge. Thus, even in the simple example case we use, there can occasionally be problems with bridge sampling.

Last, we can compute the posterior model probabilities. The ratio of posterior model probabilities, \(p(H1|y)/p(H0|y)\), can be obtained by multiplying the Bayes factor (\texttt{BF10\_SBC}) with the prior ratio of model probabilities (which is \(p(H1)/p(H0) = 0.5/0.5 = 1\) in our example): \(p(H1|y)/p(H0|y) = BF_{10} \times p(H1)/p(H0)\):

\begin{Shaded}
\begin{Highlighting}[]
\NormalTok{postModelRat \textless{}{-}}\StringTok{ }\NormalTok{BF10\_SBC }\OperatorTok{*}\StringTok{ }\NormalTok{priorsHypothesis[}\DecValTok{2}\NormalTok{]}\OperatorTok{/}\NormalTok{priorsHypothesis[}\DecValTok{1}\NormalTok{]}
\end{Highlighting}
\end{Shaded}

This posterior ratio can be used to compute the posterior probabilities for the null hypothesis and for the alternative hypothesis:

\begin{Shaded}
\begin{Highlighting}[]
\NormalTok{postModelProbsH1 \textless{}{-}}\StringTok{ }\NormalTok{postModelRat}\OperatorTok{/}\NormalTok{(postModelRat}\OperatorTok{+}\DecValTok{1}\NormalTok{)}
\NormalTok{postModelProbsH0 \textless{}{-}}\StringTok{ }\DecValTok{1}\OperatorTok{/}\NormalTok{(postModelRat}\OperatorTok{+}\DecValTok{1}\NormalTok{)}
\end{Highlighting}
\end{Shaded}

This code computes the posterior model probabilities for alll \(500\) simulation runs.
As the last step, across the \(500\) simulation runs, we average the posterior probabilities for each model, i.e., by computing the mean posterior probability across all \(500\) runs: \(\mu_{\mathcal{M}_x}^{post} = \frac{1}{n_{sim}} \sum_{i=1}^{n_{sim}} p(\mathcal{M}_x \mid y_i^{sim})\), where each \(y_i^{sim}\) is one out of \(n_{sim} = 500\) simulated data sets, \(\mathcal{M}_x\) is one selected model, and \(\mu_{\mathcal{M}_x}^{post}\) is the average posterior probability for model \(x\), or simply in R: \texttt{mean(postModelProbsH1)}.

If one wanted to make decisions based on the continuous evidence, e.g., to compute things like FDR or TDR, then one would need to specify thresholds on Bayes factors or posterior probabilities, such that Bayes factors/posterior probabilities larger or smaller than these thresholds would indicate evidence for the H0, for the H1, or for neither hypothesis. However, one key aspect of Bayesian data analysis is that it provides continuous estimates of posterior probabilities.

Now, we can investigate our question of interest in SBC: we can look at how likely each model was chosen a posteriori on average and compare these average posterior model probabilities (see below, ``means''; in addition, their 95\% binomial confidence intervals) to the prior model probabilities that were in fact used to simulate the data (i.e., 50\% each).

\begin{Shaded}
\begin{Highlighting}[]
\CommentTok{\# Obtain 95\% confidence intervals from a logistic linear model }
\CommentTok{\# and transform confidence intervals into probabilities}
\NormalTok{BM  \textless{}{-}}\StringTok{ }\KeywordTok{glm}\NormalTok{(postModelProbsH1}\OperatorTok{\textasciitilde{}}\DecValTok{1}\NormalTok{,}\DataTypeTok{family=}\StringTok{"binomial"}\NormalTok{)}
\NormalTok{CIs \textless{}{-}}\StringTok{ }\DecValTok{1}\OperatorTok{/}\NormalTok{(}\DecValTok{1}\OperatorTok{+}\KeywordTok{exp}\NormalTok{(}\OperatorTok{{-}}\KeywordTok{confint}\NormalTok{(BM)))}
\NormalTok{ME  \textless{}{-}}\StringTok{ }\KeywordTok{as.numeric}\NormalTok{(}\DecValTok{1}\OperatorTok{/}\NormalTok{(}\DecValTok{1}\OperatorTok{+}\KeywordTok{exp}\NormalTok{(}\OperatorTok{{-}}\KeywordTok{coef}\NormalTok{(BM))))}
\CommentTok{\# Show the average posterior probability for H1 with 95\% confidence intervals}
\KeywordTok{t}\NormalTok{(}\KeywordTok{data.frame}\NormalTok{(}\DataTypeTok{pH1=}\KeywordTok{round}\NormalTok{(}\DecValTok{100}\OperatorTok{*}\KeywordTok{c}\NormalTok{(}\DataTypeTok{CI=}\NormalTok{CIs[}\DecValTok{1}\NormalTok{], }\DataTypeTok{mean=}\NormalTok{ME, }\DataTypeTok{CI=}\NormalTok{CIs[}\DecValTok{2}\NormalTok{]),}\DecValTok{2}\NormalTok{)))}
\end{Highlighting}
\end{Shaded}

\begin{verbatim}
##     CI.2.5 %  mean CI.97.5 %
## pH1    45.53 49.91      54.3
\end{verbatim}

The results show that the average posterior model probability for the H1 versus the H0 was at roughly 50\%. This result directly corresponds to the prior model probability of 50\%. The confidence intervals include the prior of 50\%. This SBC analysis therefore, for this specific and simple example case, did not indicate any signs of significant bias. This is important calibration information for the bridge sampling approach, since it has not been clear so far whether bridge sampling yields unbiased estimates for the types of multilevel models studied here and often used in research practice. These results are therefore encouraging and support the application of bridge sampling for computation of Bayes factors and posterior probabilities for our case study. However, much more extensive simulation studies are required to investigate this point more generally, which is outside the scope of this paper.

In addition to these SBC results, we can also investigate additional calibration questions of interest by looking at posterior model probabilities as a function of which prior hypothesis (model) was sampled in a given run. For each simulation we know whether the data was simulated based on the H0 or the H1, that is, we know whether for a given simulated data set, the H0 or the H1 is ``true''. This information is stored in the vector \texttt{hypothesis\_samples}. For each ``true'' hypothesis, we can now look at how much posterior probability mass is allocated to the two models by the Bayesian analysis. If the artificial data were simulated based on the H0, how high is the posterior probability for the H0? Is it higher than chance? And if so, by how much. Moreover, if the artificial data were simulated based on the H1, what is the posterior probability for the H1?

\begin{Shaded}
\begin{Highlighting}[]
\NormalTok{true\_hypothesis \textless{}{-}}\StringTok{ }\KeywordTok{ifelse}\NormalTok{(hypothesis\_samples}\OperatorTok{==}\DecValTok{1}\NormalTok{, }\StringTok{"H1"}\NormalTok{, }\StringTok{"H0"}\NormalTok{)}
\NormalTok{tabSBC \textless{}{-}}\StringTok{ }\KeywordTok{data.frame}\NormalTok{(postModelProbsH0, postModelProbsH1, true\_hypothesis) }\OperatorTok{\%\textgreater{}\%}\StringTok{ }
\StringTok{  }\KeywordTok{group\_by}\NormalTok{(true\_hypothesis) }\OperatorTok{\%\textgreater{}\%}\StringTok{ }
\StringTok{  }\KeywordTok{summarize}\NormalTok{(}\DataTypeTok{pH0=}\KeywordTok{round}\NormalTok{(}\KeywordTok{mean}\NormalTok{(postModelProbsH0, }\DataTypeTok{na.rm=}\OtherTok{TRUE}\NormalTok{)}\OperatorTok{*}\DecValTok{100}\NormalTok{), }
            \DataTypeTok{pH1=}\KeywordTok{round}\NormalTok{(}\KeywordTok{mean}\NormalTok{(postModelProbsH1, }\DataTypeTok{na.rm=}\OtherTok{TRUE}\NormalTok{)}\OperatorTok{*}\DecValTok{100}\NormalTok{)) }\OperatorTok{\%\textgreater{}\%}
\StringTok{  }\KeywordTok{as.data.frame}\NormalTok{()}
\end{Highlighting}
\end{Shaded}

\begin{table}

\caption{\label{tab:tabSBC1}Average posterior probabilities for the H0 (pH0) and for the H1 (pH1) as a function of the true hypothesis in the simulations (H0 versus H1).}
\centering
\begin{tabular}[t]{lrr}
\toprule
True hypothesis & pH0 & pH1\\
\midrule
H0 & 73 & 27\\
H1 & 28 & 72\\
\bottomrule
\end{tabular}
\end{table}

The results (see Table~\ref{tab:tabSBC1}) in the first row show that if the H0 was used to simulate artificial data, then the Bayesian procedure allocated an average of 73\% posterior probability to the H0. Thus, the chance to support the null hypothesis correctly is clearly better than 50/50, i.e., better than chance, in this set of simulated data and model.
Moreover, the second row of the table shows that if H1 was used to simulate the artificial data, then the posterior probability for H1 was an average of 72\%. Thus, the alternative hypothesis is also somewhat likely to be correctly supported in the present setting. Taken together, this analysis shows that the data and the model on average provide some evidence for the hypotheses of interest. Note that this result completely depends on things such as the effect size or experimental design (including sample size), and the posterior probabilities of the true model may be higher if stronger effects or larger samples are investigated.

Importantly, the SBC analysis supported the Bayes factor estimates and could not detect a difference of average posterior model probabilities from the prior model probabilities, suggesting that the Bayes factor estimates for this analysis are valid. However, this will not always be the outcome of SBC. We will next investigate a case where SBC shows a problem with posterior model probabilities.

\hypertarget{sbc-for-bridge-sampling-an-example-where-bayes-factor-estimates-are-not-accurate-due-to-model-mis-specification}{%
\subsubsection{SBC for bridge sampling: An example where Bayes factor estimates are not accurate due to model mis-specification}\label{sbc-for-bridge-sampling-an-example-where-bayes-factor-estimates-are-not-accurate-due-to-model-mis-specification}}

Next, we again investigate the same experimental design as above, with the same priors for the parameters and the same observational model as in the previous section. Again, we are interested in the hypotheses H0 and H1 and in using bridge sampling to compute Bayes factors. We again use SBC to investigate whether average posterior model probabilities correspond to the prior model probabilities to test whether Bayes factor estimates capture the true Bayes factor. The only thing that differs in this analysis is that we leave out the random slopes in the estimation procedure. In R-formula syntax the H1 can thus be written as \texttt{simrt\ \textasciitilde{}\ x\ +\ (1\textbar{}subj)} and the H0 can be written as \texttt{simrt\ \textasciitilde{}\ 1\ +\ (1\textbar{}subj)}. It is known from analysis of frequentist tools that leaving out random slopes from a linear mixed-effects model can lead to an increase in type I (\(\alpha\)) error (Barr, Levy, Scheepers, \& Tily, 2013; Matuschek, Kliegl, Vasishth, Baayen, \& Bates, 2017). Here, we are interested in whether leaving out random slopes from a corresponding Bayesian multilevel model leads to a bias in posterior model probabilities. In line with the frequentist results, we expect that posterior model probabilities for the H1 should be inflated when neglecting random slopes. To investigate this, we perform the exact same SBC analysis as before, but only leaving out the random slopes from the fitted models.

\begin{verbatim}
##     CI.2.5 %  mean CI.97.5 %
## pH1    54.21 58.57     62.84
\end{verbatim}

The result shows that now, as expected, the average posterior probability for the alternative hypothesis (H1) of 58.57\% is higher than its prior probability of 50\%. This increase is supported by the 95\% confidence intervals, which do not overlap with 50\%. Moreover, a frequentist intercept-only logistic regression shows that the average posterior model probability highly significantly differs from the prior probability of 50\% (\(b = 0.35, SE = 0.09, z = 3.81, p = .0001\)). This shows that leaving out random slopes from a multilevel model when the data do contain random variation of the effect across subjects can lead to severe biases in the posterior model probabilities as pointed out by Barr et al. (2013) for frequentist linear mixed-effects models. SBC is a useful tool that can detect such biases. It is therefore highly recommended to use SBC to calibrate one's Bayes factor estimates for a specific study, model, and priors.

We again look at the average posterior probabilities as a function of which hypothesis was true in the simulations.

\begin{table}

\caption{\label{tab:tabSBC2}Average posterior probabilities for the H0 (pH0) and for the H1 (pH1) as a function of the true hypothesis in the simulations (H0 versus H1) when using a mis-specified model.}
\centering
\begin{tabular}[t]{lrr}
\toprule
True hypothesis & pH0 & pH1\\
\midrule
H0 & 66 & 34\\
H1 & 17 & 83\\
\bottomrule
\end{tabular}
\end{table}

The results (see Table~\ref{tab:tabSBC2}) show that the average posterior probability was largest for the correct model. When the H0 was true in the simulations (first row), then the average posterior probability for the H0 was 66\%. By contrast, when the H1 was true in the simulations (second row), then the average posterior probability for the H1 was 83\%. Thus, while the correct hypothesis still had higher average posterior probability, the increase in posterior evidence for the H1 is still visible (83\% is larger than 66\%).

The key problem in this analysis is that the models generating the data (\texttt{H0} / \texttt{H1}; which included variation in random slopes) were different from the models used to analyze the data (\texttt{pH0} /\texttt{pH1}; which assumed no variation in random slopes). Thus, we can see in the SBC that if our models are wrong, then our inferences can be misleading.

\hypertarget{sbc-for-the-savage-dickey-method}{%
\subsubsection{SBC for the Savage-Dickey method}\label{sbc-for-the-savage-dickey-method}}

The Savage--Dickey method (Dickey et al., 1970) can be used to estimate the Bayes factor for nested models, where one (or more) of the model parameters of the full model (e.g., a regression coefficient) is set to a fixed value (such as e.g., zero) to obtain a nested null model. In these cases, the Bayes factor can be obtained by computing the ratio between the densities of the posterior and the prior at the value for the model parameters of zero. This is a very interesting and elegant result in Bayesian modeling. Unfortunately, however, the implementation of the Savage--Dickey method in \texttt{brms} can give very unreliable results when the posterior is far away from zero, and any kind of Savage--Dickey ratio based on MCMC samples may have this limitation. The reason for this is that if the posterior is far from the fixed value being investigated, then only very few MCMC draws will fall close to this value. Therefore, the estimate of the posterior density at the value of zero will be very noisy. The results will be consistent in showing very little support for the null model. However, the exact amount of support for the alternative hypothesis (H1) cannot be measured in a precise way.

Here, we study Bayes factor estimates from the Savage--Dickey method using SBC to test whether the Bayes factor estimates of the method are accurate on average. Again, we run the same SBC analysis as before, now adding random slopes, but using the Savage--Dickey method to compute Bayes factors instead of using bridge sampling.

\begin{verbatim}
##     CI.2.5 %  mean CI.97.5 %
## pH1    43.02 47.38     51.77
\end{verbatim}

The result shows that the average posterior probability for the H1 is close to the prior model probability of 50\%. A frequentist intercept-only logistic regression shows that there is no evidence that the posterior model probability differs from the prior probability of 50\% (\(b = -0.10, SE = 0.09, z = -1.2, p = .242\)). Note, however, that our SBC analysis uses a limited number of \(500\) SBC iterations, and a SBC with a larger number of iterations might reveal divergences of smaller size than we can currently resolve.

Again, we also look at the average posterior probabilities as a function of which hypothesis was actually true, i.e., which model was used to simulate the data.

\begin{table}

\caption{\label{tab:tabSBC3}Average posterior probabilities for the H0 (pH0) and for the H1 (pH1) as a function of the true hypothesis in the simulations (H0 versus H1) when using the Savage-Dickey method.}
\centering
\begin{tabular}[t]{lrr}
\toprule
True hypothesis & pH0 & pH1\\
\midrule
H0 & 72 & 28\\
H1 & 31 & 69\\
\bottomrule
\end{tabular}
\end{table}

The results (see Table~\ref{tab:tabSBC3}) show that with the Savage--Dickey method, there is again higher posterior probability for the correct hypothesis (72\% for the correct H0 and 69\% for the correct H1).

\hypertarget{sbc-using-the-savage-dickey-method-an-example-with-invalid-average-posterior-probabilities}{%
\subsubsection{SBC using the Savage-Dickey method: An example with invalid average posterior probabilities}\label{sbc-using-the-savage-dickey-method-an-example-with-invalid-average-posterior-probabilities}}

Again, we provide an example where average posterior model probabilities are incorrect as determined by SBC. As for the bridge sampling, we again fit a model where the random slopes are excluded from the model. Everything else remains the same as in the study above.

\begin{verbatim}
##     CI.2.5 % mean CI.97.5 %
## pH1    54.25 58.6     62.87
\end{verbatim}

The result shows that as in the SBC with bridge sampling, the average posterior probability for the H1 is also strongly inflated when using the Savage--Dickey method to estimate Bayes factors. The posterior probability for the H1 is an average of 58.60\%, and significantly different from the prior 50\% (\(b=0.35, SE=0.09, z=3.82, p = .0001\)). This again shows that posterior probabilites are not estimated accurately when random slopes are ignored in a Bayesian linear mixed-effects model.

\begin{table}

\caption{\label{tab:tabSBC4}Average posterior probabilities for the H0 (pH0) and for the H1 (pH1) as a function of the true hypothesis in the simulations (H0 versus H1) when using the Savage-Dickey method and a mis-specified model.}
\centering
\begin{tabular}[t]{lrr}
\toprule
True hypothesis & pH0 & pH1\\
\midrule
H0 & 63 & 37\\
H1 & 19 & 81\\
\bottomrule
\end{tabular}
\end{table}

Again, the average posterior probability as a function of the true hypothesis reveals (see Table~\ref{tab:tabSBC4}) that the correct hypothesis has higher posterior probability than the incorrect hypothesis, but that the posterior probability is higher for the correct H1 (81\%) than for the H0 (63\%), reflecting the overall bias towards the H1 in the estimation.

\hypertarget{unstable-bayes-factor-estimates-due-to-the-effective-number-of-posterior-samples}{%
\subsubsection{Unstable Bayes factor estimates due to the effective number of posterior samples}\label{unstable-bayes-factor-estimates-due-to-the-effective-number-of-posterior-samples}}

An important issue that we have glossed over in the previous examples is that the number of posterior samples chosen in the Hamiltonian Markov Chain Monte Carlo (MCMC) sampler (called by \texttt{brms}) can have a strong impact on the results of the Bayes factor estimators. This is true for both bridge sampling and also for the Savage--Dickey method. Bridge sampling is a form of density estimation for which we have no good theoretical guarantees of MCMC sampling. Note that in the analyses presented above, we set the number of MCMC draws to a large number of \(n_{iter} = 10,000\) iterations. The SBC analysis therefore took a considerable amount of time.

Running the same \texttt{brms} models with less MCMC draws will induce some instability in the Bayes factor estimates based on the bridge sampling, such that running the same analysis twice would yield different results for the Bayes factor. Moreover, bridge sampling in itself may be unstable and may return different results for different bridge sampling runs on the same posterior MCMC draws (just because of different starting values). This is very concerning, as the results reported in a paper might not be stable if the number of posterior samples or effective sample size is not large enough. Indeed, the default number of posterior samples in \texttt{brms} is \texttt{iter\ =\ 2000} (and the default number of warmup samples is \texttt{warmup\ =\ 1000}). It is important to note that these defaults were not set to support bridge sampling (nor the Savage-Dickey method), i.e., they were not defined for computation of densities to support Bayes factors. Instead, they are valid for posterior inference on expectations (e.g., posterior means) for models that are not too complex. However, when using these defaults for estimation of densities and the computation of Bayes factors, then instabilities can arise.

For illustration, we perform the SBC analysis again, now using the default number of iterations in \texttt{brms} (\(s = 2,000\) samples, \(s = 1,000\) warm-up samples). We first run the model using bridge sampling, and then look at SBC for the Savage--Dickey method.

\begin{verbatim}
##     CI.2.5 %  mean CI.97.5 %
## pH1    44.58 48.96     53.35
\end{verbatim}

The results for bridge sampling show that even with a smaller number of MCMC draws, the average posterior does not diverge from the prior 50\%, suggesting accurate average estimation of posterior probabilities. This is very interesting information and in our example analysis shows that just because the Bayes factor estimator is more noisy with smaller number of MCMC draws, this does not seem to mean that it gets biased, at least in the case that we study here.

\begin{table}

\caption{\label{tab:tabSBC5}Average posterior probabilities for the H0 (pH0) and for the H1 (pH1) as a function of the true hypothesis in the simulations (H0 versus H1) using bridge sample with the small default number of MCMC draws (s = 2,000).}
\centering
\begin{tabular}[t]{lrr}
\toprule
True hypothesis & pH0 & pH1\\
\midrule
H0 & 72 & 28\\
H1 & 31 & 69\\
\bottomrule
\end{tabular}
\end{table}

Also, there is still quite a bit of information captured by the Bayes factor estimates as average posterior probabilities are larger for the correct hypothesis (see Table~\ref{tab:tabSBC5}). Thus, in this example case, average posterior probabilities perform well even when using a smaller number of bridge samples. Importantly, this may be the case because we are using a very small simulated data set with only 15 subjects and only 4 data points per subject. When using more realistic and larger data sets, which may possibly also include variation across items, such a small number of MCMC draws might be much more problematic and larger numbers of MCM iterations may be needed.

Next, we perform the same analysis for the Savage--Dickey method, again now with the default number of MCMC iterations in brms.

\begin{verbatim}
##     CI.2.5 %  mean CI.97.5 %
## pH1    47.44 51.82     56.18
\end{verbatim}

Again, the average posterior model probabilities do not diverge from the prior 50\%.

\begin{table}

\caption{\label{tab:tabSBC6}Average posterior probabilities for the H0 (pH0) and for the H1 (pH1) as a function of the true hypothesis in the simulations (H0 versus H1) using the Savage-Dickey method with the small default number of s = 2,000 MCMC draws.}
\centering
\begin{tabular}[t]{lrr}
\toprule
True hypothesis & pH0 & pH1\\
\midrule
H0 & 72 & 28\\
H1 & 26 & 74\\
\bottomrule
\end{tabular}
\end{table}

Again (see Table~\ref{tab:tabSBC6}), the average posterior probabilities are largest for the correct hypotheses.

Note that SBC cannot tell us how stable the posterior probabilities are estimated in the individual analysis run. What we need for this is to use the same data set and to estimate Bayes factors again and again, by only varying the MCMC samples on which the Bayes factor estimates are based, but leaving all other aspects of the data and the model constant. Thus, we want to know how stable the Bayes factor estimates are with respect to the MCMC chains.

To investigate this, we use the same experimental design that we introduced above. We simulate data from a linear mixed-effects model with a full variance covariance matrix for random effects (i.e., a maximal linear mixed-effects model). Importantly, this time, we do not sample model parameters from prior distributions. Instead, we set them to fixed values. We use values for the fixed effects for the intercept of \(6\) and for the effect of \texttt{x} of \(-1\). For the random effects we assume standard deviations of \(0.5\) and a correlation of \(0.3\). The residual noise is set to \(0.5\). We simulate the data using the R function \texttt{simLMM()} and make use of the functionality \texttt{empirical=TRUE}, which makes sure that the fixed effects in the data correspond precisely to the indicated values (i.e., the intercept is exactly \(6\) and the effect of \texttt{x} exactly \(-1\)).

\begin{Shaded}
\begin{Highlighting}[]
\NormalTok{design \textless{}{-}}\StringTok{ }\KeywordTok{fixed.factor}\NormalTok{(}\StringTok{"x"}\NormalTok{, }\DataTypeTok{levels=}\KeywordTok{c}\NormalTok{(}\StringTok{"{-}1"}\NormalTok{, }\StringTok{"1"}\NormalTok{), }\DataTypeTok{replications=}\DecValTok{2}\NormalTok{) }\OperatorTok{+}
\StringTok{  }\KeywordTok{random.factor}\NormalTok{(}\StringTok{"subj"}\NormalTok{, }\DataTypeTok{instances=}\DecValTok{15}\NormalTok{)}
\NormalTok{fakedata \textless{}{-}}\StringTok{ }\KeywordTok{design.codes}\NormalTok{(design)}
\NormalTok{fakedata}\OperatorTok{$}\NormalTok{x \textless{}{-}}\StringTok{ }\KeywordTok{as.numeric}\NormalTok{(}\KeywordTok{as.character}\NormalTok{(fakedata}\OperatorTok{$}\NormalTok{x))}
\CommentTok{\# simulate data}
\NormalTok{fakedata}\OperatorTok{$}\NormalTok{fakert \textless{}{-}}\StringTok{ }\KeywordTok{simLMM}\NormalTok{(}\DataTypeTok{formula=}\OperatorTok{\textasciitilde{}}\StringTok{ }\NormalTok{x }\OperatorTok{+}\StringTok{ }\NormalTok{(x }\OperatorTok{|}\StringTok{ }\NormalTok{subj), }
                           \DataTypeTok{dat   =}\NormalTok{fakedata, }
                           \DataTypeTok{Fixef =} \KeywordTok{c}\NormalTok{(}\DecValTok{6}\NormalTok{, }\FloatTok{{-}1.0}\NormalTok{),}
                           \DataTypeTok{VC\_sd =} \KeywordTok{list}\NormalTok{(}\KeywordTok{c}\NormalTok{(}\FloatTok{0.5}\NormalTok{, }\FloatTok{0.5}\NormalTok{), }\FloatTok{0.5}\NormalTok{),}
                           \DataTypeTok{CP    =} \FloatTok{0.3}\NormalTok{,}
                           \DataTypeTok{empirical=}\OtherTok{TRUE}\NormalTok{, }\DataTypeTok{verbose=}\OtherTok{FALSE}\NormalTok{)}
\CommentTok{\# save fake data}
\KeywordTok{saveRDS}\NormalTok{(fakedata, }\StringTok{"dataR/SBC\_BF\_stab\_fakeDat.RDS"}\NormalTok{)}
\CommentTok{\# frequentist linear mixed{-}effects model using lmer() }
\CommentTok{\# shows fixed effects estimates are precisely as indicated}
\KeywordTok{round}\NormalTok{(}\KeywordTok{coef}\NormalTok{(}\KeywordTok{summary}\NormalTok{(}\KeywordTok{lmer}\NormalTok{(fakert }\OperatorTok{\textasciitilde{}}\StringTok{ }\NormalTok{x }\OperatorTok{+}\StringTok{ }\NormalTok{(x }\OperatorTok{|}\StringTok{ }\NormalTok{subj), }\DataTypeTok{data=}\NormalTok{fakedata))),}\DecValTok{3}\NormalTok{)}
\end{Highlighting}
\end{Shaded}

\begin{verbatim}
##             Estimate Std. Error df t value Pr(>|t|)
## (Intercept)        6      0.115 14  52.186        0
## x                 -1      0.114 14  -8.773        0
\end{verbatim}

For this fixed simulated data set, we estimate the exact same model 100 times, i.e., each time performing new MCMC sampling using the same data, model and priors. We compute Bayes factors for each of the 100 models, and then investigate whether the 100 Bayes factors are the same in each run. We run this analysis using bridge sampling and also using the Savage--Dickey method. Moreover, we run the analyses using the default number of \(2,000\) samples, and also using the larger number of \(10,000\) samples.

\begin{figure}

{\centering \includegraphics{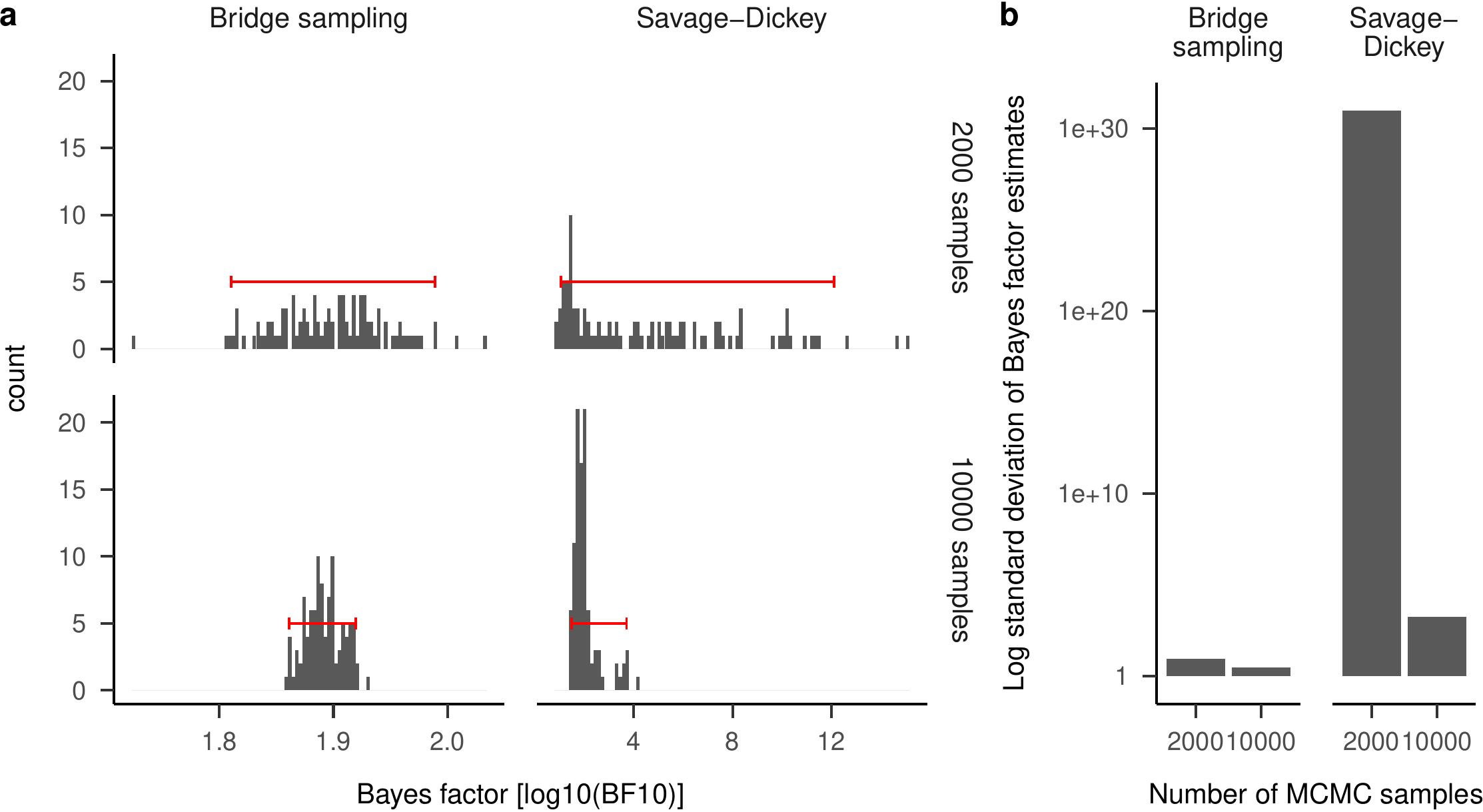} 

}

\caption{Stability of Bayes factor estimates against different MCMC chains. a) Histograms of 100 Bayes factor estimates obtained from the same data and model, where only the MCMC chains differ between runs. The histograms are shown for bridge sampling (left panels) and for the Savage--Dickey method (right panels), as well as for the default number of 2,000 samples (upper panels) and for a larger number of 10,000 samples (lower panels). Red horizontal error bars indicate 95 percent quantiles. b) Bars show the log10 of the standard deviation across 100 Bayes factor estimates displayed in (a) for each method and number of sample separately.}\label{fig:BFniter}
\end{figure}

The results, displayed in Figure~\ref{fig:BFniter}, show that Bayes factor estimates were quite stable when using bridge sampling with a large number of posterior samples. However, Bayes factors estimates become more unstable when using a smaller number of MCMC samples. And they became quite unstable and variable when estimation was done using the Savage--Dickey method.

These results demonstrate that bridge sampling with a large number of samples is required to obtain stable Bayes factor estimates. Of course, if Bayes factors are not estimated in a stable way, but depend on random noise in the MCMC chain, then these Bayes factor estimates cannot accurately represent information that is contained in the data. Thus, the instability observed for the Savage--Dickey method here may be one important reason, why in the SBC analyses above, there was very little information contained in the posterior probabilities based on the Savage--Dickey method.

As we have discussed above, one might think that the performance of bridge sampling with only 2,000 MCMC iterations does not look to bad. As noted above, this may be a result of the very small data set that we used for these simulations. A small number of bridge samples may be much more problematic when using larger data sets with possibly more complicated random effects structures.

We note that the stability of bridge sampling needed in any given application depends on how how small a difference between marginal likelihood one wants to be able to resolve. If the two models compared are very different, then even large variation might be acceptable. However, but if the two models are very close to each other, then small variation might be problematic.

Next, we studied the stability of Bayes factors when in the true data simulation process, the H0 was actually true. We used the same simulated data set as in the previous analysis, with the only difference that we set the critical fixed effect estimate to zero.

\begin{figure}

{\centering \includegraphics{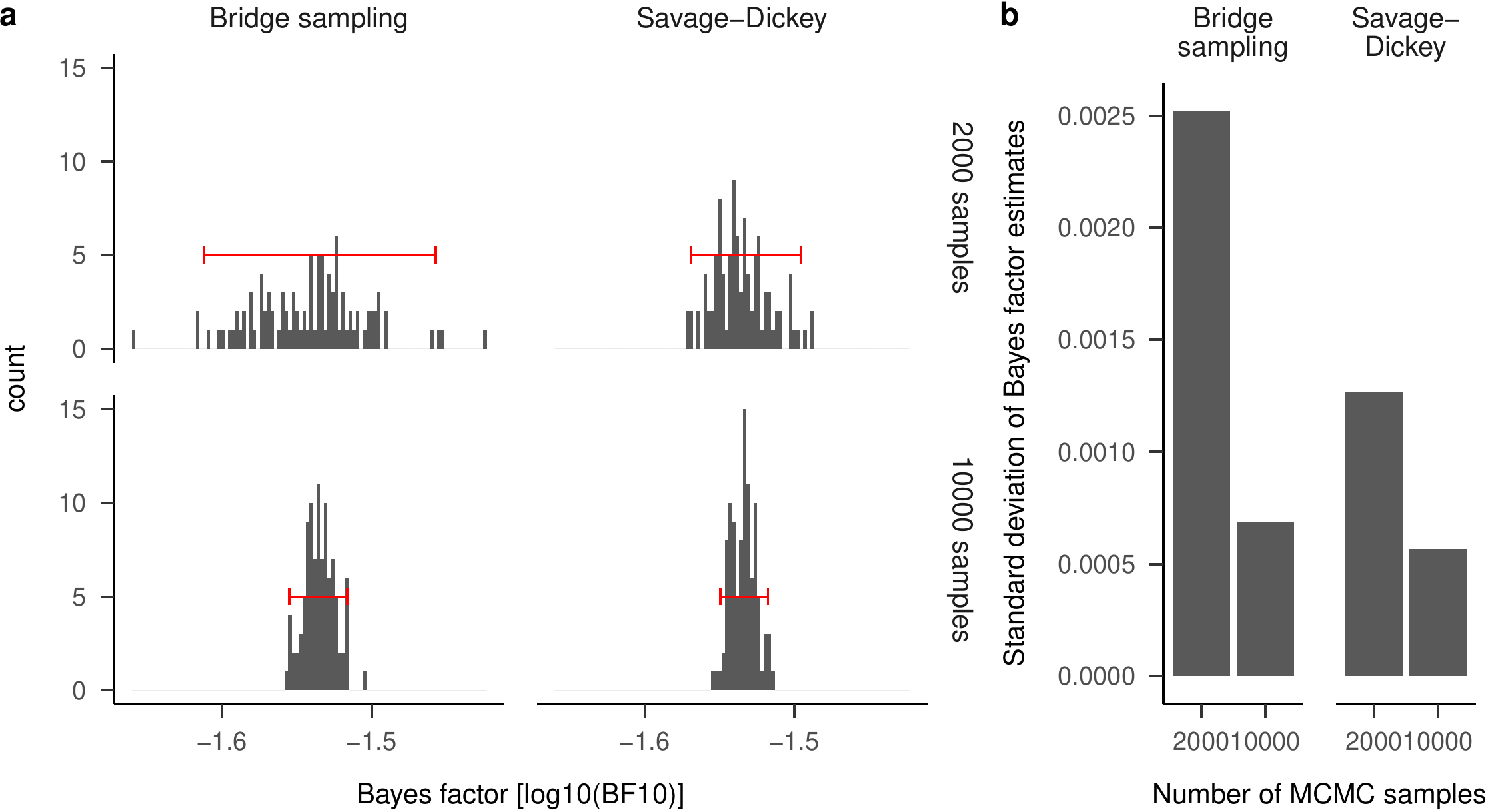} 

}

\caption{Stability of Bayes factor estimates against different MCMC chains in a situation where the H0 is the true model. a) Histograms of 100 Bayes factor estimates obtained from one same data and model, where only the MCMC chains differ between runs. The histograms are shown for bridge sampling (left panels) and for the Savage--Dickey method (right panels), as well as for the default number of 2,000 samples (upper panels) and for a larger number of 10,000 samples (lower panels). Red horizontal error bars indicate 95 percent quantiles. b) Bars show the log10 of the standard deviation across 100 Bayes factor estimates displayed in (a) for each method and number of sample separately.}\label{fig:BFniter2}
\end{figure}

The results show (see Fig.~\ref{fig:BFniter2}) that the Bayes factors from the Savage--Dickey method are now estimated in a much more stable way. The reason for this is that the Savage--Dickey method relies on the estimated posterior density for the critical fixed effect at the value of zero. In the first set of simulations, the posterior was far away from zero, and the estimation was therefore very unstable. However, in the second set of simulations, where the true fixed effect was zero, the posterior samples were also close to zero, yielding a quite stable estimation of the Bayes factor. Note that for bridge sampling and for the Savage--Dickey method, larger number of samples yielded more stable Bayes factors.

In general the instability of Bayes factor estimates against the MCMC draws (and starting values of the bridge sampler) demonstrates that it is necessary to use a large number of iterations when computing Bayes factors using \texttt{brms} and \texttt{bridge\_sampler()}. Moreover, it shows that we have to check for each data set that we analyze, whether our Bayes factor estimate is stable. It is possible to do this by running the analysis a few times (at least twice) to test whether the obtained Bayes factor estimates are stable.

That said, it's important to note that stability doesn't mean accuracy. Bridge sampling with large number of samples is returning low variability estimates. However, it is not clear from the stability analysis how those estimates relate to the true Bayes factors! I.e., whether the Bayes factor estimates are unbiased. SBC (as we have performed above) is needed to judge this aspect.

\hypertarget{continuously-varying-prior-probabilities}{%
\subsubsection{Continuously varying prior probabilities}\label{continuously-varying-prior-probabilities}}

The above SBC-based tests of whether bridge sampling performs unbiased approximations of Bayes factors relied on our a priori assumption that prior model probabilities for the H0 and the H1 were both 50\%. Here, we investigate a larger range of prior probabilities, by systematically varying the prior probability for the H0 from zero to one across 500 simulations. Based on this, we performed the same SBC analysis as above.

\begin{figure}

{\centering \includegraphics{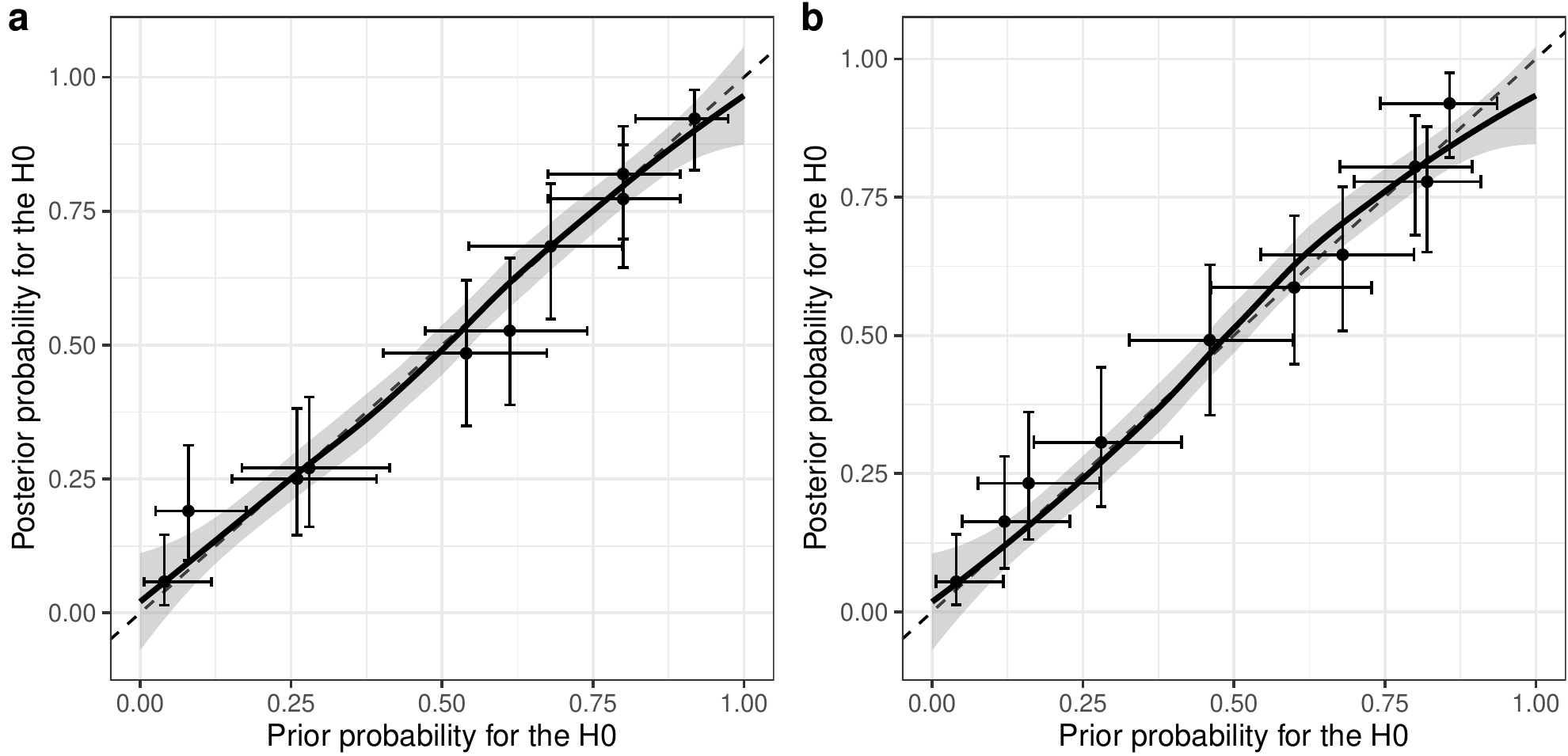} 

}

\caption{Posterior probabilities for the H0 are plotted as a function of prior probabilities for the H0. If the Bayes factor estimate is unbiased, then the data should be aligned along the diagonal (see dashed black line). The points are average posterior probabilities as a function of a priori selected hypotheses for 50 simulation runs each. Errorbars represent 95 percent binomial confidence intervals. a) Bridge sampling with 10,000 MCMC draws. b) Savage-Dickey method with 2,000 MCMC draws.}\label{fig:SBC3plot}
\end{figure}

The results of this analysis are shown in Figure~\ref{fig:SBC3plot}, which plots the posterior probability for the H0 as a function of the prior probability for the H0. Accurate Bayes factor estimates are obtained if all data points lie exactly on the diagonal. The results show that the local regression line is very close to the diagonal, and that the data points (each summarizing results from 50 simulations, with means and confidence intervals) also lie close to the diagonal, demonstrating that the estimated averaged posterior model probabilities are close to their a priori values. This result shows that posterior model probabilities, which are based on the Bayes factors approximations from the bridge sampling, are unbiased for a large range of different a priori model probabilities.

In fact, we think there is reason to argue that if the Bayes factor estimator works okay for one model prior it will work for all model priors, which is what the analysis demonstrates. Therefore, the analysis with using different model priors is not really needed as part of a normal workflow on Bayes factor analyses. However, varying prior model probabilities may still be helpful to reveal some potential problems. Consider for example a situation where a bug in the Bayesian estimation software leads to shrinkage of all posterior model probabilities towards a value of 0.5. Such a potential bias could be detected by varying the true prior probability across simulations as shown above. Thus, while we do not consider this analysis as part of the standard workflow, it can still be interesting to perform such analyses.

These results on Bayesian model inferences based on Bayes factors in theory suggest that stable and accurate Bayes factors can be computed when using large numbers of posterior MCMC draws (i.e., effective sample size). Moreover, the SBC results show that the resulting Bayes factor estimates deviated from the true Bayes factor in some of the example cases (specifically when the model formulation was incorrect), demonstrating that SBC is needed to judge whether Bayes factor estimates are accurate. Based on these results on the average theoretical performance of Bayes factor estimation, we next turn to a different issue - i.e., how Bayes factors depend on and vary with varying data, leading to bad performance in individual cases despite good average performance.

\hypertarget{DataSensitivity}{%
\subsection{Data and prior sensitivity}\label{DataSensitivity}}

\hypertarget{variation-associated-with-the-data-subjects-items-and-residual-noise}{%
\subsubsection{Variation associated with the data (subjects, items, and residual noise)}\label{variation-associated-with-the-data-subjects-items-and-residual-noise}}

A second, and very different, source limiting robustness of Bayes factor estimates derives from the variability that is observed with the data, i.e., among subjects, items, and residual noise. Thus, when repeating an experiment a second time in a replication analysis, using different subjects and items, will lead to different outcomes of the statistical analysis every time a new replication run is conducted. This limit to robustness is well known in frequentist analyses, as the ``dance of p-values'' (Cumming, 2014), where over repeated replications, p-values are not consistently significant across studies. Instead, the results yield highly different p-values each time a study is replicated, and this can even be observed when simulating data from some known truth and re-running analyses on simulated data sets.
This same type of variability should also be present in Bayesian analyses (also see \url{https://daniellakens.blogspot.com/2016/07/dance-of-bayes-factors.html}). Here we investigate this type of variability in Bayes factor analyses.

\hypertarget{variability-of-the-bayes-factor-prior-predictive-simulations}{%
\subsubsection{Variability of the Bayes factor: Prior predictive simulations}\label{variability-of-the-bayes-factor-prior-predictive-simulations}}

In the section implementing SBC above, we have seen that SBC provided evidence whether a model supported accurate estimation of Bayes factors. Moreover, we looked at average posterior probabilities when the H0 or the H1 was true in the simulations, and thus obtained some hint on whether the data contained a lot of (posterior) information about the true hypotheses in question.
Here, we take a closer look at an SBC simulation set where the diagnostics showed stable and accurate results: we take the SBC simulations where we used bridge sampling with many (\(10,000\)) draws to estimate Bayes factors.
Based on these SBC simulations, we can now take a look at how the posterior probabilities vary across individual simulation runs. Thus, while the average performance may look promising, this leaves unclear whether inferences from individual data sets are stable or highly variable, thus providing information in how far one can rely on individual data sets for reliable inference.

\begin{figure}

{\centering \includegraphics{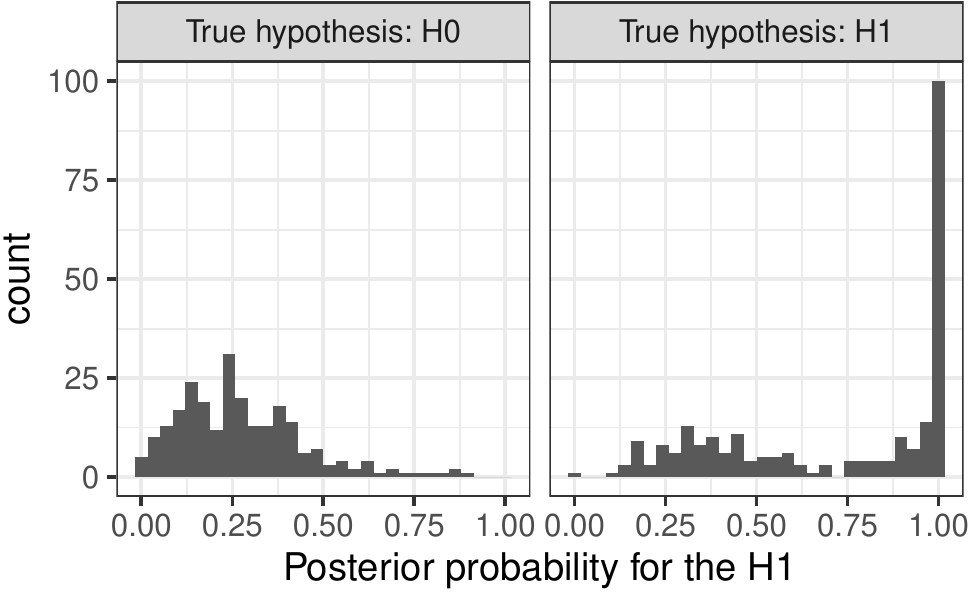} 

}

\caption{Histograms of posterior probabilities for the H1 across 500 simulated data sets, where either the H0 (left panel) or the H1 (right panel) was the true hypothesis in the data simulations. Estimation was using bridge sampling based on 10,000 MCMC draws.}\label{fig:SBCvar1}
\end{figure}

As is shown in Figure~\ref{fig:SBCvar1}, the posterior probabilities widely varied across individual data sets.
We can see that when the H0 was the true hypothesis in the simulations (Fig.~\ref{fig:SBCvar1}, left panel), then posterior model probabilities for the H1 varied quite a bit from a value of \(0\) up to values larger than \(0.5\), and approaching \(1\). Thus, posterior evidence could support either the H0 or the H1, depending on the individual data set. Note, however, that a large proportion of data sets showed a posterior probability for the H1 of smaller than \(0.5\), suggesting they indeed provided more support for the H0.
By contrast, when the H1 was true in the simulations (Fig.~\ref{fig:SBCvar1}, right panel), then quite a few data sets seemed to provide strong evidence for the H1, with a posterior probability for the H1 of close to \(1\). However, there was also a relatively large proportion of data sets that actually provided more support for the H0 (i.e., with posterior probabilities for the H1 of smaller than \(0.5\)), indicating they supported an incorrect hypothesis.
This shows that for the present experimental design, priors, and effect size, an individual data set may provide evidence that is inconsistent with the true hypothesis either for or against the effect. Thus, these individual data sets only have a limited ability to inform inference based on them, and larger data sets and/or larger effect sizes might be needed for reliable inferences based on an individual study.

Thus, results show widely varying posterior model probabilities for multiple data sets which were all simulated from the same prior truth. One option would be to follow up on looking at variability in the context of prior predictive analyses. However, we are here interested in how much information is contained in a typical cognitive data set, which may contain less variation compared to the prior predictive analyses performed above. Therefore, we simulate data based on a posterior model fit for a fairly typical cognitive study.

\hypertarget{variability-of-the-bayes-factor-posterior-simulations}{%
\subsubsection{Variability of the Bayes factor: Posterior simulations}\label{variability-of-the-bayes-factor-posterior-simulations}}

One way to investigate how variable the outcome of Bayes factor analyses can be (given that the Bayes factor is computed in a stable and accurate way), is to run posterior simulations based on a fitted model. That is, one can assume that the truth is approximately known (as approximated by the posterior model fit), and that based on this ``truth'' several artificial data sets are simulated. Computing the Bayes factor analysis again on the simulated data can provide some insight into how variable the Bayes factor will be in a situation where the ``true'' data generating process is always the same, and where variations in Bayes factor results have to be attributed to random noise in participants, items, residual variation, and to uncertainty about the precise true parameter values. Note that we already performed these simulations in the previous prior predictive analyses. However, here we perform more extensive analyses using posterior predictive simulations. We switch from prior predictive analyses to posterior predictive analyses because we are interested in how much information is contained in a typical data set from the cognitive sciences - which may be more information (i.e., less variation) compared to postulating a priori variation in a prior informed by domain expertise only.

\hypertarget{example-inhibitory-and-facilitatory-interference-effects}{%
\subsubsection{Example: Inhibitory and facilitatory interference effects}\label{example-inhibitory-and-facilitatory-interference-effects}}

For this, we will look at some fairly typical experimental example studies from the cognitive sciences. We look at studies that investigated cognitive mechanisms underlying a well-studied phenomenon in sentence comprehension. The example we consider here is the agreement attraction configuration below, where the ungrammatical sentence (2) seems more grammatical than the equally ungrammatical sentence (1):

\begin{enumerate}
\def\labelenumi{(\arabic{enumi})}
\tightlist
\item
  The key to the cabinet are in the kitchen.
\item
  The key to the cabinets are in the kitchen.
\end{enumerate}

Both sentences are ungrammatical because the subject does not agree with the verb in number: The verb (``are'') does not agree in number with the subject of the sentence (``key''). Sentences such as (2) are often found to have shorter reading times at the verb (``are'') compared to (1) (for a meta analysis see Jäger et al., 2017). Such shorter reading times are sometimes referred to as ``facilitatory interference'' (Dillon, 2011). ``Inhibitory interferece'' refers to longer reading times at the verb. One proposal explaining the shorter reading times is that the attractor word (here, cabinets) agrees locally in number with the verb, leading to an illusion of grammaticality. This is an interesting phenomenon as the plural versus singular feature of the attractor noun (``cabinet/s'') is not the subject, and therefore does not need to agree with the verb. That agreement attraction effects are consistently observed indicates that some non-compositional processes are taking place. An account of agreement attraction effects in language processing, that is based on a full computational implementation (which is in the ACT-R framework; Taatgen, Lebiere, \& Anderson, 2006), explains such agreement attraction effects in ungrammatical sentences as a result of retrieval-based working memory mechanisms (Engelmann, Jäger, \& Vasishth, 2019; cf.~Hammerly, Staub, \& Dillon, 2019). Agreement attraction in ungrammatical sentences has been investigated many times in similar experimental setups with different dependent measures such as self-paced reading and eye-tracking. It is generally believed to be a robust empirical phenomenon, and we choose it for analysis here because it provides an example of a relatively robust effect in cognitive science.

\hypertarget{overview-of-the-analyses}{%
\subsubsection{Overview of the analyses}\label{overview-of-the-analyses}}

In this section, we look at the data variability of Bayes factors (and associated effect estimates) using posterior predictive simulations across several different scenarios. First, we investigate a study by Lago, Shalom, Sigman, Lau, and Phillips (2015) using priors (in the model fitting, not in the simulation of data) derived from a meta analysis, where the prior mean differs from zero, and where the data provide some evidence for an effect. Then, we look at this same data set using a more neutral prior that is centered on zero. Next, we use data from a study where the overall effect of interest is close to zero. As one common characteristic, these two data sets are of rather small size (30-60 subjects), which is often the case and rather typical in the cognitive sciences. Here, we also investigate the variability of Bayes factors (and associated effect estimates) in a large sample study, which used 181 subjects, and yields much more stable Bayes factor estimates. Next, we go one step further by looking not at simulated replications of a study, but at ten real empirical replication studies of the same experimental effect. As in the theoretical simulations, also the real empirical data results show strong variability of the Bayes factor across studies, little evidence within each single study, but strong evidence when pooling across individual studies.

\hypertarget{LagoFit}{%
\subsubsection{Case 1: Lago et al.~(2015)}\label{LagoFit}}

First, we investigate facilitatory agreement attraction effects by looking at a self-paced reading study by Lago et al. (2015). We estimate a fixed effect for the experimental condition agreement attraction (\texttt{x}; i.e., sentence type), against a null model where the fixed effect of sentence type is excluded. Note that for the agreement attraction effect of sentence type, we use sum contrast coding (i.e., -1 and +1).
We run a multilevel model with the following formula in \texttt{brms}: \texttt{rt\ \textasciitilde{}\ 1+x\ +\ (1+x\textbar{}subj)\ +\ (1+x\textbar{}item)}, where \texttt{rt} is reading time, we have random variation associated with subjects and with items, and we assume that reading times follow a log-normal distribution: \texttt{family\ \ =\ lognormal()}.

As a next step, we determine priors for the analysis of these data. For this we use results from a meta analysis (Jäger et al., 2017) to obtain priors for the effect size of the factor \texttt{x} agreement attraction. We describe how we obtained the priors in detail below in the section showing an example for a Bayes factor workflow. Here, we simply show the prior that is derived from the meta analysis using \texttt{brms} code:

\begin{Shaded}
\begin{Highlighting}[]
\NormalTok{priors \textless{}{-}}\StringTok{ }\KeywordTok{c}\NormalTok{(}\KeywordTok{set\_prior}\NormalTok{(}\StringTok{"normal(6, 0.5)"}\NormalTok{, }\DataTypeTok{class =} \StringTok{"Intercept"}\NormalTok{),}
            \KeywordTok{set\_prior}\NormalTok{(}\StringTok{"normal({-}0.03, 0.009)"}\NormalTok{, }\DataTypeTok{class =} \StringTok{"b"}\NormalTok{),}
            \KeywordTok{set\_prior}\NormalTok{(}\StringTok{"normal(0, 0.5)"}\NormalTok{, }\DataTypeTok{class =} \StringTok{"sd"}\NormalTok{),}
            \KeywordTok{set\_prior}\NormalTok{(}\StringTok{"normal(0, 1)"}\NormalTok{, }\DataTypeTok{class =} \StringTok{"sigma"}\NormalTok{),}
            \KeywordTok{set\_prior}\NormalTok{(}\StringTok{"lkj(2)"}\NormalTok{, }\DataTypeTok{class =} \StringTok{"cor"}\NormalTok{))}
\end{Highlighting}
\end{Shaded}

Next, using these priors, we fit the Bayesian model using brms:

\begin{Shaded}
\begin{Highlighting}[]
\CommentTok{\# run alternative model}
\NormalTok{m1\_lagoE1 \textless{}{-}}\StringTok{ }\KeywordTok{brm}\NormalTok{(rt }\OperatorTok{\textasciitilde{}}\StringTok{ }\DecValTok{1}\OperatorTok{+}\NormalTok{x }\OperatorTok{+}\StringTok{ }\NormalTok{(}\DecValTok{1}\OperatorTok{+}\NormalTok{x}\OperatorTok{|}\NormalTok{subj)}\OperatorTok{+}\StringTok{ }\NormalTok{(}\DecValTok{1}\OperatorTok{+}\NormalTok{x}\OperatorTok{|}\NormalTok{item),}
                 \DataTypeTok{data    =}\NormalTok{ lagoE1,}
                 \DataTypeTok{family  =} \KeywordTok{lognormal}\NormalTok{(),}
                 \DataTypeTok{prior   =}\NormalTok{ priors,}
                 \DataTypeTok{warmup  =} \DecValTok{2000}\NormalTok{,}
                 \DataTypeTok{iter    =} \DecValTok{10000}\NormalTok{,}
                 \DataTypeTok{cores   =} \DecValTok{4}\NormalTok{,}
                 \DataTypeTok{save\_pars =} \KeywordTok{save\_pars}\NormalTok{(}\DataTypeTok{all =} \OtherTok{TRUE}\NormalTok{),}
                 \DataTypeTok{control =} \KeywordTok{list}\NormalTok{(}\DataTypeTok{adapt\_delta =} \FloatTok{0.99}\NormalTok{,}
                                \DataTypeTok{max\_treedepth=}\DecValTok{15}\NormalTok{))}
\end{Highlighting}
\end{Shaded}

We skip a careful checking of the model here (also see Schad et al., 2021; Betancourt, 2020b), and show these analyses later in the section discussing an example for the Bayes factor workflow.

Next, we take a look at the population-level results from the Bayesian modeling.

\begin{Shaded}
\begin{Highlighting}[]
\KeywordTok{round}\NormalTok{(}\KeywordTok{fixef}\NormalTok{(m1\_lagoE1),}\DecValTok{3}\NormalTok{)}
\end{Highlighting}
\end{Shaded}

\begin{verbatim}
##           Estimate Est.Error   Q2.5  Q97.5
## Intercept    6.015     0.056  5.903  6.127
## x           -0.031     0.008 -0.046 -0.015
\end{verbatim}

They show that for the fixed effect \texttt{x}, capturing the agreement attraction effect, the 95\% credible interval does not overlap with zero. This indicates that the effect may have the expected negative direction, reflecting shorter reading times in the plural condition.
As discussed above, such estimation does not answer the question: How much evidence do we have in support for an effect at all? They may hint that the predictor may be needed to explain the data, but they are not really answering this question how much evidence there is that the parameter is needed to explain the data (see Wagenmakers et al., 2019; Rouder et al., 2018). We cannot draw such an inference, because we did not specify the null hypothesis of zero effect explicitly. Instead, Bayes factors are needed that compare a null model without the effect/parameter to an alternative model that contains the parameter.

To this end, we run the model again, now without the parameter of interest, i.e., the null model, which essentially fixes \(\beta\) to exactly zero: \texttt{rt\ \textasciitilde{}\ 1\ +\ (1+x\textbar{}subj)\ +\ (1+x\textbar{}item)}.

Now everything is ready for computing the log marginal likelihood, that is, the probability of the data given the model, after integrating out the model parameters, which we estimate using bridge sampling as before (Gronau et al., 2017b, 2020). We perform this integration using the function \texttt{bridge\_sampler()} for each of the two models:

\begin{Shaded}
\begin{Highlighting}[]
\CommentTok{\# run bridge sampler}
\NormalTok{lml\_m1\_lagoE1 \textless{}{-}}\StringTok{ }\KeywordTok{bridge\_sampler}\NormalTok{(m1\_lagoE1, }\DataTypeTok{silent =} \OtherTok{TRUE}\NormalTok{)}
\NormalTok{lml\_m0\_lagoE1 \textless{}{-}}\StringTok{ }\KeywordTok{bridge\_sampler}\NormalTok{(m0\_lagoE1, }\DataTypeTok{silent =} \OtherTok{TRUE}\NormalTok{)}
\end{Highlighting}
\end{Shaded}

This gives us the marginal log likelihoods for each of the models. From these, we can compute the Bayes factors.

\begin{Shaded}
\begin{Highlighting}[]
\NormalTok{h\_lagoE1 \textless{}{-}}\StringTok{ }\KeywordTok{bayes\_factor}\NormalTok{(lml\_m1\_lagoE1, lml\_m0\_lagoE1)}
\end{Highlighting}
\end{Shaded}

We use the command \texttt{bayes\_factor(lml\_m1\_lagoE1,\ lml\_m0\_lagoE1)} to specify that we want to compute the Bayes factor between the full model, where the effect of agreement attraction is included, and the null model, where the effect of agreement attraction is absent. It computes the Bayes factor \(BF_{10}\), that is, the evidence of the alternative over the null:

\begin{Shaded}
\begin{Highlighting}[]
\NormalTok{h\_lagoE1}\OperatorTok{$}\NormalTok{bf}
\end{Highlighting}
\end{Shaded}

\begin{verbatim}
## [1] 6.744471
\end{verbatim}

It shows a Bayes factor of \(6\), suggesting that there is some support for the alternative model, which contains the fixed effect of agreement attraction. That is, this provides evidence for the alternative hypothesis that there is a difference between the experimental conditions, i.e., a facilitatory effect in the plural condition of the size derived from the meta analysis.
Under the criteria shown in Table~\ref{tab:BFs}, the Bayes factor provides moderate evidence for an effect of sentence type on reading times.

However, our current purpose is to perform posterior predictive analyses. We can now take the Bayesian hierarchical model fitted to the data above (Lago et al., 2015) and run posterior predictive simulations. In these simulations, in each simulation run (i) one takes a posterior sample for the model parameters (i.e., \(p(\Theta \mid y)\)) and then (ii) uses this sample of model parameters to simulate new data \(\tilde{y}\) from the model \(p(\tilde{y} \mid \Theta)\). That is, posterior predictive simulations are a Bayesian way to perform artificial data simulation. Posterior predictive simulations from the fitted \texttt{brms} model can be performed using the brms-function \texttt{posterior\_predict()}.

\begin{Shaded}
\begin{Highlighting}[]
\NormalTok{pred\_lagoE1 \textless{}{-}}\StringTok{ }\KeywordTok{posterior\_predict}\NormalTok{(m1\_lagoE1)}
\end{Highlighting}
\end{Shaded}

\begin{figure}

{\centering \includegraphics{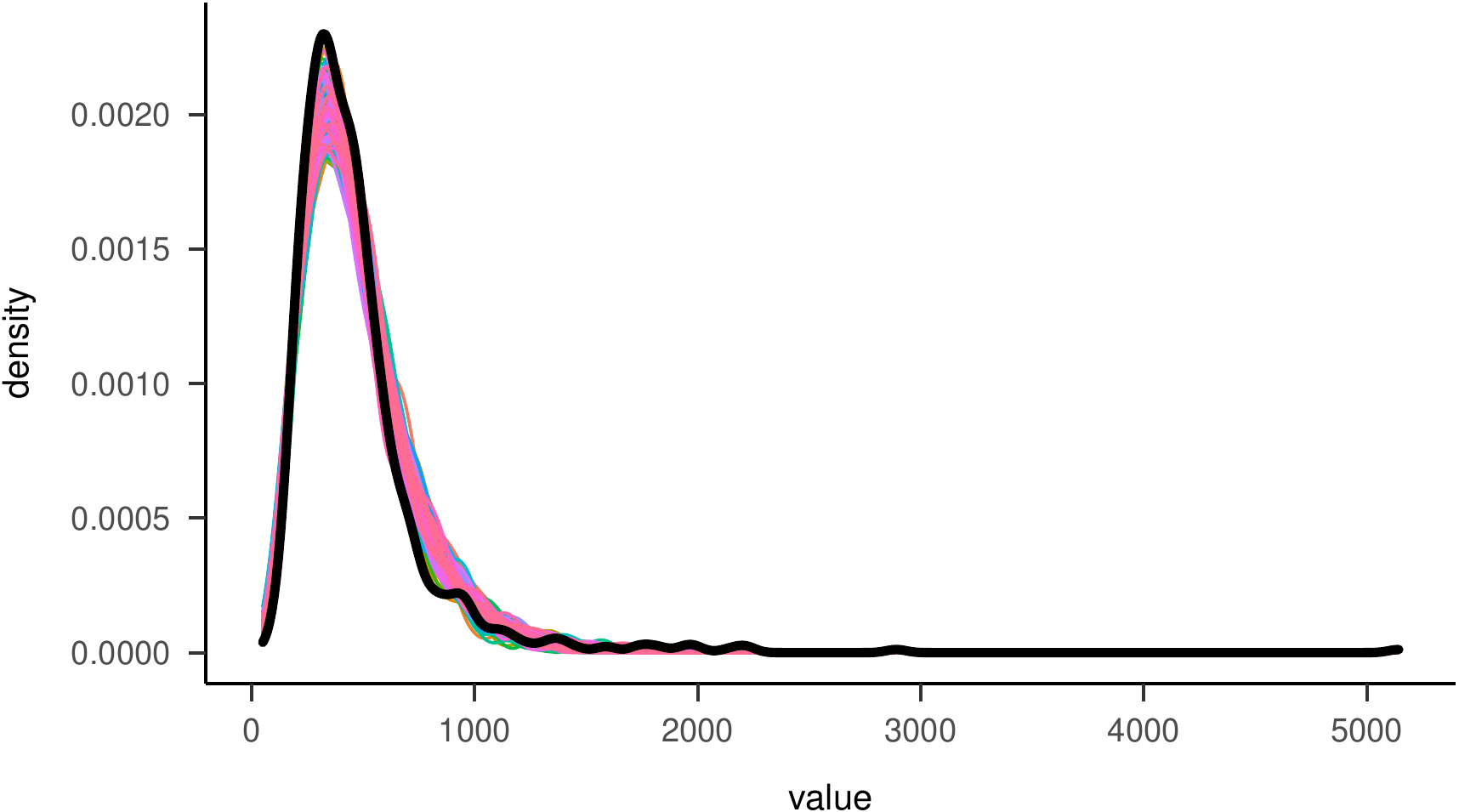} 

}

\caption{Density plots for observed data (black) and for 100 posterior artificial data sets simulated from a fitted Bayesian model (shown in color/grey).}\label{fig:posteriorPredictiveSimulations}
\end{figure}

Figure~\ref{fig:posteriorPredictiveSimulations} visualizes the simulated data via density plots for the observed data (black) and for \(100\) posterior simulated data sets (shown in color/grey). It shows that the simulated data seem fairly well in line with the empirically observed data, at least when investigating the marginal distribution.

The question that we are interested in is, how much information is contained in this posterior simulated data. That is, we can run Bayesian models on this posterior simulated data and compute Bayes factors to test whether in the simulated data there is evidence for agreement attraction effects. Of great interest to us is then the question of how variable the results of these Bayes factor analyses will be across different simulated replications of the same study.
Note that this effectively constitutes a prior predictive simulation with the prior informed by a previous experiment.

We now perform this analysis for \(50\) different artificial data sets simulated from the posterior predictive distribution. For each of these data sets, we can proceed in exactly the same way as we did for the real observed experimental data. That is, \(50\) times, we again fit the same \texttt{brms} model, now to the simulated data, and using the same prior as before. For each simulated data-set, we use bridge sampling to compute the Bayes factor of the alternative model compared to a null model where the agreement attraction effect (fixed effect predictor of sentence type, \texttt{x}) is set to \(0\). For each simulated posterior predictive data set, we store the resulting Bayes factor. We again use the prior from the meta analysis.

We can now visualize the distribution of Bayes factors (\(BF_{10}\)) across posterior predictive distributions by plotting a histogram. Values larger than one in this histogram indicate evidence for the alternative model (H1) that agreement attraction effects exist (i.e., the sentence type effect is different from zero), and Bayes factor values smaller than one indicate evidence for the null model (H0) that no agreement attraction effect exists (i.e., the difference in reading times between experimental conditions is zero).

\begin{figure}

{\centering \includegraphics{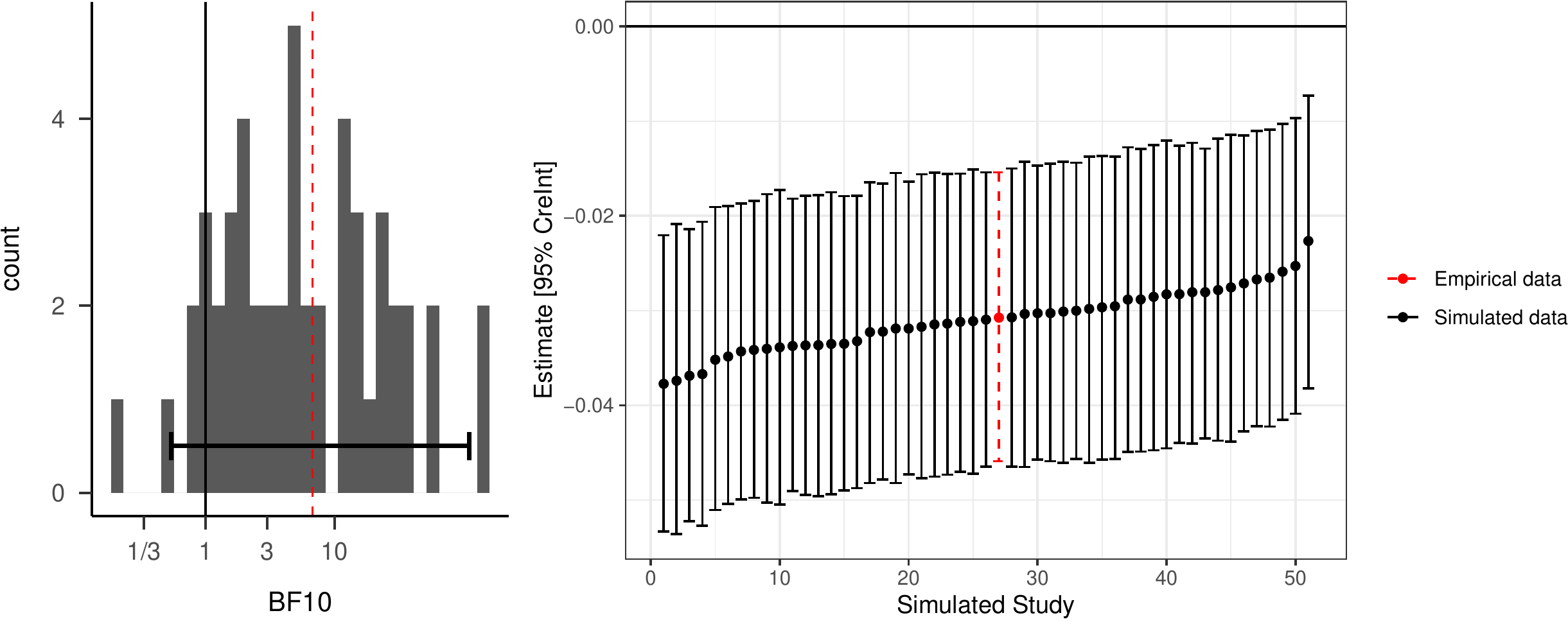} 

}

\caption{Left panel: Histogram of Bayes factors (BF10) of the alternative model over the null model in 50 simulated data sets. The vertical solid black line shows equal evidence for both hypotheses; the dashed red line shows the Bayes factor computed from the empirical data; the horizontal errorbar shows 95 percent of all Bayes factors. Right panel: Estimates of the facilitatory effect of retrieval interference and 95 percent credible intervals across all simulations (black, solid lines) and the empirically observed data (red, dashed line).}\label{fig:plotBFdistrn}
\end{figure}

The results (see Fig.~\ref{fig:plotBFdistrn}) show that the Bayes factors are quite variable. Although all data sets are simulated from the same posterior predictive distribution, the Bayes factor results are as different as providing moderate evidence for the null model (\(BF_{10} < 1/3\)) or providing strong evidence for the alternative model (\(BF_{10} > 10\)). The bulk of the simulated data sets provide moderate or anecdotal evidence for the alternative model. That is, much like the ``dance of p-values'' (Cumming, 2014), this analysis reveals a ``dance of the Bayes factors'' with simulated repetitions of the same study. The variability in these results shows that a typical cognitive or psycholinguistic data set is not necessarily highly informative for drawing firm conclusions about the hypotheses in question.

What is driving these differences in the Bayes factors between simulated data sets? One obvious reason why the outcomes may be so different is that the difference in reading times between the two sentence types, that is, the experimental effect that we wish to make inferences about, may vary based on the noise and uncertainty in the posterior predictive simulations. It is therefore interesting to plot the Bayes factors from this simulated data set as a function of the difference in simulated reading times between the two sentence types as estimated in the Bayesian model. That is, we extract the estimated mean difference in reading times at the verb between plural and singular attractor conditions from the fixed effects of the Bayesian model, and plot the Bayes factor as a function of this difference (together with 95\% credibility intervals).

\begin{figure}

{\centering \includegraphics{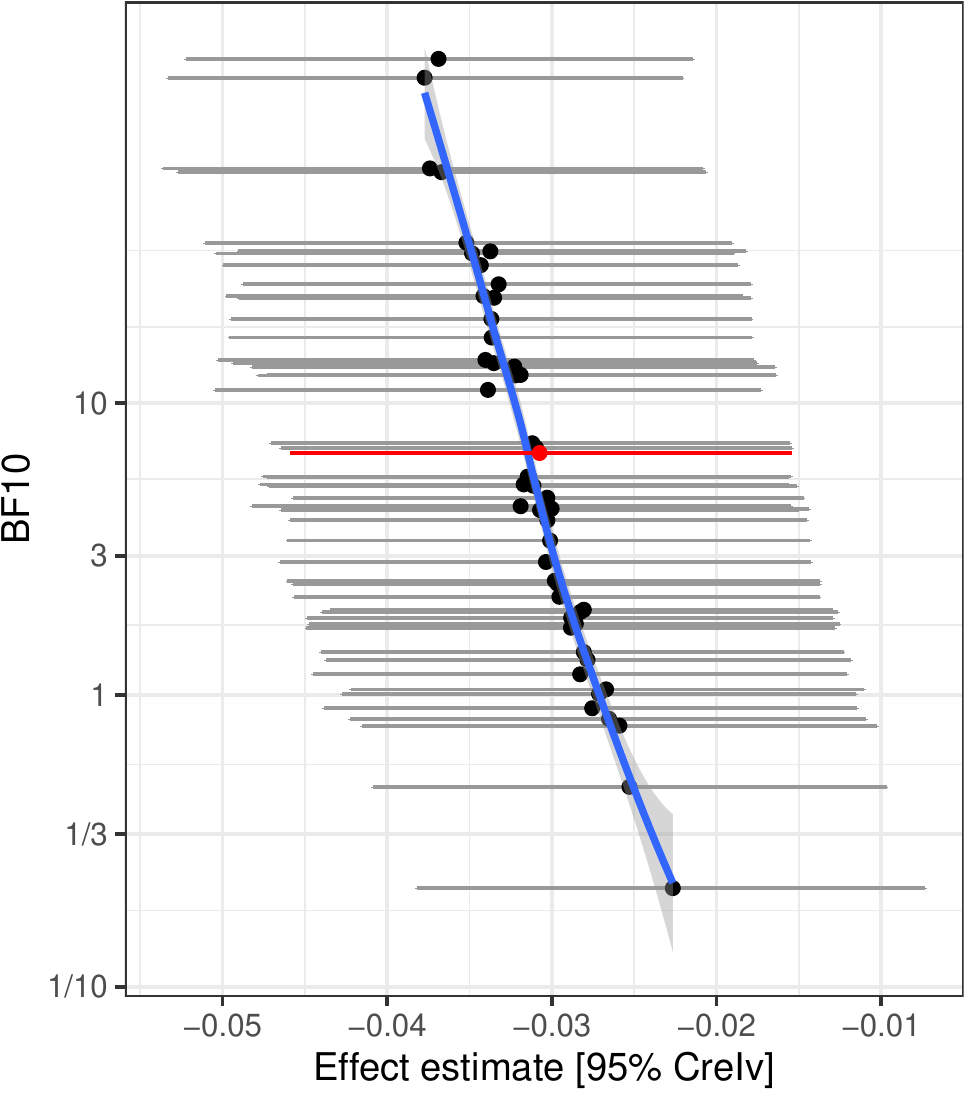} 

}

\caption{Bayes factor (BF10) as a function of the estimate (with 95 percent credible intervals) of the facilitatory effect of retrieval interference across 50 simulated data sets. The prior is from a meta analysis. Analysis results from the empirical data are shown in red.}\label{fig:BFregression}
\end{figure}

The results (displayed in Figure~\ref{fig:BFregression}) show that the mean difference in reading times between experimental conditions varies across posterior predictive simulations. This indicates that the experimental data and design contain a limited amount of information about the effect of interest. Of course, if the (simulated) data is not stable, Bayes factor analyses based on this simulated data cannot be stable across simulations either. Accordingly, as is clear from Figure~\ref{fig:BFregression}, the magnitude of the difference in mean reading times between experimental conditions is indeed a main driving force for the Bayes factor calculations.

One important thing to note in Figure~\ref{fig:BFregression} is that as the difference between reading times becomes more negative,
that is, as the plural noun condition (i.e., ``cabinets'' in the example; sentence 2) is read faster than the singular noun condition (i.e., ``cabinet''; example sentence 1), the Bayes factor BF10 increases to larger and larger values, indicating that the evidence in favor of the alternative model increases. When the difference between reading times becomes \textit{less} negative, by contrast, i.e., the plural condition (sentence 2) is not read much faster than the singular condition (sentence 1), then the Bayes factor BF10 decreases to values smaller than 1. Importantly, this behavior occurs because we are using our informative priors from the meta analysis, where the prior mean for the agreement attraction effect is not centered at a mean of zero, but has a negative value (i.e., a prior mean of \(-0.027\) on the log millisecond scale). Therefore, differences in reading times that are \textit{less} negative / more positive than this prior mean are more in line with a null model of no effect. This also leads to the striking observation that the 95\% credible intervals are quite consistent and all do not overlap with zero, whereas the Bayes factor results are far more variable. Note that computing Bayes factors for such a prior with a non-zero mean asks the very specific question of whether the data provide more evidence for the effect size obtained from the meta analysis compared to the absence of any effect.

\hypertarget{case-2-using-a-prior-centered-on-zero}{%
\subsubsection{Case 2: Using a prior centered on zero}\label{case-2-using-a-prior-centered-on-zero}}

As an alternative, we can also use a centered prior, where the prior mean is zero, that is, we are agnostic with respect to the direction of the effect. We repeat the simulations here with such a mean-centered prior for the agreement attraction effect (using a prior standard deviation of \(0.3\); see below sensitivity analysis).

\begin{figure}

{\centering \includegraphics{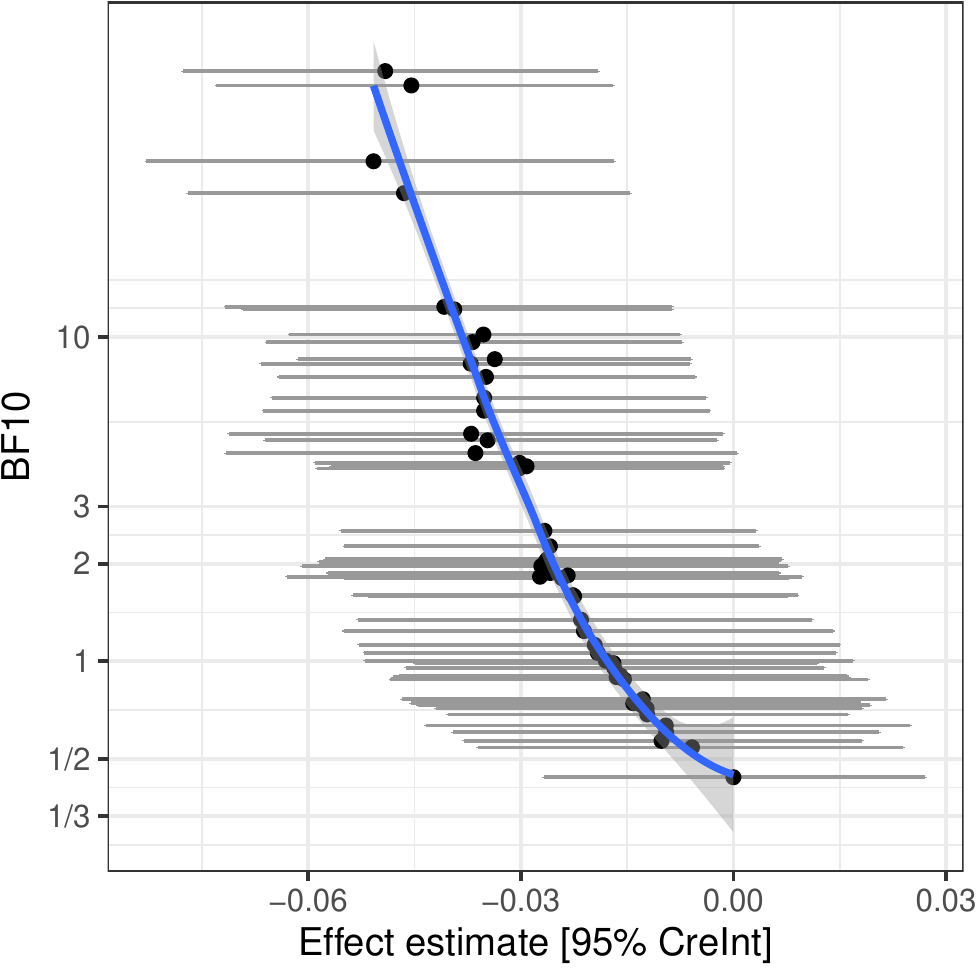} 

}

\caption{Centered prior with mean 0 and standard deviation 0.3. Bayes factor (BF10) as a function of the effect estimate (with 95 percent credible intervals) for 50 simulated studies.}\label{fig:BFregressionPriorN}
\end{figure}

For this changed, centered, prior, the Bayes factors now show a slightly different result. As is displayed in Figure~\ref{fig:BFregressionPriorN}, Bayes factors now follow a hockey-stick function (which would turn into a u-shape in case more positive effect sizes would be present in the data). For large negative differences between reading times in agreement attraction conditions (i.e., left side of the plot), Figure~\ref{fig:BFregressionPriorN} again shows positive values for the Bayes factor BF10, indicating evidence in favor of the alternative hypothesis (i.e., the model including the fixed-effect of agreement attraction). Moreover, when the estimated difference between reading times approaches zero, Bayes factors show values smaller than one (but close to one), indicating support for the null hypothesis. However, support for the null hypothesis is now much less pronounced (only anecdotal, i.e., \(BF_{10} > 1/3\)). This is the case as the alternative hypothesis (H1) now specifies a prior mean of zero for the effect size, such that even an estimated effect size of zero can still be somewhat explained by the alternative model, and the null model is not so much better in explaining the data.

\hypertarget{case-3-a-study-with-an-effect-size-close-to-zero}{%
\subsubsection{Case 3: A study with an effect size close to zero}\label{case-3-a-study-with-an-effect-size-close-to-zero}}

\begin{figure}

{\centering \includegraphics{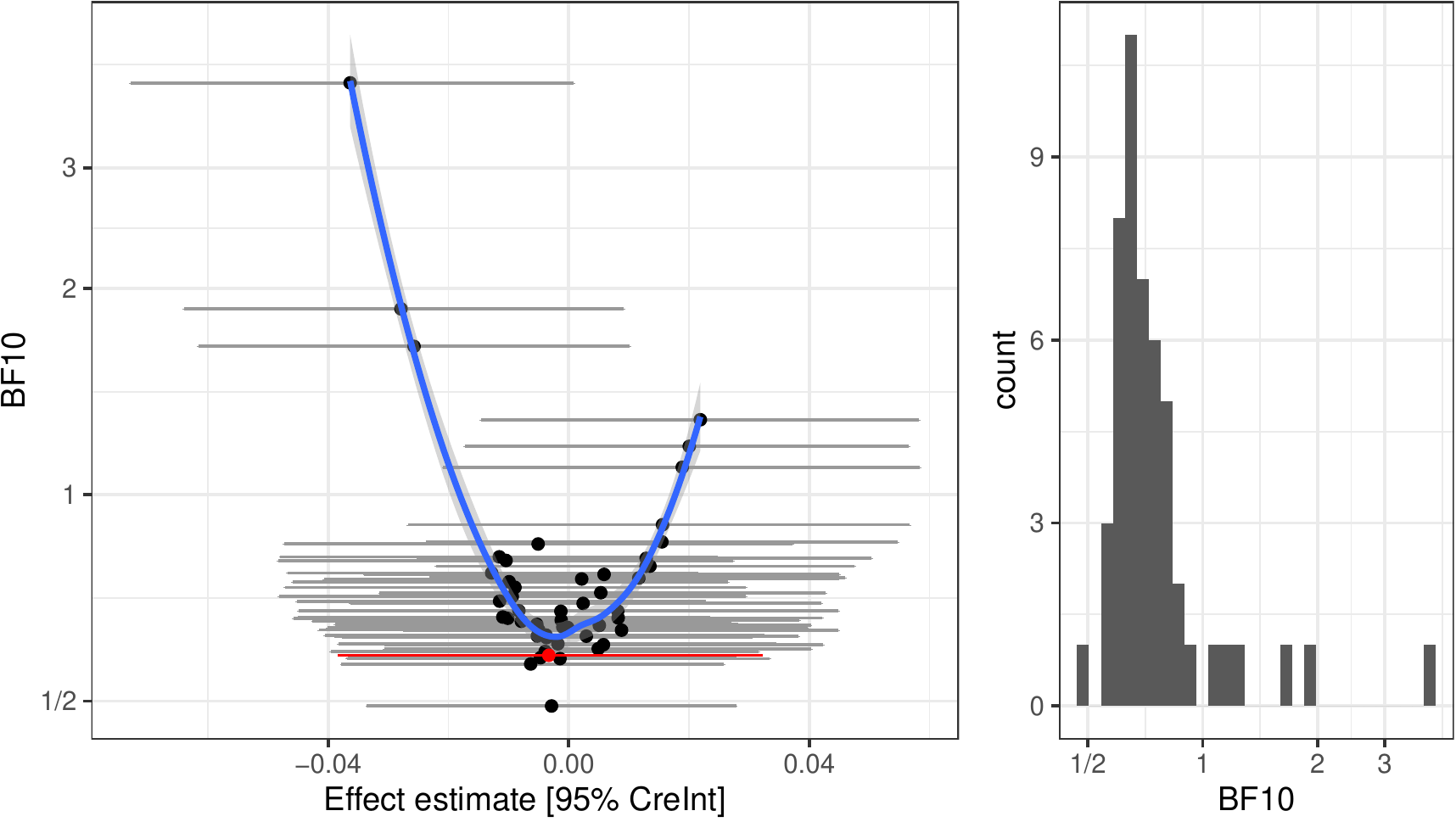} 

}

\caption{Results from a study with no facilitatory effect (Wagers et al., 2009, Exp. 3, singular). Centered prior with mean 0 and standard deviation 0.3. Bayes factor (BF10) as a function of the effect estimate (with 95 percent credible intervals) for 50 simulated studies.}\label{fig:BFregressionPriorWagersE3sg}
\end{figure}

Next, we show an example case where the effect size in the original study is very close to zero, i.e., there is no difference between experimental conditions (Wagers, Lau, \& Phillips, 2009, experiment 3, singular). Therefore, the simulated effect sizes vary between positive and negative values. We again use the mean-centered prior (prior mean = 0, prior standard deviation = 0.3). Figure~\ref{fig:BFregressionPriorWagersE3sg} shows that the Bayes factor gets positive, providing some support for the alternative model not only for negative estimated effect sizes, but also for positive estimated effect sizes. Any difference between experimental conditions - negative or positive - can support the alternative model. Note that this only happens for a centered prior, where the prior mean is zero. For our informative prior based on the meta analysis, which is not centered on zero, by contrast, increasingly positive effect estimates would lead to increasing evidence for the null.

\hypertarget{case-4-a-large-sample-study}{%
\subsubsection{Case 4: A large sample study}\label{case-4-a-large-sample-study}}

\begin{figure}

{\centering \includegraphics{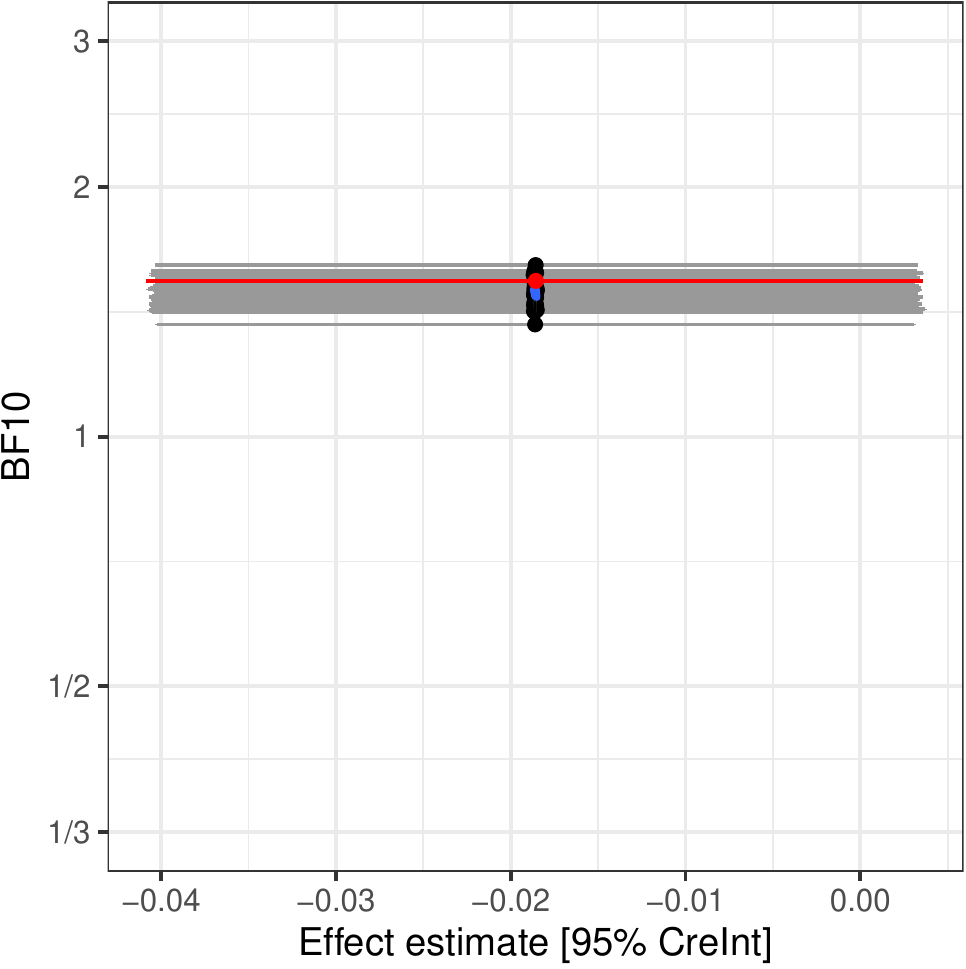} 

}

\caption{Analysis for a large-sample study (Jaeger et al., 2020). Centered prior with mean 0 and standard deviation 0.3. Bayes factor (BF10) as a function of the effect estimate (with 95 percent credible intervals) for 50 simulated studies.}\label{fig:BFregressionPriorJaeger}
\end{figure}

Last, the previous studies had relatively limited sample sizes, e.g., the study by Lago et al. (2015) (experiment 1) had 32 subjects, and the study by Wagers et al. (2009) (experiment 3, singular) had data from 60 subjects. We now want to see how stable Bayes factors are in a situation where the sample size is relatively large. For this, we perform the same Bayes factor analysis for data from a study by Jäger, Mertzen, Van Dyke, and Vasishth (2020), which contains data from 181 subjects. The results are displayed in Figure~\ref{fig:BFregressionPriorJaeger}. They show that with 181 subjects, the Bayes factor is quite stable. That is, across 50 posterior simulated data sets, the Bayes factor computed on the simulated data always ranges between \(1.3\) and \(1.7\). Thus, in large sample studies the ``dance of the Bayes factor'' is very limited to a narrow range, and Bayes factors are quite stable. This may in part be the case because the posterior predictive distribution was so narrow that all the simulated data sets were very much alike.

\hypertarget{sensitivity-analysis}{%
\subsubsection{Sensitivity analysis}\label{sensitivity-analysis}}

In the above example, there was good prior information about the free model parameter \(\beta\) from a meta analysis. However, what happens if we are not sure about the prior for the model parameter? It might happen that we compare the null model with a very ``bad'' alternative model, because our prior for \(\beta\) is not appropriate.

To deal with this situation, many authors use or recommend default prior distributions, where the priors for the model parameters are fixed (e.g., at the scale of an effect size), and are independent of the scientific problem in question, and of potential subjective perspectives on it (Morey \& Rouder, 2011; Navarro, 2015; Rouder, Speckman, Sun, Morey, \& Iverson, 2009; Zellner \& Siow, 1980). While Rouder et al. (2009) provide default priors that are appropriate to generic situations, they also (p.~235) state: ``simply put, principled inference is a thoughtful process that cannot be performed by rigid adherence to defaults.'' In other words, they point out that it is important to consider alternative values for the prior; a sensitivity analysis is necessary.

Sensitivity analysis refers to examining a lot of different alternative models, where each model uses different prior assumptions. This way, it's possible to investigate the extent to which the Bayes factor results depend on, or are sensitive to, the prior assumptions. This is called a sensitivity analysis. Recall that the model is the likelihood \emph{and} the priors. We can therefore compare models that only differ in the prior (for an example see Nicenboim, Vasishth, \& Rösler, 2020). We will next perform a sensitivity analysis for effects of agreement attraction for the data by Lago et al. (2015). This involves running the \texttt{brms} model on the actual observed data using different priors.

What we do is to examine Bayes factors for several models. Each model has the same likelihood but a different prior for \(\beta\) (i.e., the effect of sentence type \texttt{x}). For all of the priors we assume a normal distribution with a mean of zero. Assuming a mean of zero means that we do not make any assumption a priori about the direction of the effect. If the effect should differ from zero, we want the data to tell us that. What differs between the different priors is their standard deviation. That is, what differs is the amount of uncertainty about the effect size that we allow for in the prior. A large standard deviation allows for very large effect sizes a priori, whereas a small standard deviation implies the assumption that we expect the effect not to be very large a priori. Note that while a model with a wide prior (i.e., large standard deviation) also allocates prior probability density to small effect sizes, it allocates much less probability density to small effect sizes. Thus, if the effect size in the observed data is actually small, then a model with a narrow prior (small standard deviation) will have a better chance of detecting the effect.

Note that a sensitivity analysis is a case of inference over model space (i.e., with many different models), where one reports the entire model posterior instead of choosing any particular model (i.e., a particular prior). Importantly, the difference between inference and decision making is critical here. Thus, the posterior provides continuous evidence about models with different priors, but it does not support decision making, i.e., selection of individual models, without using a utility function. We will discuss this critical distinction further below.

Here, we use the same priors as used in the previous analysis of these data, now again with a centered prior mean for the effect of \texttt{x}. That is, the normal distribution for the prior for the agreement attraction effect (i.e., difference in reading times between sentence types, \texttt{x}) now has a mean of zero. Moreover, we now vary its standard deviation, using different values ranging from SD = \(0.005\) to SD = \(0.4\).

We run the \texttt{brms} model for each of the 10 different priors (which differ only in the prior standard deviation for the experimental factor). Then, we compute the Bayes factor using bridge sampling against the null model with \texttt{x} = \(0\), and we store the resulting Bayes factor for each model.

\begin{figure}

{\centering \includegraphics{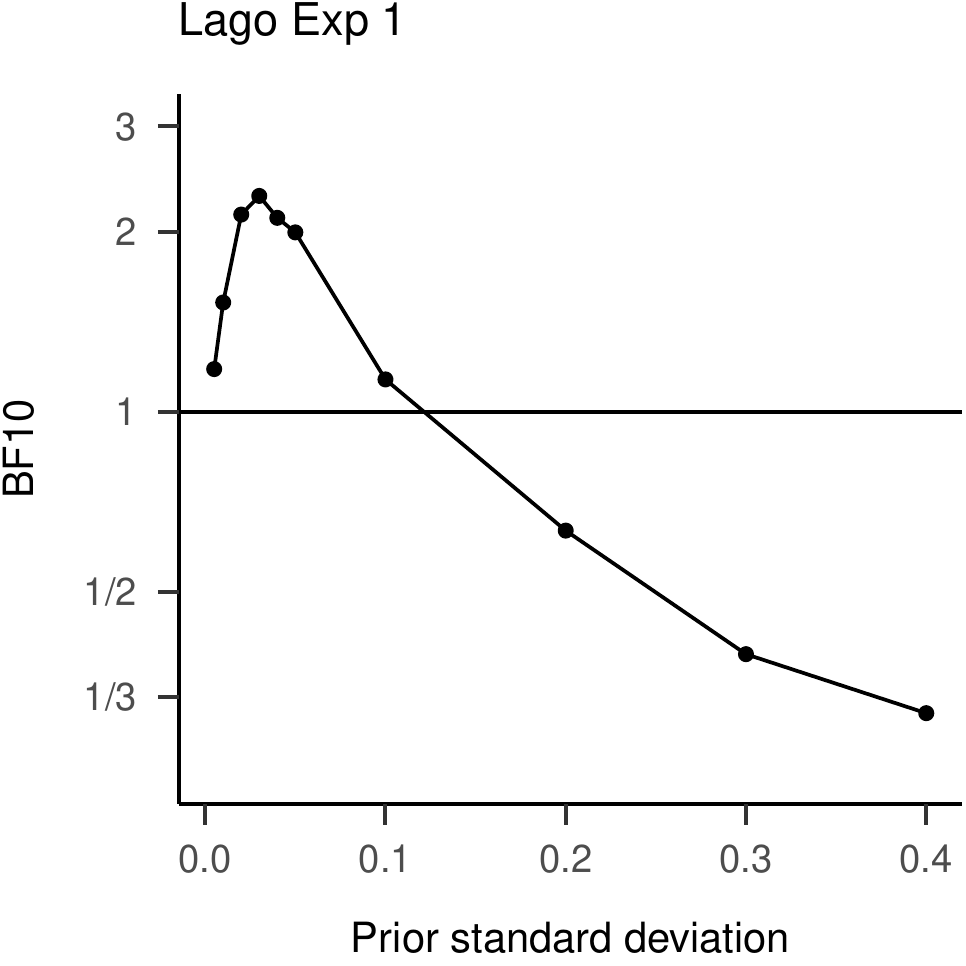} 

}

\caption{Sensitivity analysis: Bayes factor (BF10) as a function of the prior standard deviation.}\label{fig:sensAnalysPlot}
\end{figure}

Finally, we plot the Bayes factors as a function of the prior standard deviation (see Fig.~\ref{fig:sensAnalysPlot}). We show the \(BF_{10}\), that is, the evidence for the alternative model over the null model.

The results show that there is very little evidence for an agreement attraction effect in the sensitivity analysis. The Bayes factors provide little evidence for the alternative model for small prior standard deviations, i.e., for small effect sizes (the maximum lies at a standard deviation of \(0.03\)), but this evidence is only anecdotal. At the same time there is anecdotal evidence against agreement attraction effects for models with a larger prior standard deviation, that is, there is anecdotal evidence against agreement attraction effects with large effect sizes.
Conceptually, the data do not fully support such big effect sizes, but start to favor the null model relatively more when such big effect sizes are tested against the null.
Indeed, Bayes factors explicitly penalize models with wide priors if the data aren't consistent with large effect sizes. This is the effect of Occam's razor that we discussed above (see Fig.~\ref{fig:OccamFactor}).
Note that these results do not directly support a decision to pick the model with standard deviation of \(0.03\) as the best model (without using utility functions) - the evidence in posterior inference is continuous rather than discrete.

The reason that the conclusion differs (sometimes dramatically) as a function of the prior is that priors are never uninformative when it comes to Bayes factors. The wide priors specify that we expect very large effect sizes (with some considerable probability), and there is relatively little evidence in the data for such large effect sizes.

Indeed, very recently, Uri Simonsohn criticized Bayes factors because they might provide evidence in favor of the null and against a very specific alternative model, when the researchers only knew the direction of the effect (see \url{https://datacolada.org/78a}). This can happen when very uninformative vague priors are used, and provide a major motivation for using more informed prior distributions.

Overall, we think that the outcome of the sensitivity analysis of weak evidence is quite reasonable. The result basically shows that the Bayes factors lie somewhere between 3 and 1/3, which all indicate inconclusive results. Given that all results are inconclusive, it doesn't really matter whether the Bayes factors are larger or smaller than one, since no conclusions can be drawn from them anyway.

The above example also again shows the impact of the prior. First, we had performed an analysis with an informative prior derived from a meta analysis with a prior mean of \(-0.027\), and had obtained strong evidence for the alternative hypothesis (\(BF_{10} = 6\)). Here, we use mildly informative priors with prior means of \(0\). The results show that the data does not contain enough information to counteract this mildly informative prior. Thus, conclusive evidence is only obtained under prior beliefs that the average effect is smaller than zero, but not under more agnostic prior beliefs.

\hypertarget{how-consistent-is-the-bayes-factor-across-multiple-studies}{%
\subsubsection{How consistent is the Bayes factor across multiple studies?}\label{how-consistent-is-the-bayes-factor-across-multiple-studies}}

The analyses performed above (section ``Visualize distribution of Bayes factors'') quantified the uncertainty measured via Bayes factors that is inherent in posterior predicted data. This analysis thus showed that even if we hold the true data generating process constant (i.e., we simulate from a given posterior), we observe variability in what inferences the simulated data support.

Here, we go one step further. Instead of relying on simulated replications of the same experiment, we take real data from real empirical replications of the same type of experimental study. This allows us to investigate in how far the results from Bayes factor analyses vary from study to study, even if the same experimental effect is investigated. In particular, we obtained the experimental data from a set of different studies that have one thing in common: they all investigate (inter alia) agreement attraction in ungrammatical sentences (Dillon, Mishler, Sloggett, \& Phillips, 2013; Lago et al., 2015; Wagers et al., 2009).

All of these data sets use similar experimental manipulations to study agreement attraction effects during sentence processing. They all investigate reading time on a target word, measured via self-paced reading or via eye-tracking. There are of course important differences between the studies: some investigate English, others Spanish; and the syntactic configurations differ across studies. However, they all investigate the same basic type of effect, agreement attraction in ungrammatical configurations, and are therefore trying to estimate the same basic effect. Importantly, agreement attraction is generally thought to be a robust empirical phenomenon (Phillips, Wagers, \& Lau, 2011); this example therefore provides an example case for an empirically well-established effect in the cognitive sciences.

For all of these studies, we focus on the question of whether there is evidence for a difference in mean reading times between sentence types. Again, we use Bayesian modeling using \texttt{brms} for posterior estimation, we assume a zero-centered prior for the agreement attraction effect, and we compute Bayes factors using bridge sampling. We now perform a sensitivity analysis for each of the data sets separately.

\begin{figure}

{\centering \includegraphics{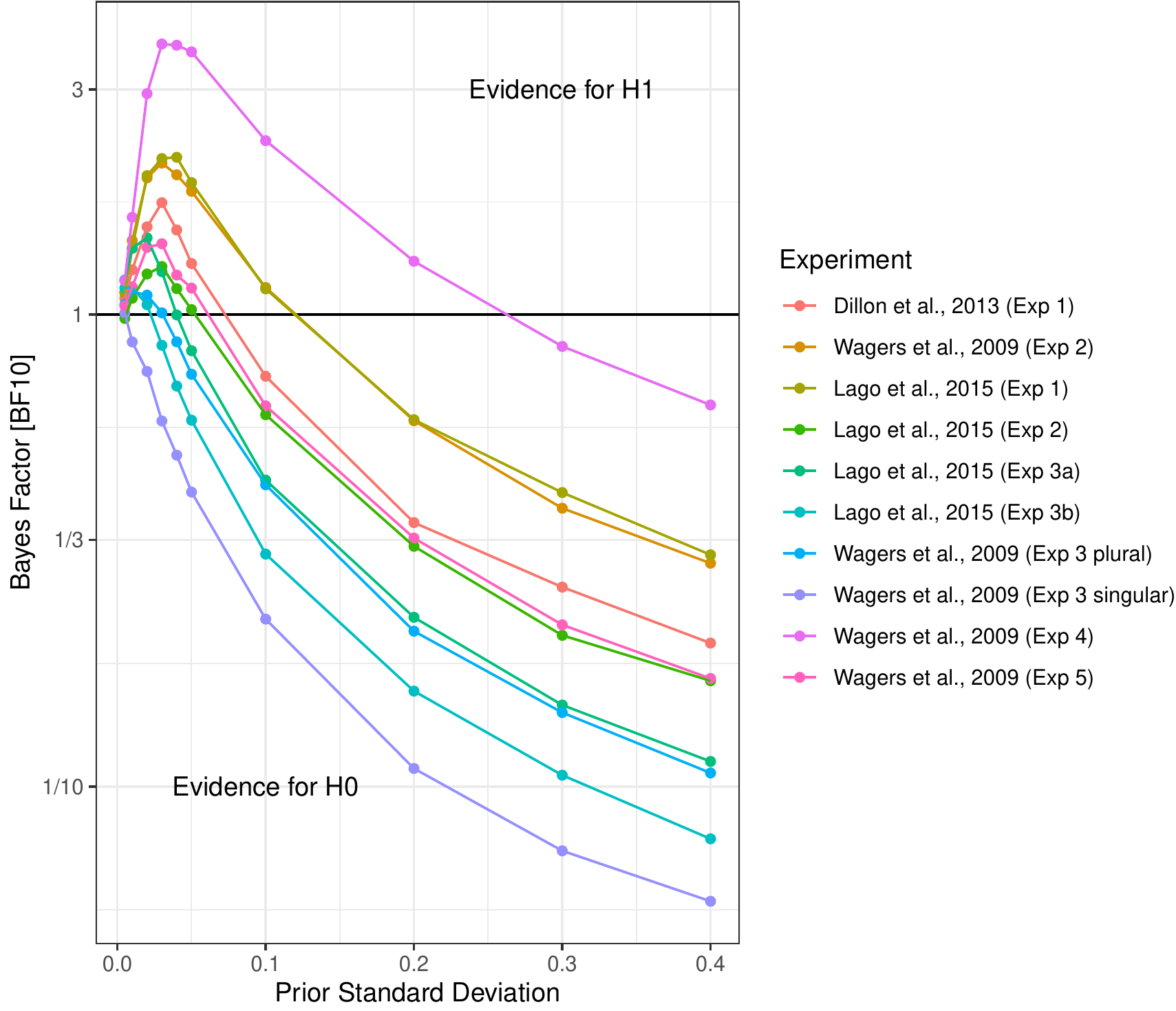} 

}

\caption{Prior sensitivity analyses for different empirical data sets, each implementing a replication study of interference effects of number attraction in sentence comprehension. For each empirical study (indicated by different colors), the Bayes factor (BF10) of the alternative model against the null model is shown for different prior standard deviations for the size of the experimental effect.}\label{fig:plotSens}
\end{figure}

Figure~\ref{fig:plotSens} visualizes the results of this analysis. It shows that the evidence in support of agreement attraction effects is very weak for every single analyzed study. One study (Wagers et al., 2009, Experiment 4) shows at least moderate evidence (\(BF_{10} > 3\)) for an interference effect of small size (the prior standard deviation of \(0.040\) shows the largest Bayes factor). Moreover, several of the other studies also show some evidence for small agreement attraction effects, but this evidence is anecdotal at best, with maximal Bayes factors ranging between 1 and 3.

What the analysis consistently shows, however, is that (i) in all studies the estimated effect is in the expected direction, i.e., it is negative, and (ii) all studies provide evidence against a large agreement attraction effect. For the largest studied prior standard deviation of \(0.4\), the results show at least moderate evidence for the null model and against the alternative for 9 out of the 10 data sets, and two data sets actually provide strong evidence (\(BF_{10} < 1/10\)) against such a large prior effect size.

Moreover, the analysis reveals large variability in the results across data sets. While analysis of some data sets suggest tentative evidence for the alternative model, supporting agreement attraction effects in sentence comprehension (e.g., Wagers et al., 2009, Experiment 4), other data sets show no evidence for agreement attraction effects at all (e.g., Wagers et al., 2009, Experiment 3, singular).

This analysis shows that Bayes factors used in a sensitivity analysis can quantify the evidence in favor of a range of different hypotheses. The evidence varies considerably with the data set even though we investigate different experimental investigations of the same phenomenon of agreement attraction. However, the prior sensitivity analyses of the different data sets are consistent in that for small expected effect sizes, none of them provides strong evidence, neither in support of the H1 nor in support of the H0, whereas all data sets provide some evidence against large effect sizes.

That no data set provides strong evidence for the H1 might be quite surprising to the reader, given that agreement attraction effects are generally thought to be a robust empirical phenomenon. What we illustrate here is that individual studies may in fact carry quite limited information about a fairly standard experimental effect. Indeed, standard experimental designs and sample sizes may be insufficiently sensitive to accurately detect a typical cognitive effect such as agreement attraction. Evidence synthesis through
meta analyses will be needed to make clear inferences about the effect of agreement attraction (cf.~Nicenboim et al., 2020).

Meta analyses can be performed using Bayesian modeling, and again, Bayes factors can be used to quantify the evidence a meta analysis provides in favor of some hypothesis. We illustrate this point here. First, we run frequentist linear mixed effects models for each of the data sets (using the R function \texttt{lmer()}). From each analysis, we save the estimated agreement attraction effect and its associated standard error.

Based on these estimates, we perform a Bayes factor meta analysis:

\begin{Shaded}
\begin{Highlighting}[]
\NormalTok{priorsM \textless{}{-}}\StringTok{ }\KeywordTok{c}\NormalTok{(}\KeywordTok{set\_prior}\NormalTok{(}\StringTok{"normal(0, 1)"}\NormalTok{, }\DataTypeTok{class =} \StringTok{"b"}\NormalTok{),}
             \KeywordTok{set\_prior}\NormalTok{(}\StringTok{"normal(0, 0.5)"}\NormalTok{, }\DataTypeTok{class =} \StringTok{"sd"}\NormalTok{))}

\CommentTok{\# run alternative model H1 with different priors}
\NormalTok{lml\_brm1 \textless{}{-}}\StringTok{ }\KeywordTok{list}\NormalTok{()}
\ControlFlowTok{for}\NormalTok{ (j }\ControlFlowTok{in} \DecValTok{1}\OperatorTok{:}\KeywordTok{length}\NormalTok{(priSD)) \{}
  
\NormalTok{  priorsM[}\DecValTok{1}\NormalTok{,] \textless{}{-}}\StringTok{ }\KeywordTok{set\_prior}\NormalTok{(}\KeywordTok{paste0}\NormalTok{(}\StringTok{"normal(0, "}\NormalTok{,priSD[j],}\StringTok{")"}\NormalTok{),}\DataTypeTok{class =} \StringTok{"b"}\NormalTok{)}
  
\NormalTok{  m.brm1 \textless{}{-}}\StringTok{ }\KeywordTok{brm}\NormalTok{(b }\OperatorTok{|}\StringTok{ }\KeywordTok{se}\NormalTok{(SE) }\OperatorTok{\textasciitilde{}}\StringTok{ }\DecValTok{0}\OperatorTok{+}\NormalTok{Intercept }\OperatorTok{+}\StringTok{ }\NormalTok{(}\DecValTok{1}\OperatorTok{|}\NormalTok{expt),}
             \DataTypeTok{data =}\NormalTok{ lmerResults,}
             \DataTypeTok{prior =}\NormalTok{ priorsM, }\DataTypeTok{save\_pars =} \KeywordTok{save\_pars}\NormalTok{(}\DataTypeTok{all =} \OtherTok{TRUE}\NormalTok{),}
             \DataTypeTok{iter =} \DecValTok{4000}\NormalTok{, }\DataTypeTok{control=}\KeywordTok{list}\NormalTok{(}\DataTypeTok{adapt\_delta=}\FloatTok{0.95}\NormalTok{))}
\NormalTok{  lml\_brm1[[j]] \textless{}{-}}\StringTok{ }\KeywordTok{bridge\_sampler}\NormalTok{(m.brm1, }\DataTypeTok{silent =} \OtherTok{TRUE}\NormalTok{)}
\NormalTok{\}}

\CommentTok{\# run null model H0}
\NormalTok{m.brm0 \textless{}{-}}\StringTok{ }\KeywordTok{brm}\NormalTok{(b }\OperatorTok{|}\StringTok{ }\KeywordTok{se}\NormalTok{(SE) }\OperatorTok{\textasciitilde{}}\StringTok{ }\DecValTok{0} \OperatorTok{+}\StringTok{ }\NormalTok{(}\DecValTok{1}\OperatorTok{|}\NormalTok{expt),}
             \DataTypeTok{data =}\NormalTok{ lmerResults,}
             \DataTypeTok{prior =}\NormalTok{ priorsM[}\OperatorTok{{-}}\DecValTok{1}\NormalTok{,], }\DataTypeTok{save\_pars =} \KeywordTok{save\_pars}\NormalTok{(}\DataTypeTok{all =} \OtherTok{TRUE}\NormalTok{),}
             \DataTypeTok{iter =} \DecValTok{4000}\NormalTok{, }\DataTypeTok{control=}\KeywordTok{list}\NormalTok{(}\DataTypeTok{adapt\_delta=}\FloatTok{0.95}\NormalTok{))}
\NormalTok{lml\_brm0 \textless{}{-}}\StringTok{ }\KeywordTok{bridge\_sampler}\NormalTok{(m.brm0, }\DataTypeTok{silent =} \OtherTok{TRUE}\NormalTok{)}

\NormalTok{BF\_ln \textless{}{-}}\StringTok{ }\KeywordTok{c}\NormalTok{()}
\ControlFlowTok{for}\NormalTok{ (j }\ControlFlowTok{in} \DecValTok{1}\OperatorTok{:}\KeywordTok{length}\NormalTok{(priSD)) }\CommentTok{\# j \textless{}{-} 1}
\NormalTok{  (BF\_ln[j] \textless{}{-}}\StringTok{ }\KeywordTok{bayes\_factor}\NormalTok{(lml\_brm1[[j]], lml\_brm0)}\OperatorTok{$}\NormalTok{bf)}

\NormalTok{metaResults \textless{}{-}}\StringTok{ }\KeywordTok{data.frame}\NormalTok{(priSD,BF\_ln)}
\end{Highlighting}
\end{Shaded}

\begin{figure}

{\centering \includegraphics{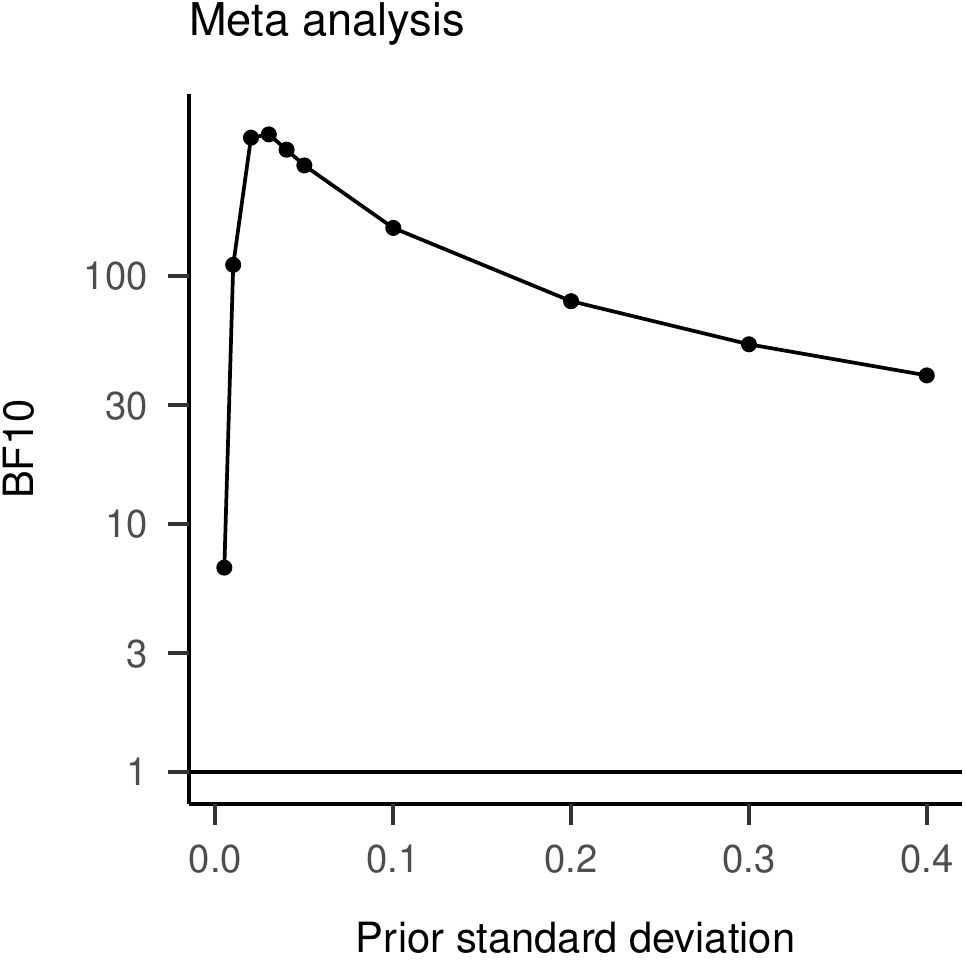} 

}

\caption{Sensitivity analysis for the Bayesian meta analysis: the Bayes factor (BF10) as a function of the prior standard deviation provides extreme evidence in favor of the effect.}\label{fig:metaAnalysPlot}
\end{figure}

The results from this meta analysis using a sensitivity analysis with the Bayes factor (see Figure~\ref{fig:metaAnalysPlot}) shows that across studies, there is extreme evidence (\(BF_{10} > 100\)) for the alternative hypothesis that agreement attraction effects exist. Thus, while the individual studies, each considered separately, do not provide much evidence for the effect, combining studies into a Bayesian meta analysis clearly shows that the effect exists.

Note that because we have all the raw data available for all 10 studies here, instead of running a Bayesian meta analysis based on frequentist test statistics, we can also run one large hierarchical Bayesian model that captures the data from all 10 studies at the same time and treats experiment as a random effect. The formula for this model could be: \texttt{rt\ \textasciitilde{}\ 1+x\ +\ (1+x\textbar{}subj)\ +\ (1+x\textbar{}item)\ +\ (1+x\textbar{}expt)}. And we can use this multilevel model to perform meta analytic Bayes factor analyses. Note that due to the large amount of data involved in analyzing all data sets simultaneously, this analysis needs a very large number of MCMC draws for computing stable Bayes factors, and we therefore skip it here for brevity.

\hypertarget{PoorlyCalibratedDecisions}{%
\subsection{Poorly calibrated decisions}\label{PoorlyCalibratedDecisions}}

Above (in the section on Estimation Error), we have used SBC to calibrate the accuracy of Bayes factor computations. Interestingly, these simulated data sets can also be used to calibrate decisions based on the Bayesian evidence, which is what we turn to here.

\hypertarget{using-sbc-simulations-to-calibrate-decisions}{%
\subsubsection{Using SBC simulations to calibrate decisions}\label{using-sbc-simulations-to-calibrate-decisions}}

Above, we had used SBC to calibrate the continuous evidence obtained by computing Bayes factors. An alternative way to look at the results from the SBC analysis is to use thresholds on the Bayes factor to make discrete decisions, such as the decision to declare discovery. Frequently, such decisions are made by relying on conventions, such as e.g., declaring discovery when a Bayes factor is larger than \(10\). However, note that to perform such decisions in a principled way, utility functions are needed that define the utility of each decision in light of the underlying truth. We illustrate this below (see section ``Principled decisions using utility functions'').

In a first approach, we here use one threshold sometimes used in practice, i.e.~that a Bayes factor larger than 10 provides (strong) evidence for the alternative hypothesis (H1), a Bayes factor smaller than 1/10 provides (strong) evidence for the null hypothesis (H0), and a Bayes factor between 1/10 and 10 provides ``moderate'' or ``anecdotal'' evidence for either hypothesis. For simplicity, we here use the thresholds of 10 and 1/10. We can look at what decisions the Bayes factors support by looking at the simulated data from the SBC (see section on ``Estimation error'' above, subsection ``Simulation-based calibration: Recovering the prior from the data''; i.e., the simulation using bridge sampling with many MCMC draws). We investigate decisions based on whether the H0 or the H1 was actually used to simulate the artificial data.

\begin{Shaded}
\begin{Highlighting}[]
\NormalTok{pDatBF \textless{}{-}}\StringTok{ }\KeywordTok{data.frame}\NormalTok{(}\DataTypeTok{Evidence\_H0=}\NormalTok{BF10\_SBC}\OperatorTok{\textless{}=}\DecValTok{1}\OperatorTok{/}\DecValTok{10}\NormalTok{, }
                     \DataTypeTok{Evidence\_H1=}\NormalTok{BF10\_SBC}\OperatorTok{\textgreater{}=}\DecValTok{10}\NormalTok{,}
                     \DataTypeTok{No\_Evidence=}\NormalTok{BF10\_SBC}\OperatorTok{\textgreater{}}\DecValTok{1}\OperatorTok{/}\DecValTok{10} \OperatorTok{\&}\StringTok{ }\NormalTok{BF10\_SBC}\OperatorTok{\textless{}}\DecValTok{10}\NormalTok{,}
                     \DataTypeTok{true\_hypothesis=}\NormalTok{true\_hypothesis)}
\NormalTok{plyr}\OperatorTok{::}\KeywordTok{ddply}\NormalTok{(pDatBF, }\StringTok{"true\_hypothesis"}\NormalTok{, plyr}\OperatorTok{::}\NormalTok{summarize, }
            \DataTypeTok{Evidence\_H0=}\KeywordTok{round}\NormalTok{(}\KeywordTok{mean}\NormalTok{(Evidence\_H0,}\DataTypeTok{na.rm=}\OtherTok{TRUE}\NormalTok{)}\OperatorTok{*}\DecValTok{100}\NormalTok{),}
            \DataTypeTok{No\_Evidence=}\KeywordTok{round}\NormalTok{(}\KeywordTok{mean}\NormalTok{(No\_Evidence,}\DataTypeTok{na.rm=}\OtherTok{TRUE}\NormalTok{)}\OperatorTok{*}\DecValTok{100}\NormalTok{), }
            \DataTypeTok{Evidence\_H1=}\KeywordTok{round}\NormalTok{(}\KeywordTok{mean}\NormalTok{(Evidence\_H1,}\DataTypeTok{na.rm=}\OtherTok{TRUE}\NormalTok{)}\OperatorTok{*}\DecValTok{100}\NormalTok{))}
\end{Highlighting}
\end{Shaded}

\begin{verbatim}
##   true_hypothesis Evidence_H0 No_Evidence Evidence_H1
## 1              H0          13          87           0
## 2              H1           0          52          48
\end{verbatim}

The results show that when the H0 was actually true in the data simulation, the Bayes factor provided no strong evidence (\(10 > BF_{10} > 1/10\)) in 87\% of cases, and provided evidence for the H0 in only 13\% of simulations. However, when the H0 was true, it never decided for the H1, reflecting a false discovery rate (FDR) or zero. Likewise, when the H1 was actually true in the data simulation, the Bayes factor provided no strong evidence in 52\% simulations (i.e., the true discovery rate, TDR), and provided evidence for the H1 in 48\% of cases. However, it never provided evidence for the H0. These results show that for this example case of a small artificial data set with rather strong effect sizes, the Bayesian decision rule is often uncertain about the true hypothesis, but that it does not decide for the false hypothesis.

An alternative Bayesian decision-rule that is sometimes used in practice is to choose the model that has the highest posterior probability (\texttt{chooseH1\ \textless{}-\ postModelProbsH1\ \textgreater{}\ 0.5}; note that this does not involve the possibility to be undecided).

\begin{verbatim}
##   true_hypothesis No_Discovery Discovery
## 1              H0         0.91      0.09
## 2              H1         0.33      0.67
\end{verbatim}

For the present data set this shows a false discovery rate of 9\% and a true discovery rate of 67\%, again suggesting that the effect size and experimental design in the artificial data set were sufficient for detecting a true effect from the data with a reasonable accuracy.

\hypertarget{bayesian-calibration-of-frequentist-analysis-methods-for-the-same-data}{%
\subsubsection{Bayesian calibration of frequentist analysis methods for the same data}\label{bayesian-calibration-of-frequentist-analysis-methods-for-the-same-data}}

It is possible to compare the results from the Bayesian calibration of Bayes factor analyses with corresponding analyses of frequentist analysis tools for the simulated data. In frequentist analyses, the H0 is rejected if the p-value, i.e., the probability for obtaining an effect as extreme as observed or stronger under the null hypothesis, is small. In frequentist statistics, a finding is considered statistically significant if the p-value is smaller than some threshold, which is conventionally \(p < .05\) (see Benjamin et al., 2018, for an alternative threshold of p \textless{} .005); the H0 is not rejected otherwise, i.e., if \(p > .05\). Unlike in Bayesian data analysis, frequentist null hypothesis significance testing often favors one hard cut-off value, which is conventionally \(p = .05\). Based on such a cut-off, when the H0 was used to generate artificial simulated data, we can compute the number of times that the H0 is falsely rejected, i.e., compute the \(\alpha\) error rate. Moreover, in the cases where the H1 was used to simulate the artificial data, we can compute how often the frequentist model rejects the H0 correctly, reflecting statistical power. For this, we fit a frequentist linear mixed-effects model using the \texttt{lmer} function to each of the \(500\) simulated data sets.

We can now apply the \(p < .05\) cut-off, and compute how often the H0 is rejected when the H0 is true, and how often the H0 is rejected when the H1 is true (note that \(p > .05\) is not a good cut-off to accept the H0, but simply to fail to reject it):

\begin{verbatim}
##     hypothesis_lmer
##      H0 H1
##   H0 97  3
##   H1 41 59
\end{verbatim}

The results show that when the null hypothesis is actually true (first row of the table), the empirical alpha error is estimated as 3\%, which is reasonably close to the expected value of 5\%. When the alternative hypothesis is true (i.e., second row of the table), statistical power is estimated to lie at 59\%. This result shows that the effect size in our artificial example is large enough to detect it in the small data set with intermediate (but not with good) power.

Note that cut-offs different from \(p < .05\) could be used as well, such as \(p < .1\) or \(p < .005\). This would lead to different \(\alpha\) error rates (presumably close to \(0.1\) or to \(0.005\)), and also to other values for the statistical power, where lower values for \(\alpha\) will lead to lower power.

Importantly, this Bayesian calibration analysis is different from a standard frequentist simulation analysis of the \(\alpha\) and \(\beta\) errors. Note that in this Bayesian analysis, we assume uncertainty about the exact effect size, since the priors are specified as distributions. In frequentist analysis, one would often assume a single fixed effect size and compute \(\alpha\) and \(\beta\) errors for exactly this effect size, possibly without considering uncertainty that exist about the precise effect size.

The results from the calibration analyses show similarities and differences in the calibration between the Bayesian decision rule and corresponding frequentist null hypothesis significant testing. The Bayesian and frequentist decisions were similar as both had a fairly good chance of detecting a true H1 from the data (for the Bayes factor decision: 48\%; for the posterior probability decision: 67\%; for the frequentist decision: 59\% of the true H1 were detected). However, the Bayes factor decision-rule distinguished between cases where there was no evidence and cases where support for the H0 could be observed. Thus, the Bayes factor decision-rule (\(BF_{10} < 1/10\)) provided correct support for the H0 in 13\% of the simulations. The frequentist analysis (and the decision rule based on posterior probabilities) did not distinguish between situations of ``no evidence'' versus ``evidence for the H0''.

\hypertarget{principled-decisions-using-utility-functions}{%
\subsubsection{Principled decisions using utility functions}\label{principled-decisions-using-utility-functions}}

The decision-rules that we have studied in the previous sections (based on Bayesian and frequentist analyses) relied on conventions about thresholds that would determine a decision, e.g., on whether to declare discovery. However, as noted before, these conventions provide no principled approach on how to perform decisions in a Bayesian setup.
Therefore, utility functions can be defined to specify the value of the consequences that originate from a given decision-rule. Utility refers to the value that is associated with possible choices under given truths. For example, there may be negative value (negative utility = loss) of falsely acting based on a false hypothesis claim. To decide on actions, utility functions are needed that specify the value of each possible action taken under all possible true states of the world (true hypothesis in this case). The threshold we used above, i.e., to choose the model with the highest posterior probability, is indeed used in much of machine learning, and is optimal if all of the possible decision-truth combinations have equal utility. However, in practice, the different combinations may have different utilities, which necessitates the definition of utility functions.

Here we define an exemplary utility function to support discrete decision-making. Let's assume that we make a decision between 2 options: claiming a discovery or not claiming a discovery. Then let's assume that a true discovery (TD) has utility of \(U_{TD} = 10\) and that missing to claim a true discovery (i.e., a false rejection, FR) has utility \(U_{FR} = -5\). Moreover, we assume that a false discovery (FD) has utility (loss) of \(U_{FD} = -50\), whereas correctly rejecting a discovery (true rejection, TR) has utility \(U_{TR} = 5\).
Note that these numbers seem rather arbitrary for the kind of basic research applications in the cognitive sciences that we have in mind. Thus, it is not clear how to choose these numbers appropriately. However, note that thresholds for p-values or for labeling results from Bayes factor analyses are also quite arbitrary, but fixed by convention.
Importantly, such threshold conventions define implicit utility functions, which may or may not be relevant to a given problem!
Utility analyses explicitly quantify the consequences of different possible actions. Specific utility functions could be agreed upon by research communities and as a result of such agreement, such utility analyses could be used by editors from different journals to decide upon publication based on more liberal/risky or more conservative strategies or publication categories. One problem with utility functions is that it is currently unclear what procedure could be used to quantify such utilities. That is, how can we quantify the utility of a false positive published findings, e.g., measured by the number of false citations. Thus, future research in the cognitive sciences is needed to investigate how utilities can be quantified and linked to evidence, yielding procedures for their definition. Alternatively, given the clear and good utility functions are hard to derive, an alternative approach is to not make any decisions, but rather to communicate continuous evidence.

Next, we can compute the average expected utility given a certain decision threshold. For this, we define an index matrix \(TA\) (``truth-action''), where each column indicates one combination of truth and actions. For example, column one would indicate all cases in the simulations where the H0 was true (i.e., the data was simulated based on the H0), and the decision procedure decided to claim discovery (i.e., false discovery). Column two would indicate cases where the H0 was true and no discovery was claimed (true rejection). Column three would indicate cases where the H1 was true and discovery was claimed (true discovery), and column four indicates cases where the H1 was true and no discovery was claimed (false rejection). Each row of the index matrix \(TA\) corresponds to one simulated data set from the SBC, and for each simulated data set the matrix indicates via a \(1\) which truth-action combination was realized in the SBC simulations with a given decision-rule, whereas all other truth-action combinations are marked with a \(0\). We here define the index matrix \(TA\) in R:

\begin{Shaded}
\begin{Highlighting}[]
\NormalTok{postDat \textless{}{-}}\StringTok{ }\KeywordTok{data.frame}\NormalTok{(true\_hypothesis, chooseH1)}
\KeywordTok{levels}\NormalTok{(postDat}\OperatorTok{$}\NormalTok{true\_hypothesis) \textless{}{-}}\StringTok{ }\KeywordTok{c}\NormalTok{(}\StringTok{"TrueH0"}\NormalTok{,}\StringTok{"TrueH1"}\NormalTok{)}
\NormalTok{postDat}\OperatorTok{$}\NormalTok{act \textless{}{-}}\StringTok{ }\KeywordTok{factor}\NormalTok{(postDat}\OperatorTok{$}\NormalTok{chooseH1,}
                      \DataTypeTok{levels=}\KeywordTok{c}\NormalTok{(}\OtherTok{FALSE}\NormalTok{,}\OtherTok{TRUE}\NormalTok{),}
                      \DataTypeTok{labels=}\KeywordTok{c}\NormalTok{(}\StringTok{"NoDisc"}\NormalTok{,}\StringTok{"Disc"}\NormalTok{))}
\NormalTok{postDat}\OperatorTok{$}\NormalTok{TA\_ \textless{}{-}}\StringTok{ }\KeywordTok{paste0}\NormalTok{(postDat}\OperatorTok{$}\NormalTok{true\_hypothesis,}\StringTok{"."}\NormalTok{,postDat}\OperatorTok{$}\NormalTok{act)}
\KeywordTok{table}\NormalTok{(postDat}\OperatorTok{$}\NormalTok{TA\_)}
\end{Highlighting}
\end{Shaded}

\begin{verbatim}
## 
##   TrueH0.Disc TrueH0.NoDisc   TrueH1.Disc TrueH1.NoDisc 
##            23           222           170            83
\end{verbatim}

\begin{Shaded}
\begin{Highlighting}[]
\NormalTok{mm \textless{}{-}}\StringTok{ }\KeywordTok{data.frame}\NormalTok{(}\KeywordTok{model.matrix}\NormalTok{(}\OperatorTok{\textasciitilde{}}\StringTok{ }\DecValTok{{-}1} \OperatorTok{+}\StringTok{ }\NormalTok{TA\_, }\DataTypeTok{data=}\NormalTok{postDat))}
\KeywordTok{str}\NormalTok{(mm)}
\end{Highlighting}
\end{Shaded}

\begin{verbatim}
## 'data.frame':    498 obs. of  4 variables:
##  $ TA_TrueH0.Disc  : num  0 0 0 0 0 0 1 0 0 0 ...
##  $ TA_TrueH0.NoDisc: num  1 1 1 1 1 0 0 1 1 1 ...
##  $ TA_TrueH1.Disc  : num  0 0 0 0 0 1 0 0 0 0 ...
##  $ TA_TrueH1.NoDisc: num  0 0 0 0 0 0 0 0 0 0 ...
\end{verbatim}

Moreover, we define a vector of utilities \(u\) for these four different possible truth-action combinations:

\begin{Shaded}
\begin{Highlighting}[]
\NormalTok{utility \textless{}{-}}\StringTok{ }\KeywordTok{c}\NormalTok{(}\OperatorTok{{-}}\DecValTok{50}\NormalTok{,}\DecValTok{5}\NormalTok{,}\DecValTok{10}\NormalTok{,}\OperatorTok{{-}}\DecValTok{5}\NormalTok{)}
\KeywordTok{names}\NormalTok{(utility) \textless{}{-}}\StringTok{ }\KeywordTok{names}\NormalTok{(mm)}
\NormalTok{utility}
\end{Highlighting}
\end{Shaded}

\begin{verbatim}
##   TA_TrueH0.Disc TA_TrueH0.NoDisc   TA_TrueH1.Disc TA_TrueH1.NoDisc 
##              -50                5               10               -5
\end{verbatim}

Based on these definitions, we can now compute the average expected utility (averaged across all simulated data sets) as:

\begin{equation}
average\;expected\;utility = \frac{1}{N} \sum_{n=1}^{N} TA \times u
\end{equation}

\begin{Shaded}
\begin{Highlighting}[]
\NormalTok{(avUt \textless{}{-}}\StringTok{ }\KeywordTok{mean}\NormalTok{( }\KeywordTok{as.matrix}\NormalTok{(mm) }\OperatorTok{\%*\%}\StringTok{ }\KeywordTok{t}\NormalTok{(}\KeywordTok{t}\NormalTok{(utility)) ))}
\end{Highlighting}
\end{Shaded}

\begin{verbatim}
## [1] 2.5
\end{verbatim}

In this example, the average expected utility for a decision-rule that chooses the hypothesis with the highest posterior probability is thus 2.50. Here, we used the posterior probabilities for decision making. An alternative approach is to use decision rules based on Bayes factors instead.
Let's use the threshold \(BF_{10} \geq 10\) for a discovery claim and again compute the average expected utility for this decision-rule:

\begin{Shaded}
\begin{Highlighting}[]
\KeywordTok{levels}\NormalTok{(postDat\_SBC1x}\OperatorTok{$}\NormalTok{true\_hypothesis) \textless{}{-}}\StringTok{ }\KeywordTok{c}\NormalTok{(}\StringTok{"TrueH0"}\NormalTok{,}\StringTok{"TrueH1"}\NormalTok{)}
\NormalTok{postDat\_SBC1x}\OperatorTok{$}\NormalTok{chooseH1\_BF \textless{}{-}}\StringTok{ }\NormalTok{postDat\_SBC1x}\OperatorTok{$}\NormalTok{BF10\_SBC }\OperatorTok{\textgreater{}=}\StringTok{ }\DecValTok{10}
\NormalTok{postDat\_SBC1x}\OperatorTok{$}\NormalTok{act \textless{}{-}}\StringTok{ }\KeywordTok{factor}\NormalTok{(postDat\_SBC1x}\OperatorTok{$}\NormalTok{chooseH1\_BF, }
      \DataTypeTok{levels=}\KeywordTok{c}\NormalTok{(}\OtherTok{FALSE}\NormalTok{,}\OtherTok{TRUE}\NormalTok{), }\DataTypeTok{labels=}\KeywordTok{c}\NormalTok{(}\StringTok{"NoDisc"}\NormalTok{,}\StringTok{"Disc"}\NormalTok{))}
\NormalTok{postDat\_SBC1x}\OperatorTok{$}\NormalTok{TA\_ \textless{}{-}}\StringTok{ }\KeywordTok{paste0}\NormalTok{(postDat\_SBC1x}\OperatorTok{$}\NormalTok{true\_hypothesis,}\StringTok{"."}\NormalTok{,postDat\_SBC1x}\OperatorTok{$}\NormalTok{act)}
\NormalTok{postDat\_SBC1x}\OperatorTok{$}\NormalTok{TA\_ \textless{}{-}}\StringTok{ }\KeywordTok{factor}\NormalTok{(postDat\_SBC1x}\OperatorTok{$}\NormalTok{TA\_,}
      \DataTypeTok{levels=}\KeywordTok{c}\NormalTok{(}\StringTok{"TrueH0.Disc"}\NormalTok{,}\StringTok{"TrueH0.NoDisc"}\NormalTok{,}\StringTok{"TrueH1.Disc"}\NormalTok{,}\StringTok{"TrueH1.NoDisc"}\NormalTok{))}
\KeywordTok{table}\NormalTok{(postDat\_SBC1x}\OperatorTok{$}\NormalTok{TA\_)}
\end{Highlighting}
\end{Shaded}

\begin{verbatim}
## 
##   TrueH0.Disc TrueH0.NoDisc   TrueH1.Disc TrueH1.NoDisc 
##             0           245           121           132
\end{verbatim}

\begin{Shaded}
\begin{Highlighting}[]
\NormalTok{mm \textless{}{-}}\StringTok{ }\KeywordTok{data.frame}\NormalTok{(}\KeywordTok{model.matrix}\NormalTok{(}\OperatorTok{\textasciitilde{}}\StringTok{ }\DecValTok{{-}1} \OperatorTok{+}\StringTok{ }\NormalTok{TA\_, }\DataTypeTok{data=}\NormalTok{postDat\_SBC1x))}
\NormalTok{(avUt \textless{}{-}}\StringTok{ }\KeywordTok{mean}\NormalTok{( }\KeywordTok{as.matrix}\NormalTok{(mm) }\OperatorTok{\%*\%}\StringTok{ }\KeywordTok{t}\NormalTok{(}\KeywordTok{t}\NormalTok{(utility)) ))}
\end{Highlighting}
\end{Shaded}

\begin{verbatim}
## [1] 3.564257
\end{verbatim}

\begin{figure}

{\centering \includegraphics{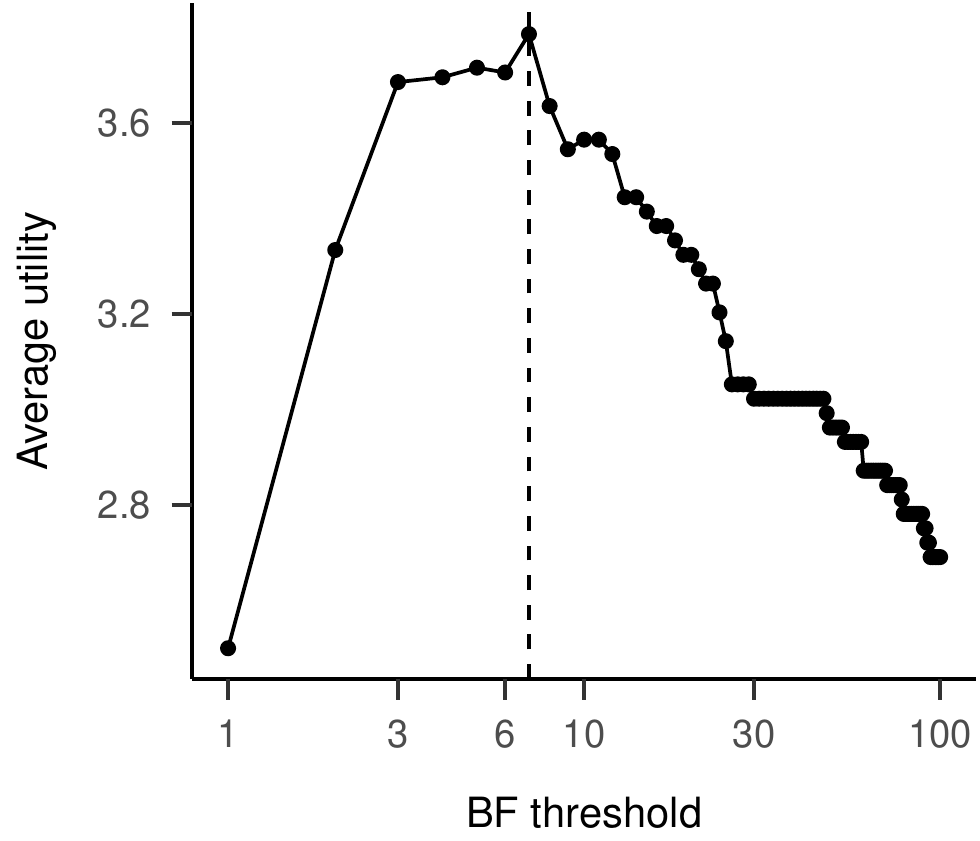} 

}

\caption{Utility for claiming discovery as a function of the critical BF cut-off. The utility for a true discovery is set to 10 and the utility for a false discovery is set to -50. Note that with more (than 500) simulations, the line should become smooth.}\label{fig:utility1}
\end{figure}

Now, the expected average utility is 3.56 and thus higher than before.
We can vary the discovery (Bayes factor) threshold to select the threshold with the highest average utility.
The analysis shows (see Fig.~\ref{fig:utility1}) that for decision-rules using a low value for the Bayes factor threshold, the average utility is low. The largest average utility is obtained for a Bayes factor threshold of 7. For thresholds larger than 7, average utility declines again. Based on this analysis, one could thus call a discovery when the Bayes factor reaches a value of at least 7. The true discovery rates (TDR) and false discovery rates (FDR) for this threshold are:

\begin{verbatim}
##   true_hypothesis No_Discovery Discovery
## 1          TrueH0          100         0
## 2          TrueH1           48        52
\end{verbatim}

This is very close to the results with a Bayes factor threshold of \(10\). Again, the false discovery rate (FDR) is \(0\). However, the true discovery rate (TDR) is now a bit larger and takes a value of 52\%.
Note that while analysis of TDR and FDR provide a good first approach, more elaborate rates can be defined when a ``no decision'' option is possible.

\hypertarget{Example1}{%
\section{Example: Inhibitory and facilitatory interference effects}\label{Example1}}

In the following, we will illustrate the Bayesian workflow using a concrete example from the cognitive sciences.
We again investigate the example on inhibitory and facilitatory interference effects that we described above in the section on data variability. We have described the observational model above. Therefore, we start by describing how we obtained the priors from a meta analysis, and then execute further steps from the Bayes factor workflow.

\hypertarget{determine-priors-using-meta-analysis}{%
\subsubsection{Determine priors using meta analysis}\label{determine-priors-using-meta-analysis}}

Building good priors is a challenging task. Indeed, it is one of the crucial steps involved in a principled Bayesian workflow (Betancourt, 2020b; Gelman et al., 2020; O'Hagan et al., 2006; Schad et al., 2021).
One good way to obtain priors for Bayesian analyses, and specifically for Bayes factor analyses, is to use results from meta analyses on the subject. Here, we take the prior for the experimental manipulation of agreement attraction from a published meta analysis (Jäger et al., 2017).\footnote{Note that this meta analysis already includes the data that we want to make inference about; thus, this meta analysis estimate is not really the right estimate to use, since it involves using the data twice. We ignore this detail here because our goal is simply to illustrate the approach.} It is important to note here that meta analyses almost always have some limitations. First, the studies included can have important differences in implementation and different sources of bias, leading to quite a lot of between-study variability. Second, most of the studies on agreement attraction are severely underpowered (Jäger et al., 2017). This has the effect that biased estimates tend to get published; this naturally biases the meta analysis estimates. As long as one remains aware of these limitations, a meta analysis based estimate can be a reasonable starting point. Moreover, the problems inherent to meta analyses can be partially compensated by widening priors. In other words prior elicitation is more robust when using meta analysis only to get a reasonable order of magnitude instead of a precise shape. Here, we use meta analysis to determine the precise effect size and its uncertainty.

The mean effect size (difference in reading time between the two experimental conditions) in the meta analysis is \(-22\) milliseconds (ms), with \(95\% \;CI = [-36, \; -9]\) (cf., Jäger et al., 2017, Table 4). This means that on average, the target word (i.e., the verb) in sentences such as (2) is read \(22\) milliseconds faster than in sentences such as (1). The size of the effect is measured on the millisecond scale, assuming a normal distribution of effect sizes across studies.

However, individual reading times usually do not follow a normal distribution. Instead, a better assumption about the distribution of reading times is a log-normal distribution. This is what we will assume in the \texttt{brms} model. Therefore, to use the prior from the meta analysis in the Bayesian analysis, we have to transform the prior values from the millisecond scale to log millisecond scale.

For this transformation, we assume an intercept in the log-normal distribution of \(\beta_0\) and a slope of \(\beta_1\) (assuming sum coding -1/+1). Based on this, we know that the difference in reading times between agreement attraction conditions is 22 ms (i.e., \(\text{effSize} = 22\;ms\)), and that this difference can be computed based on \(\beta_0\) and \(\beta_1\) from the log-normal distribution. We can write:

\begin{equation}
22\;ms = \text{effSize} = \exp(\beta_0 + \beta_1) - \exp(\beta_0 - \beta_1)
\end{equation}

What we want to know is the value of \(\beta_1\) that we can assume for our prior. The equation shows that the slope \(\beta_1\) in log-normally distributed data depends on the intercept term \(\beta_0\). Here, we assume an intercept in log-space of \(\beta_0 = 6.0\). This yields a plausible expectation for a mean reading time of \(\exp(6) = 403\) milliseconds (cf., Schad et al., 2021). Based on this intercept term, we compute the effect size (half the difference between the two experimental conditions; reflecting sum contrast coding, i.e., singular = -1 and plural = +1) (Schad, Vasishth, Hohenstein, \& Kliegl, 2020) in log space as the \(\beta_1\) parameter, which yields a value of \(\beta_1 = 0.027\)\footnote{Given an effect size of \texttt{effSize\ =\ -22} and an intercept term of \(\beta_0 = 6.0\), this is computed as: \(\beta_1 = \log \left ( -\text{effSize}/\exp(\beta_0) + \sqrt{(\text{effSize}/\exp(\beta_0))^2+4} \right ) - \log(2)\)}. Adding and subtracting this parameter value to/from the intercept (\(\exp(6) = 403\)) in log space and computing the difference (i.e., \(\exp(6 - 0.027) - \exp(6 + 0.027)\) gives the difference of \(-22\) ms between the two experimental conditions. This simply confirms that our transformation has been computed correctly.

However, we also want to consider the uncertainty about the effect size, given as confidence intervals in the meta analysis. To this end, we compute values of 1 standard error below or above the mean (approximating the standard error based on a normal distribution as the range of the confidence interval divided by 4, i.e., \((-36 - (-9))/4\)), and we compute the corresponding values in log space. Based on this, we take the standard deviation of the normal prior distribution in log space as the average distance between (a) the mean in log space and (b) the values of 1 standard deviation above/below the mean (measured in milliseconds, and transformed into log space). This provides our prior standard deviation (in log-space), informed by the meta analysis (Jäger et al., 2017).

Next, we set the priors for the analysis with \texttt{brms}. Based on the previous calculations, the prior for the experimental factor of interference effects is set to a normal distribution with mean = \(-0.03\) and standard deviation = \(0.009\). For the other model parameters, we use mildly informative priors based on our recent analysis of a principled Bayesian workflow (Schad et al., 2021).

\hypertarget{prior-predictive-checks}{%
\subsubsection{Prior predictive checks}\label{prior-predictive-checks}}

An additional and highly recommended way to obtain appropriate priors (Betancourt, 2020b; Gabry et al., 2019; Good, 1950; Schad et al., 2021) is to perform prior predictive checks. Here, the idea is to simulate data from the model and the priors, and then to analyze the simulated data using summary statistics. For example, it would be possible to compute the summary statistic of the difference in the reading times between agreement attraction conditions (i.e., sentences (1) versus sentences (2)). The simulations would yield a distribution of differences. Arguably, this distribution of differences, that is, the data analyses of the simulated data, are much easier to judge for plausibility than the prior parameters specifying prior distributions. That is, we might find it easier to judge whether a difference in reading times between sentence types is plausible rather than judging the parameters of the model.

Here, we implement exemplary prior predictive checks. We start with the prior \(\beta \sim Normal(-0.03,0.009)\), which was derived from the meta analysis. Moreover, we add prior assumptions for the intercept, the residual standard deviation, for random effects variances, and random effects correlations. For these additional assumptions see the following prior specification, specified using the brms package:

\begin{Shaded}
\begin{Highlighting}[]
\NormalTok{priors \textless{}{-}}\StringTok{ }\KeywordTok{c}\NormalTok{(}\KeywordTok{set\_prior}\NormalTok{(}\StringTok{"normal(6, 0.5)"}\NormalTok{, }\DataTypeTok{class =} \StringTok{"Intercept"}\NormalTok{),}
            \KeywordTok{set\_prior}\NormalTok{(}\StringTok{"normal({-}0.03, 0.009)"}\NormalTok{, }\DataTypeTok{class =} \StringTok{"b"}\NormalTok{),}
            \KeywordTok{set\_prior}\NormalTok{(}\StringTok{"normal(0, 0.5)"}\NormalTok{, }\DataTypeTok{class =} \StringTok{"sd"}\NormalTok{),}
            \KeywordTok{set\_prior}\NormalTok{(}\StringTok{"normal(0, 1.0)"}\NormalTok{, }\DataTypeTok{class =} \StringTok{"sigma"}\NormalTok{),}
            \KeywordTok{set\_prior}\NormalTok{(}\StringTok{"lkj(2)"}\NormalTok{, }\DataTypeTok{class =} \StringTok{"cor"}\NormalTok{))}
\end{Highlighting}
\end{Shaded}

We load the experimental design by Lago et al. (2015) and use this for our prior predictive checks. To this end, we first repeatedly simulate parameters from the priors. For this, we use the custom R function \texttt{SimFromPrior()} (taken from Schad et al., 2021).

\begin{Shaded}
\begin{Highlighting}[]
\NormalTok{nsimPP \textless{}{-}}\StringTok{ }\DecValTok{500}
\NormalTok{beta0 \textless{}{-}}\StringTok{ }\NormalTok{beta1 \textless{}{-}}\StringTok{ }\NormalTok{sigma\_u0 \textless{}{-}}\StringTok{ }\NormalTok{sigma\_u1 \textless{}{-}}\StringTok{ }\NormalTok{sigma\_w0 \textless{}{-}}\StringTok{ }
\StringTok{  }\NormalTok{sigma\_w1 \textless{}{-}}\StringTok{ }\NormalTok{rho\_u \textless{}{-}}\StringTok{ }\NormalTok{rho\_w \textless{}{-}}\StringTok{ }\NormalTok{sigma \textless{}{-}}\StringTok{ }\OtherTok{NA}
\KeywordTok{set.seed}\NormalTok{(}\DecValTok{123}\NormalTok{)}
\ControlFlowTok{for}\NormalTok{ (i }\ControlFlowTok{in} \DecValTok{1}\OperatorTok{:}\NormalTok{nsimPP) \{}
\NormalTok{  beta0[i]    \textless{}{-}}\StringTok{ }\KeywordTok{SimFromPrior}\NormalTok{(priors,}\DataTypeTok{class=}\StringTok{"Intercept"}\NormalTok{,}\DataTypeTok{coef=}\StringTok{""}\NormalTok{)}
\NormalTok{  beta1[i]    \textless{}{-}}\StringTok{ }\KeywordTok{SimFromPrior}\NormalTok{(priors,}\DataTypeTok{class=}\StringTok{"b"}\NormalTok{)}
\NormalTok{  sigma\_u0[i] \textless{}{-}}\StringTok{ }\KeywordTok{SimFromPrior}\NormalTok{(priors,}\DataTypeTok{class=}\StringTok{"sd"}\NormalTok{)}
\NormalTok{  sigma\_u1[i] \textless{}{-}}\StringTok{ }\KeywordTok{SimFromPrior}\NormalTok{(priors,}\DataTypeTok{class=}\StringTok{"sd"}\NormalTok{)}
\NormalTok{  sigma\_w0[i] \textless{}{-}}\StringTok{ }\KeywordTok{SimFromPrior}\NormalTok{(priors,}\DataTypeTok{class=}\StringTok{"sd"}\NormalTok{)}
\NormalTok{  sigma\_w1[i] \textless{}{-}}\StringTok{ }\KeywordTok{SimFromPrior}\NormalTok{(priors,}\DataTypeTok{class=}\StringTok{"sd"}\NormalTok{)}
\NormalTok{  rho\_u[i]    \textless{}{-}}\StringTok{ }\KeywordTok{SimFromPrior}\NormalTok{(priors,}\DataTypeTok{class=}\StringTok{"cor"}\NormalTok{)}
\NormalTok{  rho\_w[i]    \textless{}{-}}\StringTok{ }\KeywordTok{SimFromPrior}\NormalTok{(priors,}\DataTypeTok{class=}\StringTok{"cor"}\NormalTok{)}
\NormalTok{  sigma[i]    \textless{}{-}}\StringTok{ }\KeywordTok{SimFromPrior}\NormalTok{(priors,}\DataTypeTok{class=}\StringTok{"sigma"}\NormalTok{)}
\NormalTok{\}}
\end{Highlighting}
\end{Shaded}

Next, we use these simulated parameters to simulate artificial reading times.

\begin{Shaded}
\begin{Highlighting}[]
\NormalTok{rtfakemat \textless{}{-}}\StringTok{ }\KeywordTok{matrix}\NormalTok{(}\OtherTok{NA}\NormalTok{,}\KeywordTok{nrow}\NormalTok{(lagoE1),nsim)}
\ControlFlowTok{for}\NormalTok{ (i }\ControlFlowTok{in} \DecValTok{1}\OperatorTok{:}\NormalTok{nsim)}
\NormalTok{  rtfakemat[,i] \textless{}{-}}\StringTok{ }\KeywordTok{exp}\NormalTok{(}\KeywordTok{simLMM}\NormalTok{(}
      \DataTypeTok{formula=}\OperatorTok{\textasciitilde{}}\StringTok{ }\NormalTok{x }\OperatorTok{+}\StringTok{ }\NormalTok{(x }\OperatorTok{|}\StringTok{ }\NormalTok{subj) }\OperatorTok{+}\StringTok{ }\NormalTok{(x }\OperatorTok{|}\StringTok{ }\NormalTok{item), }
      \DataTypeTok{dat=}\NormalTok{lagoE1, }
      \DataTypeTok{Fixef=}\KeywordTok{c}\NormalTok{(beta0[i], beta1[i]), }
      \DataTypeTok{VC\_sd=}\KeywordTok{list}\NormalTok{(}\KeywordTok{c}\NormalTok{(sigma\_u0[i], sigma\_u1[i]), }
                 \KeywordTok{c}\NormalTok{(sigma\_w0[i], sigma\_w1[i]), }
\NormalTok{                 sigma[i]),}
      \DataTypeTok{CP=}\KeywordTok{c}\NormalTok{(rho\_u[i], rho\_w[i]), }\DataTypeTok{empirical=}\OtherTok{FALSE}\NormalTok{))}
\end{Highlighting}
\end{Shaded}

\begin{figure}

{\centering \includegraphics{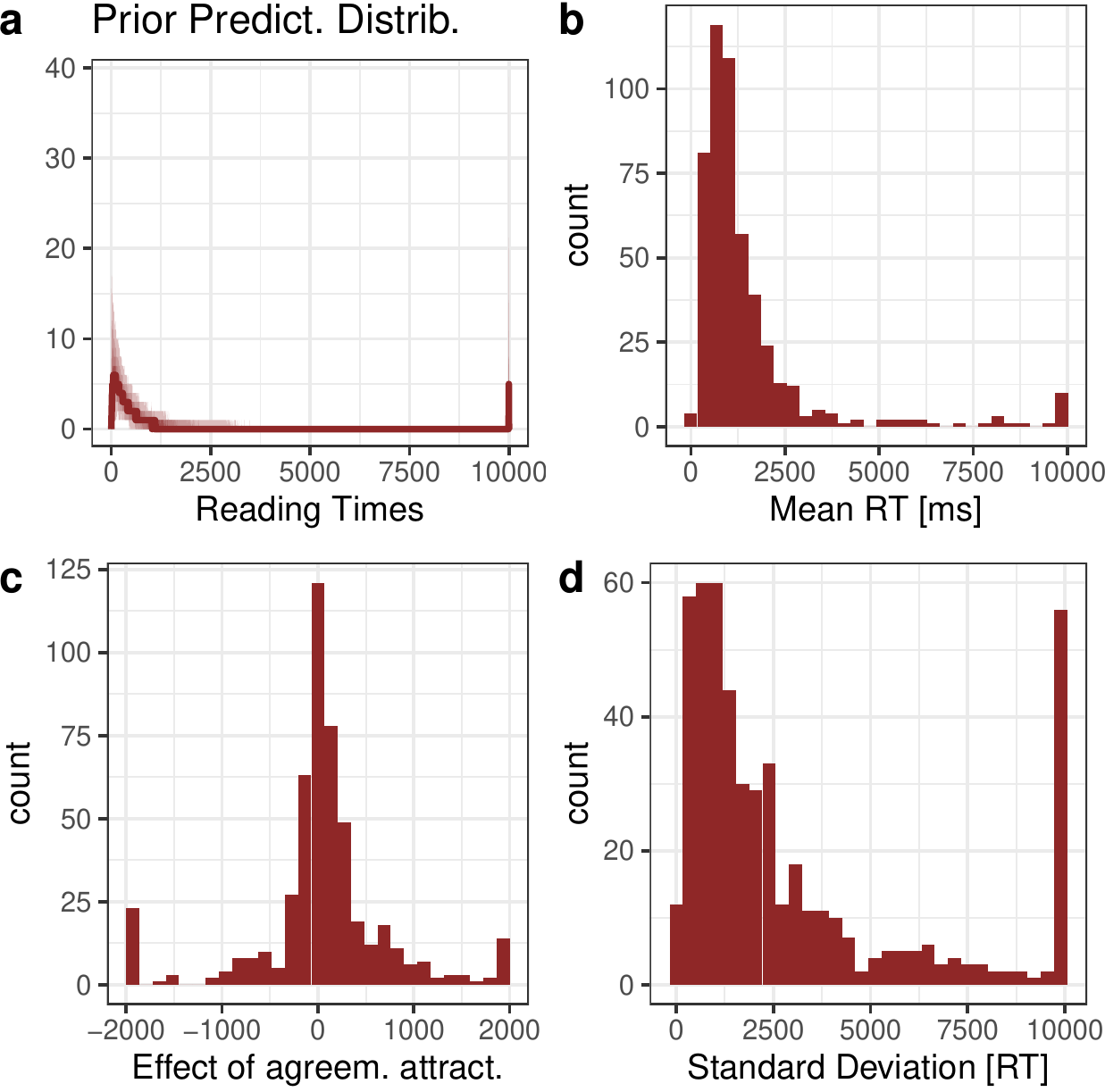} 

}

\caption{Prior predictive checks for the agreement attraction reading time data by Lago et al. (2015). Distributions are over simulated hypothetical data. a) Multivariate summary statistic: Distribution of histograms of reading times. Shaded areas correspond to 10–90 percent, 20–80 percent, 30–70 percent, and 40–60 percent quantiles across histograms; the solid line (in the middle of the shaded area) indicates the median across hypothetical data-sets. The distribution of reading times shows reading times are mostly below 1000 ms, but also shows a long tail with longer reading times, possibly due to the log-normal distribution. b)-d) Scalar summary statistics. b) Distribution of average reading times shows that a priori reading times are in the expected range (between 0 and 2000 ms). c) Distribution of differences in reading times between agreement attraction conditions shows that effect sizes are reasonably expected in the range of +/- 500 ms. d) Distribution of standard deviations of reading times shows a reasonable range of variation, between 0 and 2500 ms, but some outliers of standard deviations larger than 10,000. a)+b)+d) Values > 10,000 are plotted as 10,000. c) Values < -2,000 are plotted as -2,000 and values > 2,000 are plotted as 2,000.}\label{fig:xFigPPC}
\end{figure}

Based on the simulated data, we compute several summary statistics (see Figure~\ref{fig:xFigPPC}).
The results from the prior predictive checks show that the simulated data is in a plausible range. Specifically, the distribution of histograms of reading times shows many values in the range between 0 and 1000 ms (Fig.~\ref{fig:xFigPPC}a). However, there is also a small number of unplausibly long reading times. Moreover, we plot the distribution of mean reading times (Fig.~\ref{fig:xFigPPC}b), which shows reasonable reading times in the range of 0 to 2000 ms. Crucially, the difference in reading times between agreement attraction conditions shows a distribution (Fig.~\ref{fig:xFigPPC}c) where effect sizes are reasonably expected in the range of +/- 500 ms. Last, the distribution of standard deviations of reading times shows a reasonable range of variation, between 0 and 2500 ms (Fig.~\ref{fig:xFigPPC}d), however, there are again some outliers with very large standard deviations.

Overall, the variation is somewhat high, but the summary statistics of the a priori simulated data generally show results that are in a plausible range, which supports the use of the priors on which these simulations are built (Schad et al., 2021). We therefore proceed with using these priors to analyze the real experimentally observed reading time data.

\hypertarget{fitting-the-brms-model}{%
\subsubsection{Fitting the brms model}\label{fitting-the-brms-model}}

The next step would be to fit the model to the empirical data and to estimate Bayes factors. Note that we have already performed this model fitting and Bayes factor estimation above. We here show the brms-code used to do the model fitting:

\begin{Shaded}
\begin{Highlighting}[]
\CommentTok{\# run alternative model}
\NormalTok{m1\_lagoE1 \textless{}{-}}\StringTok{ }\KeywordTok{brm}\NormalTok{(rt }\OperatorTok{\textasciitilde{}}\StringTok{ }\DecValTok{1}\OperatorTok{+}\NormalTok{x }\OperatorTok{+}\StringTok{ }\NormalTok{(}\DecValTok{1}\OperatorTok{+}\NormalTok{x}\OperatorTok{|}\NormalTok{subj)}\OperatorTok{+}\StringTok{ }\NormalTok{(}\DecValTok{1}\OperatorTok{+}\NormalTok{x}\OperatorTok{|}\NormalTok{item),}
                 \DataTypeTok{data    =}\NormalTok{ lagoE1,}
                 \DataTypeTok{family  =} \KeywordTok{lognormal}\NormalTok{(),}
                 \DataTypeTok{prior   =}\NormalTok{ priors,}
                 \DataTypeTok{warmup  =} \DecValTok{2000}\NormalTok{,}
                 \DataTypeTok{iter    =} \DecValTok{10000}\NormalTok{,}
                 \DataTypeTok{cores   =} \DecValTok{4}\NormalTok{,}
                 \DataTypeTok{save\_pars =} \KeywordTok{save\_pars}\NormalTok{(}\DataTypeTok{all =} \OtherTok{TRUE}\NormalTok{),}
                 \DataTypeTok{control =} \KeywordTok{list}\NormalTok{(}\DataTypeTok{adapt\_delta =} \FloatTok{0.99}\NormalTok{,}
                                \DataTypeTok{max\_treedepth=}\DecValTok{15}\NormalTok{))}
\CommentTok{\# run null model}
\NormalTok{m0\_lagoE1 \textless{}{-}}\StringTok{ }\KeywordTok{brm}\NormalTok{(rt }\OperatorTok{\textasciitilde{}}\StringTok{ }\DecValTok{1} \OperatorTok{+}\StringTok{ }\NormalTok{(}\DecValTok{1}\OperatorTok{+}\NormalTok{x}\OperatorTok{|}\NormalTok{subj)}\OperatorTok{+}\StringTok{ }\NormalTok{(}\DecValTok{1}\OperatorTok{+}\NormalTok{x}\OperatorTok{|}\NormalTok{item),}
                 \DataTypeTok{data    =}\NormalTok{ lagoE1,}
                 \DataTypeTok{family  =} \KeywordTok{lognormal}\NormalTok{(),}
                 \DataTypeTok{prior   =}\NormalTok{ priors[}\OperatorTok{{-}}\DecValTok{2}\NormalTok{,],}
                 \DataTypeTok{warmup  =} \DecValTok{2000}\NormalTok{,}
                 \DataTypeTok{iter    =} \DecValTok{10000}\NormalTok{,}
                 \DataTypeTok{cores   =} \DecValTok{4}\NormalTok{,}
                 \DataTypeTok{save\_pars =} \KeywordTok{save\_pars}\NormalTok{(}\DataTypeTok{all =} \OtherTok{TRUE}\NormalTok{),}
                 \DataTypeTok{control =} \KeywordTok{list}\NormalTok{(}\DataTypeTok{adapt\_delta =} \FloatTok{0.99}\NormalTok{,}
                                \DataTypeTok{max\_treedepth=}\DecValTok{15}\NormalTok{))}
\CommentTok{\# run bridge sampler}
\NormalTok{lml\_m1\_lagoE1 \textless{}{-}}\StringTok{ }\KeywordTok{bridge\_sampler}\NormalTok{(m1\_lagoE1, }\DataTypeTok{silent =} \OtherTok{TRUE}\NormalTok{)}
\NormalTok{lml\_m0\_lagoE1 \textless{}{-}}\StringTok{ }\KeywordTok{bridge\_sampler}\NormalTok{(m0\_lagoE1, }\DataTypeTok{silent =} \OtherTok{TRUE}\NormalTok{)}
\CommentTok{\# compute Bayes factor}
\NormalTok{h\_lagoE1 \textless{}{-}}\StringTok{ }\KeywordTok{bayes\_factor}\NormalTok{(lml\_m1\_lagoE1, lml\_m0\_lagoE1)}
\end{Highlighting}
\end{Shaded}

We show the results from the posterior analyses:

\begin{Shaded}
\begin{Highlighting}[]
\KeywordTok{round}\NormalTok{(}\KeywordTok{fixef}\NormalTok{(m1\_lagoE1),}\DecValTok{3}\NormalTok{)}
\end{Highlighting}
\end{Shaded}

\begin{verbatim}
##           Estimate Est.Error   Q2.5  Q97.5
## Intercept    6.015     0.056  5.903  6.127
## x           -0.031     0.008 -0.046 -0.015
\end{verbatim}

They show that for the fixed effect \texttt{x}, capturing the agreement attraction effect, the 95\% credible interval does not overlap with zero. This indicates that there is some hint that the effect may have the expected negative direction, reflecting shorter reading times in the plural condition. As mentioned earlier, this does not provide direct evidence of the hypothesis that the effect exists and is not zero. This is not inferred here, because we did not specify the null hypothesis of zero effect explicitly. We can, however, investigate this null hypothesis by using the Bayes factor. We now look at the Bayes factor of the alternative model compared to the null model (\(BF_{10}\)):

\begin{Shaded}
\begin{Highlighting}[]
\NormalTok{h\_lagoE1}\OperatorTok{$}\NormalTok{bf}
\end{Highlighting}
\end{Shaded}

\begin{verbatim}
## [1] 6.744471
\end{verbatim}

Next, we check the estimated model.
We first check whether the posterior fit was successful. We look at the \(\hat{R}\) statistic.

\begin{Shaded}
\begin{Highlighting}[]
\KeywordTok{rhat}\NormalTok{(m1\_lagoE1)[}\StringTok{"b\_x"}\NormalTok{]}
\end{Highlighting}
\end{Shaded}

\begin{verbatim}
##       b_x 
## 0.9999915
\end{verbatim}

\begin{figure}

{\centering \includegraphics{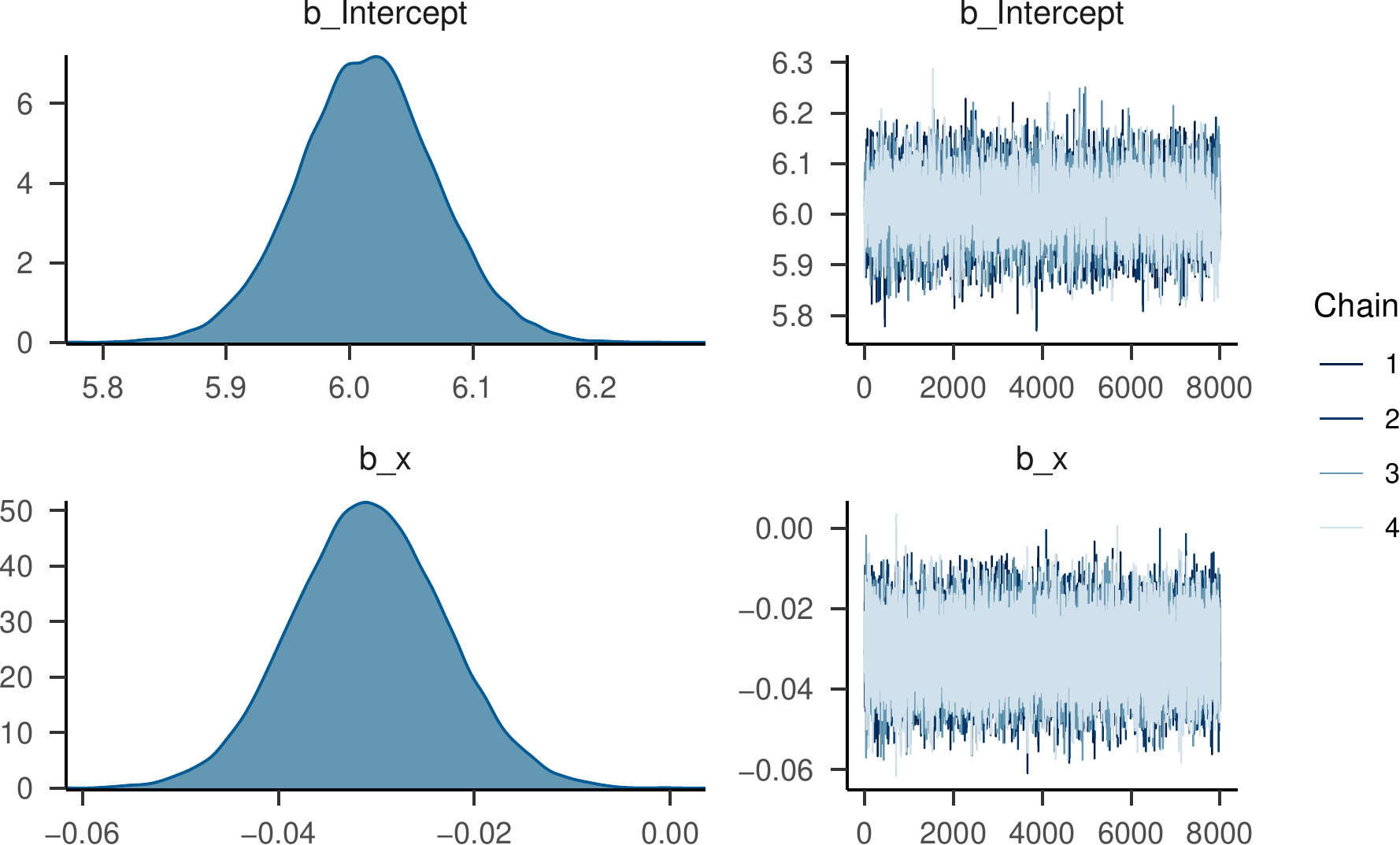} 

}

\caption{Density plots (left panels) and trace plots (right panels) for the intercept parameter (upper panels) and for the effect of agreement attraction (labelled as x; lower panels).}\label{fig:TracePlots1}
\end{figure}

It is very close to one, indicating no problem with convergence of the parameter mean. Moreover, the parameters \texttt{Bulk\_ESS\ =\ 54816} and \texttt{Tail\_ESS\ =\ 24746} for factor \texttt{x} suggested a large enough effective sample size to estimate the effect of agreement attraction.
Also, the model showed no warnings for divergent transitions, indicating no problems in the posterior fit.
Next, we look at posterior densities and at trace plots for the intercept parameter and for the parameter estimating the critical effect of agreement attraction. Figure~\ref{fig:TracePlots1} shows that posterior samples look reasonable and do not suggest any problems with the posterior model fit.

\hypertarget{posterior-predictive-checks}{%
\subsubsection{Posterior predictive checks}\label{posterior-predictive-checks}}

We next perform posterior predictive checks to see whether the model captures the structure in the data well. Posterior predictive simulations can be performed using the brms function \texttt{posterior\_predict()}.

\begin{Shaded}
\begin{Highlighting}[]
\NormalTok{pred\_lagoE1 \textless{}{-}}\StringTok{ }\KeywordTok{posterior\_predict}\NormalTok{(m1\_lagoE1)}
\end{Highlighting}
\end{Shaded}

\begin{figure}

{\centering \includegraphics{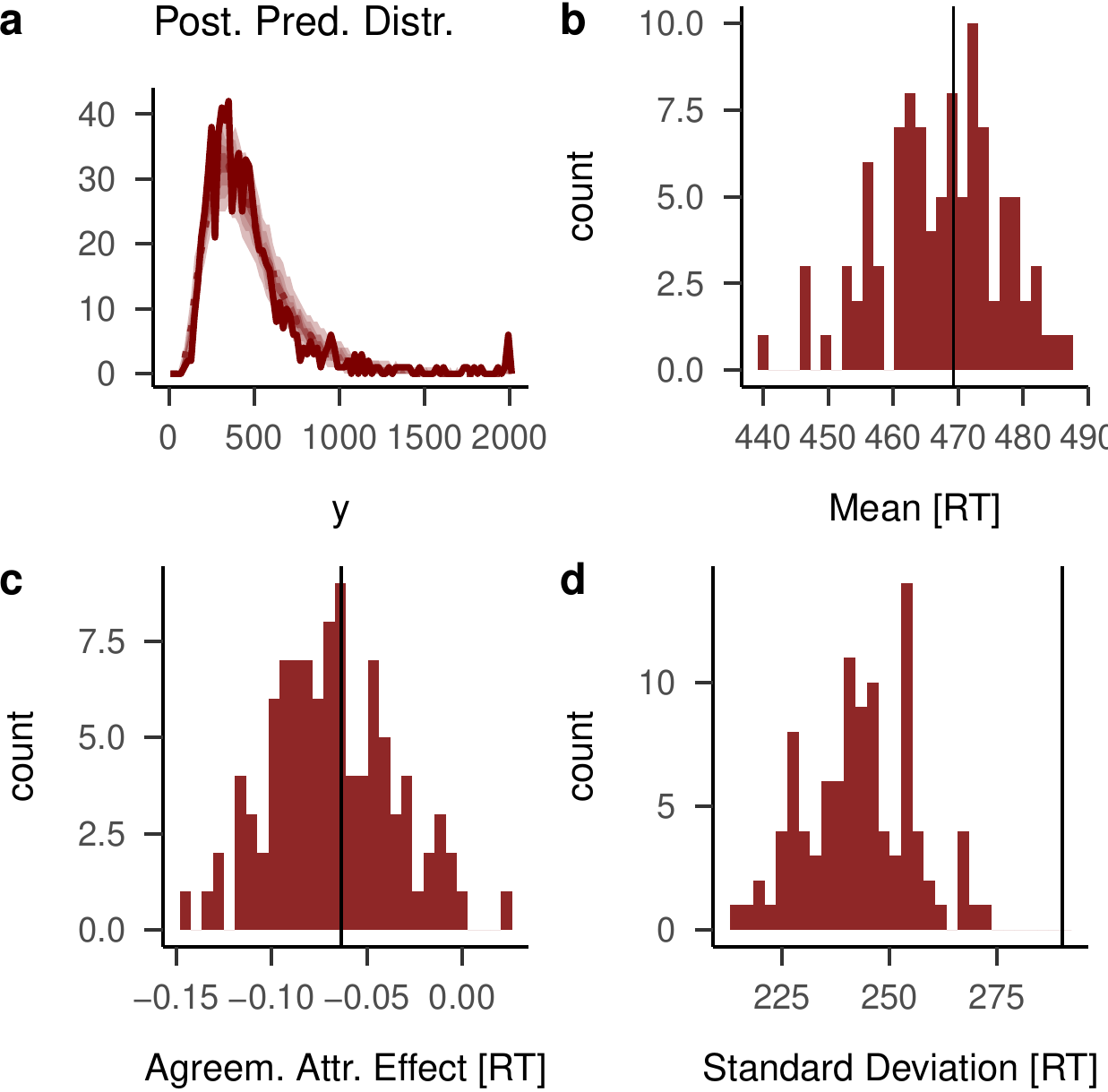} 

}

\caption{Posterior predictive checks. Distributions are over posterior predictive simulated data. a) Histograms of reading times. 10-90 percent, 20-80 percent, 30-70 percent, and 40-60 percent quantiles across histograms are shown as shaded areas; the median is shown as a dotted line and the observed data as a solid line. For illustration, values > 2000 are plotted as 2000; modeling was done on the original data. b) Average reading times. c) Differences in reading times between agreement attraction conditions. d) Standard deviations of reading times.}\label{fig:FigPost}
\end{figure}

The results displayed in Figure~\ref{fig:FigPost} show that the histogram of reading times was well captured by the brms model (Fig.~\ref{fig:FigPost}a). Likewise, the mean reading time across subjects (Fig.~\ref{fig:FigPost}b) and the average agreement attraction effect (Fig.~\ref{fig:FigPost}c) were well captured by the model. Only the distribution of standard deviations of reading times in the model was smaller than in the empirical data (Fig.~\ref{fig:FigPost}d), suggesting that the model had some difficulty capturing all the variability in the data well. Overall we are satisfied with the results from the posterior predictive checks and proceed further with our Bayes factor workflow.

\hypertarget{stability-of-bayes-factors-against-mcmc-draws}{%
\subsubsection{Stability of Bayes factors against MCMC draws}\label{stability-of-bayes-factors-against-mcmc-draws}}

To make sure that we are using enough MCMC draws to support stable Bayes factor estimates, we estimate the H0 and the H1 models four times on the empirical data.

The results from the four runs show that the Bayes factor estimates are all fairly close to each other:

\begin{Shaded}
\begin{Highlighting}[]
\KeywordTok{round}\NormalTok{(BF\_lagoE1,}\DecValTok{2}\NormalTok{)}
\end{Highlighting}
\end{Shaded}

\begin{verbatim}
## [1] 6.57 6.60 6.49 6.39
\end{verbatim}

Based on these rather similar results, we conclude that the Bayes factor estimation is stable enough for our current purposes.

We then go to the next step, which is to test the accuracy of Bayes factor estimates using SBC.

\hypertarget{simulation-based-calibration}{%
\subsubsection{Simulation-based calibration}\label{simulation-based-calibration}}

We perform 500 runs of SBC for the data set by Lago et al. (2015) based on the priors for the parameters from the meta analysis, and based on a priori model probabilities for the H0 and H1 of each 50\%.

\begin{verbatim}
##     CI.2.5 %  mean CI.97.5 %
## pH1    45.95 50.33      54.7
\end{verbatim}

The results show that the average posterior probability for the H1 (versus H0) is \(50.33\), and thus very close to the prior value of \(50\), and 95\% confidence intervals clearly include the 50\%.
This SBC analysis therefore shows that posterior inference on model probabilities based on Bayes factors is accurate and not biased, at least for the specific and simple case, model, and experimental design that we investigate here.

In addition to these SBC results, we can also investigate additional calibration questions of interest by looking at posterior model probabilities as a function of which prior hypothesis (model) was sampled in a given run. For each ``true'' hypothesis, we can now look at how much posterior probability mass is allocated to the two models by the Bayesian analysis. If the artificial data were simulated based on the H0, how high is the posterior probability for the H0? Is it higher than chance? And if so, by how much. Moreover, if the artificial data were simulated based on the H1, what is the posterior probability for the H1?

\begin{verbatim}
##   true_hypothesis pH0 pH1
## 1              H0  54  46
## 2              H1  46  54
\end{verbatim}

The results in the first row show that if the H0 was used to simulate artificial data, then the Bayesian procedure allocated an average of 54\% posterior probability to the H0. Thus, the chance to support the null hypothesis correctly is not much better than 50/50, i.e., not much better than chance, in this data set and model. Note that these probabilities are close to the prior probabilities for the hypotheses, which were set to 50\% for the H0. Thus, averaged across data sets, the data do not provide a lot of information for changing the prior beliefs.

Moreover, the second row of the table shows that if the H1 was used to simulate the artificial data, then the posterior probability for the H1 was an average of 54\%. Thus, the alternative hypothesis is also not likely to be correctly supported much in the present setting. Taken together, this analysis shows that the data and the model on average provide hardly any information or evidence for the hypotheses of interest. Larger sample studies or more precise hypotheses may be needed to obtain better posterior information from the data.

Importantly, we can see that given the priors and the model that we have defined, the present experimental design does not seem to contain a lot of information for making inferences about the true hypothesis that has generated the simulated data. Thus, larger sample sizes or more informative priors may be needed to obtain clear results from the empirical data.

\hypertarget{adapting-prior-distributions}{%
\subsubsection{Adapting prior distributions}\label{adapting-prior-distributions}}

The previous simulations showed that the Bayesian models had major difficulties in determining the true model from the data. A major reason for this is that in the SBC, the a priori assumptions about model parameters (which were used to simulate data) were quite vague. For example, variance components indicated high levels of noise in the data simulations. As an alternative, we here use the posterior distributions based on the Lago et al. (2015) data (see fitted model object \texttt{m1\_lagoE1}) as the priors for the data simulation (i.e., posterior predictive analyses). Note that this is something that we would normally never do in practice. We cannot take the posterior from the analysis for some data set as a prior for the analysis of the same data. Here, we do this simply to illustrate what would happen in case we would have more informative priors.

\begin{Shaded}
\begin{Highlighting}[]
\NormalTok{priorsLagoPost \textless{}{-}}\StringTok{ }\KeywordTok{c}\NormalTok{(}
  \CommentTok{\# fixed effects}
  \KeywordTok{set\_prior}\NormalTok{(}\StringTok{"normal( 6.02  ,0.0570 )"}\NormalTok{,}\DataTypeTok{class=}\StringTok{"Intercept"}\NormalTok{),}
  \KeywordTok{set\_prior}\NormalTok{(}\StringTok{"normal({-}0.0284,0.00754)"}\NormalTok{,}\DataTypeTok{class=}\StringTok{"b"}\NormalTok{),}
  \CommentTok{\# SD parameters items}
  \KeywordTok{set\_prior}\NormalTok{(}\StringTok{"normal(0.04,0.02)"}\NormalTok{,}\DataTypeTok{class=}\StringTok{"sd"}\NormalTok{,}\DataTypeTok{coef=}\StringTok{"Intercept"}\NormalTok{,}\DataTypeTok{group=}\StringTok{"item"}\NormalTok{),}
  \KeywordTok{set\_prior}\NormalTok{(}\StringTok{"normal(0.02,0.01)"}\NormalTok{,}\DataTypeTok{class=}\StringTok{"sd"}\NormalTok{,}\DataTypeTok{coef=}\StringTok{"x"}\NormalTok{,}\DataTypeTok{group=}\StringTok{"item"}\NormalTok{),}
  \CommentTok{\# SD parameters subjects}
  \KeywordTok{set\_prior}\NormalTok{(}\StringTok{"normal(0.31,0.04)"}\NormalTok{,}\DataTypeTok{class=}\StringTok{"sd"}\NormalTok{,}\DataTypeTok{coef=}\StringTok{"Intercept"}\NormalTok{,}\DataTypeTok{group=}\StringTok{"subj"}\NormalTok{),}
  \KeywordTok{set\_prior}\NormalTok{(}\StringTok{"normal(0.03,0.02)"}\NormalTok{,}\DataTypeTok{class=}\StringTok{"sd"}\NormalTok{,}\DataTypeTok{coef=}\StringTok{"x"}\NormalTok{,}\DataTypeTok{group=}\StringTok{"subj"}\NormalTok{),}
  \CommentTok{\# residual variance + correlation}
  \KeywordTok{set\_prior}\NormalTok{(}\StringTok{"normal(0.41,0.01)"}\NormalTok{, }\DataTypeTok{class=}\StringTok{"sigma"}\NormalTok{),}
  \KeywordTok{set\_prior}\NormalTok{(}\StringTok{"lkj(2)"}\NormalTok{, }\DataTypeTok{class=}\StringTok{"cor"}\NormalTok{))}
\end{Highlighting}
\end{Shaded}

Based on these priors, we again perform simulation based calibration of the Bayes factors.

Again, we can perform SBC by looking at the average posterior probabilities (``means'' and 95\% confidence intervals) for each of the models.

\begin{verbatim}
##     CI.2.5 %  mean CI.97.5 %
## pH1    47.33 51.71     56.08
\end{verbatim}

The results for the average posterior model probabilities show - as for the previous simulations -, that posterior probability for the H1 (versus the H0) was very close to the prior model probabilities of 50\%, and that the 95\% confidence intervals clearly include the prior probability of 50\%. This analysis thus supports our earlier result that bridge sampling yields unbiased estimates of Bayes factors (and posterior model probabilities), now for a set of more informative priors. This result thus again supports the use of bridge sampling for the analysis of our multilevel (i.e., generalized linear mixed-effects) model for our present case study. However, note that this is also a relatively small ensemble (with \(n = 500\) simulations) and hence not a very strong test of bridge sampling.

Second, we can again perform additional calibration analyses by looking at the supported hypotheses given that either the H0 or the H1 was used to simulate the data.

\begin{verbatim}
##   true_hypothesis pH0 pH1
## 1              H0  70  30
## 2              H1  28  72
\end{verbatim}

The results now show much better identifiability of the hypotheses. When the H0 was the true model in the simulations (first row of the matrix), then this was correctly supported with an average posterior probability of 70\%, suggesting moderately higher evidence for the correct null model. At the same time, 30\% of the posterior probability was falsely allocated to the H1.
Moreover, when the H1 was the true model in the simulations (second row of the matrix), then this was correctly supported with an average posterior probability of 72\%, again suggesting a moderately higher evidence for the correct alternative model. Note that this identifiability is better than before, but may be still worse than often expected in frequentist analyses.

\hypertarget{data-variability}{%
\subsubsection{Data variability}\label{data-variability}}

Based on the good average performance of posterior model probabilities, we next look at how posterior model probabilities vary across posterior predictive data sets. That is, even if the average performance of inferences based on Bayes factors seems good, it is unclear how much this inference varies with variability in the data.

\begin{figure}

{\centering \includegraphics{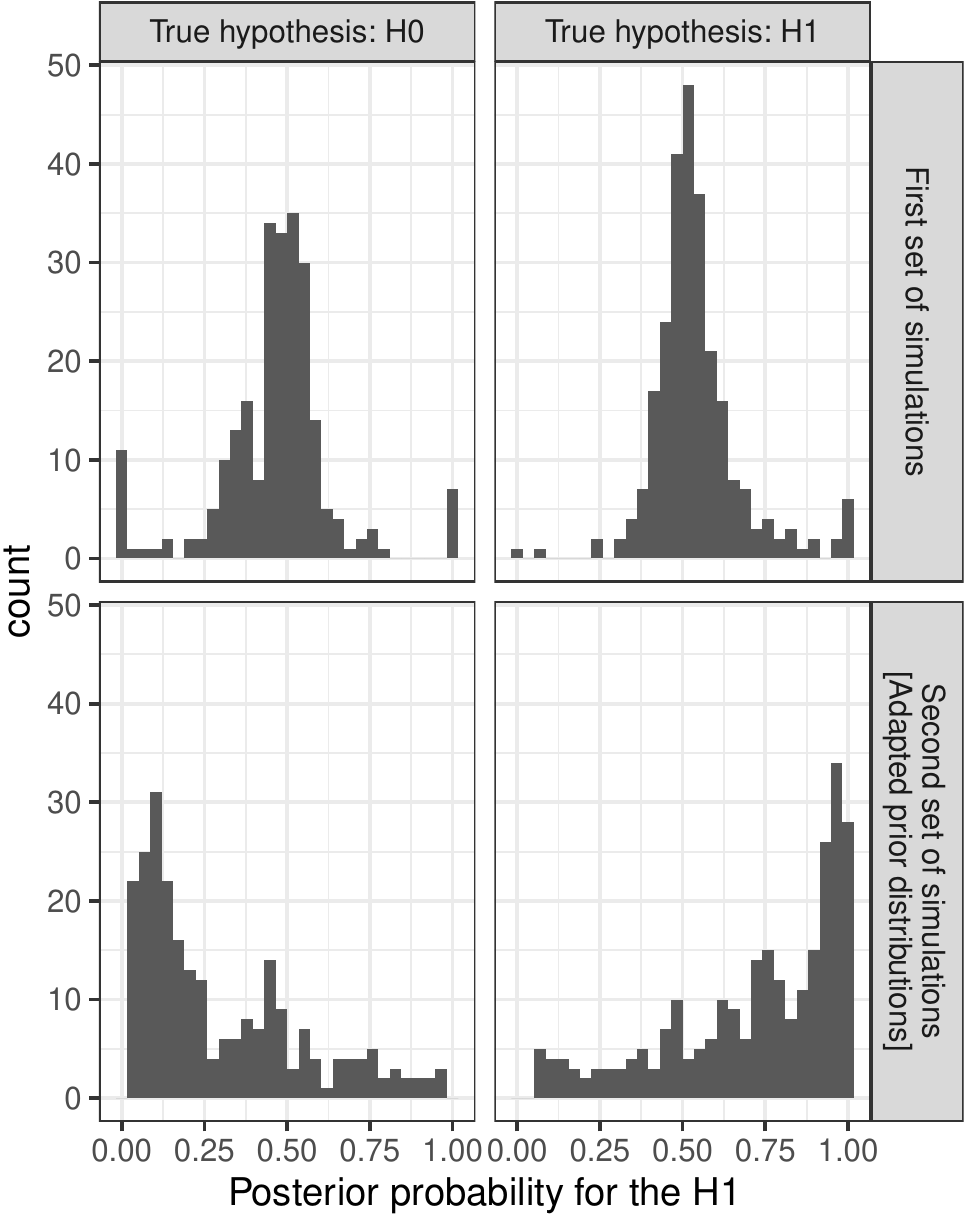} 

}

\caption{Histograms of posterior probabilities for the H1 across 500 simulated data sets, where either the H0 (left panel) or the H1 (right panel) was the true hypothesis in the data simulations.}\label{fig:SBCvar2}
\end{figure}

As is shown in Figure~\ref{fig:SBCvar2}, the posterior probabilities widely varied across individual data sets. The lower panels of Figure~\ref{fig:SBCvar2} represent the SBC using the adjusted (more informative) priors from the section above, which was based on the posterior from the Lago et al. (2015) study. We can see that in this analysis, when the true hypothesis was the H0 (left panels), then posterior model probabilities tended to be smaller and closer to 0, whereas when the true hypothesis was the H1 (right panels), then posterior model probabilities tended to be larger and closer to 1. Thus, in the second set of posterior predictive simulations, there seemed to be some information contained in the data. However, there was still a large amount of variation, and posterior probabilities could be 0 or 1 for individual data sets irrespective of which hypothesis was true in the data.

Even less information was contained in the first set of simulations (upper panels of Fig.~\ref{fig:SBCvar2}), which indicated the prior predictive analysis based only on the meta analysis. Here, the distributions of posterior model probabilities seem quite the same, irrespective of which hypothesis was true (i.e., used to simulate the data). However, while on average there was not a lot information in the data, individual data sets still seemed to provide quite some support to either the H0 or the H1, approaching posterior probabilities close to zero or one. Importantly, this support is an illusion here, since we know what the true hypothesis was in each of these cases. This shows that for the present experimental design, priors, and effect size, an individual data set may seem to provide evidence either for or against the effect, nearly independent of which hypothesis had really been true in the simulation of the data. Thus, these individual data sets are not sufficient to inform inference or even decisions based on them, and larger data sets and/or larger effect sizes might be needed for reliable inferences or decisions.

\hypertarget{using-sbc-simulations-to-calibrate-decisions-1}{%
\subsubsection{Using SBC simulations to calibrate decisions}\label{using-sbc-simulations-to-calibrate-decisions-1}}

The actions that we aim to perform are either to declare discovery or to not declare discovery. To study these decisions based on the Bayesian evidence, we first define utility functions. We use the same utilities that we had used above:

\begin{Shaded}
\begin{Highlighting}[]
\NormalTok{utility \textless{}{-}}\StringTok{ }\KeywordTok{c}\NormalTok{(}\OperatorTok{{-}}\DecValTok{50}\NormalTok{,}\DecValTok{5}\NormalTok{,}\DecValTok{10}\NormalTok{,}\OperatorTok{{-}}\DecValTok{5}\NormalTok{)}
\KeywordTok{names}\NormalTok{(utility) \textless{}{-}}\StringTok{ }\KeywordTok{c}\NormalTok{(}\StringTok{"TrueH0.Disc"}\NormalTok{,}\StringTok{"TrueH0.NoDisc"}\NormalTok{,}\StringTok{"TrueH1.Disc"}\NormalTok{,}\StringTok{"TrueH1.NoDisc"}\NormalTok{)}
\NormalTok{utility}
\end{Highlighting}
\end{Shaded}

\begin{verbatim}
##   TrueH0.Disc TrueH0.NoDisc   TrueH1.Disc TrueH1.NoDisc 
##           -50             5            10            -5
\end{verbatim}

Next, we investigate which Bayes factor threshold gives highest overall utility.

\begin{figure}

{\centering \includegraphics{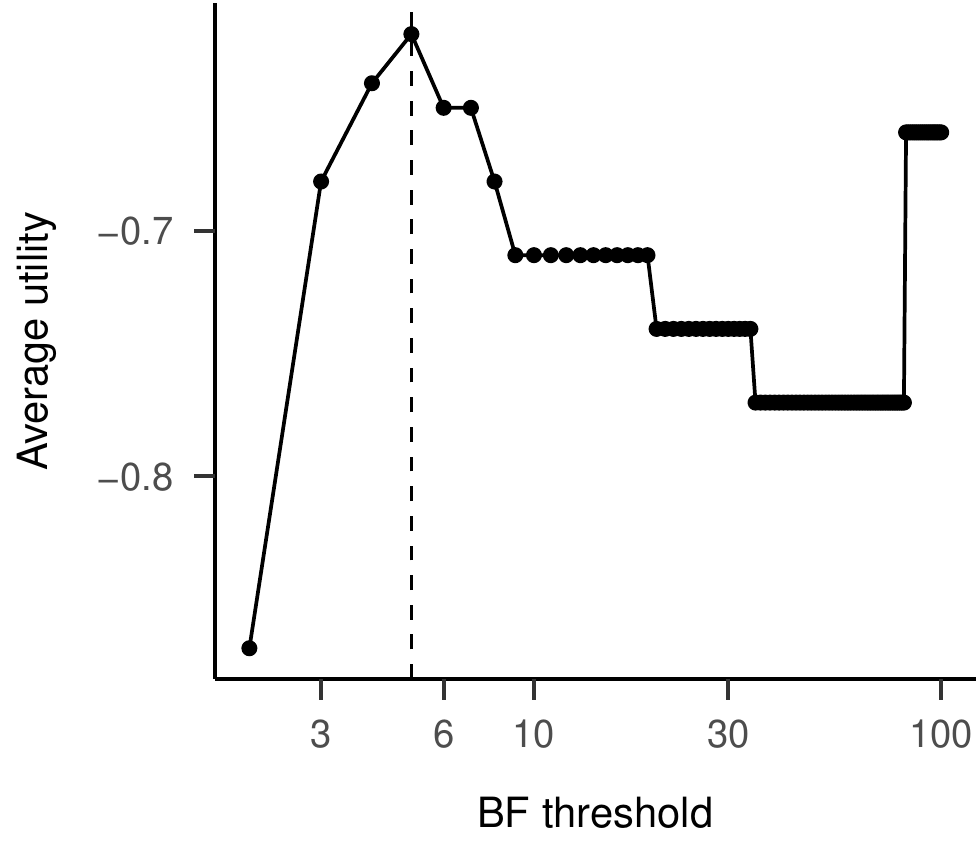} 

}

\caption{Average expected utility as a function of the critical BF threshold: prior predictive analysis of the Lago et al. (2015) data.}\label{fig:utility2}
\end{figure}

The results, displayed in Figure~\ref{fig:utility2} show that the optimal Bayes factor threshold is the value \(5\). With this value, we go back to our empirical data. Above, we had performed analyses to check the stability of Bayes factor estimates against different MCMC draws. This analysis had revealed a Bayes factor of roughly \(BF_{10} = 6.5\):

\begin{Shaded}
\begin{Highlighting}[]
\KeywordTok{round}\NormalTok{(BF\_lagoE1,}\DecValTok{2}\NormalTok{)}
\end{Highlighting}
\end{Shaded}

\begin{verbatim}
## [1] 6.57 6.60 6.49 6.39
\end{verbatim}

\begin{Shaded}
\begin{Highlighting}[]
\KeywordTok{mean}\NormalTok{(BF\_lagoE1)}
\end{Highlighting}
\end{Shaded}

\begin{verbatim}
## [1] 6.513411
\end{verbatim}

We can now compare the Bayes factor estimate of \(BF_{10} = 6.5\) to the threshold of \(5\). The result shows that the Bayes factor is larger than the threshold. Thus, we would declare discovery. That is, we would claim the discovery that facilitatory agreement attraction exists.

However, remember that our analyses of the data variability of Bayesian evidence showed that posterior evidence widely varied based on this experimental design, models, and priors. This suggests that we should be quite cautious with our discovery claim, since it seems that there might be a good chance that it originated from chance only. We look at this more closely by studying TDR and FDR:

\begin{verbatim}
##   true_hypothesis No_Discovery Discovery
## 1          TrueH0         0.97      0.03
## 2          TrueH1         0.96      0.04
\end{verbatim}

The results show a FDR of 3\% and a TDR of 4\%. Thus, even though the optimal threshold was exceeded, this analysis suggests that there is a \(3\%/(3\%+4\%)*100=43\%\) chance that the discovery claim in fact originates from false discovery. This demonstrates that the given experimental design was insufficient to constrain decisions based on our uninformative priors. We might thus consider improving the experimental design or the prior information that we take into account. We here repeat the analysis using the adjusted (more informative) prior information, which was based on the posterior from the Lago et al. (2015) study (see section ``Adapting prior distributions'' above).

We obtain a Bayes factor of 6.73.

We again use the same utilities for optimizing the Bayes factor threshold. Now, with the new more informative priors, we obtain an optimal Bayes factor threshold of \(4\). The empirical Bayes factor estimate of \(BF_{10} = 6.7\) thus again supports a discovery claim.

\begin{verbatim}
##       true_hypothesis No_Discovery Discovery
## 1 True hypothesis: H0         0.95      0.05
## 2 True hypothesis: H1         0.55      0.45
\end{verbatim}

When we look at the FDR and TDR, we can see that the more informative prior beliefs yield again a low FDR of 5\%. However, the TDR is now much higher with a value of 45\%, providing more confidence that the discovery claim based on the empirical data might be a true discovery rather than a false discovery.

\hypertarget{discussion}{%
\section{Discussion}\label{discussion}}

We provided a discussion of the Bayesian quantification of evidence in favor of one of two alternative hypotheses and investigated the performance of Bayes factors with respect to prior assumptions, effective sample size, simulation-based calibration, data variability, and utility functions. We implemented competing hypotheses in hierarchical Bayesian models using the R-package \texttt{brms}, and tested these hypotheses against each other by estimating Bayes factors approximately using bridge sampling.

The results illustrate the strong dependence of Bayes factors on the prior assumptions, which calls for the use of (1) (weakly) informative priors (cf., Schad et al., 2021) and of (2) prior sensitivity analyses, to investigate Bayes factors for different prior assumptions about the size of the effect. Our results moreover illustrate challenges and limitations in the performance of Bayes factor analyses. First, we studied theoretical aspects of Bayes factor estimation. We showed that Bayes factors can be estimated in an unstable way because a very large effective sample size (of the Hamiltonian Markov Chain Monte Carlo sampler) is needed in order to obtain stable results from the bridge sampling algorithm. Moreover, we noted that even if Bayes factor approximations are stable, because bridge sampling does not come with strong guarantees, it is unclear whether approximate Bayes factor estimates are accurate, i.e., whether approximate Bayes factor estimates correspond to the true Bayes factor. We showed how simulation-based calibration can be used to investigate whether Bayes factor estimates are accurate for a given application case (i.e., model, priors, and experimental design). We moreover performed further ordinary Bayesian calibration analyses by testing average posterior model probabilities given the data was simulated based on the H0 or the H1. Our results illustrate how a robust effect in the cognitive sciences, i.e., facilitatory agreement attraction effects, could hardly be detected from a standard experimental design and analysis, and how much stronger effect sizes (or much larger samples) are needed if the aim is to draw firm conclusions from single experimental studies.

Second, analyses of artificial and of real replication data showed that the results from Bayes factor analyses - just like p-values in frequentist analyses - can considerably vary across different repeated replication attempts. However, for a range of different real empirical studies, the results also show some robustness against drawing strong conclusions. This again suggests that some typical linguistic or psychological experiments may not be sufficiently powered to provide strong evidence for or against the small effect sizes that may be realistic to expect and that are of theoretical interest. Importantly, using Bayesian statistics and the Bayes factor does not solve this problem, since low-powered studies will most likely yield inconclusive results in a Bayes factor analysis. Studies with larger sample sizes or stronger effect sizes may therefore be needed, for example by sharing data across labs, to overcome such situations of low power.

Third, we studied decision-making based on Bayesian analyses and saw how decisions can widely vary with the data. We discussed some heuristics for performing decisions, and illustrated how utility functions can be used to obtain optimal decisions.

Based on these challenges to the robustness of Bayes factors and the resulting inferences and decisions, we here formulate a Bayes factor workflow, where simulations-based calibrations can be used to investigate these different issues for a given application case. This workflow then allows us to judge the extend to which inferences and decisions based on Bayes factors are robust for the given application case.

Taken together, Bayes factor analyses provide a useful tool that can be used to investigate evidence for different hypotheses in the cognitive sciences. We showed how Bayes factors can misbehave based on estimation error, data variation, and poor Bayesian decision-procedures, partially reflecting that fact that wide-spread limitations in experimental design can also limit the conclusions that can be drawn based on individual data sets. We propose a Bayes factor workflow to identify these potential problems for a given application case. When used with care and calibrated accordingly, Bayes factors provide a useful approach for quantifying evidence and supporting decision-making on discovery claims in the cognitive sciences.

\hypertarget{acknowledgements}{%
\section{Acknowledgements}\label{acknowledgements}}

This work was partly funded by the Deutsche Forschungsgemeinschaft (DFG), Sonderforschungsbereich 1287, Project Q (PIs: Shravan Vasishth and Ralf Engbert), project number 317633480 (Limits of Variability in Language).

\newpage

\hypertarget{references}{%
\section{References}\label{references}}

\begingroup
\setlength{\parindent}{-0.5in}
\setlength{\leftskip}{0.5in}

\hypertarget{refs}{}
\begin{cslreferences}
\leavevmode\hypertarget{ref-aitkin1991posterior}{}%
Aitkin, M. (1991). Posterior Bayes factors. \emph{Journal of the Royal Statistical Society: Series B (Methodological)}, \emph{53}(1), 111--128.

\leavevmode\hypertarget{ref-barr2013random}{}%
Barr, D. J., Levy, R., Scheepers, C., \& Tily, H. J. (2013). Random effects structure for confirmatory hypothesis testing: Keep it maximal. \emph{Journal of Memory and Language}, \emph{68}(3), 255--278.

\leavevmode\hypertarget{ref-benjamin2018redefine}{}%
Benjamin, D. J., Berger, J. O., Johannesson, M., Nosek, B. A., Wagenmakers, E.-J., Berk, R., \ldots{} others. (2018). Redefine statistical significance. \emph{Nature Human Behaviour}, \emph{2}(1), 6--10.

\leavevmode\hypertarget{ref-bennettEfficientEstimationFree1976}{}%
Bennett, C. H. (1976). Efficient estimation of free energy differences from Monte Carlo data. \emph{Journal of Computational Physics}, \emph{22}(2), 245--268. \url{https://doi.org/10.1016/0021-9991(76)90078-4}

\leavevmode\hypertarget{ref-betancourt2016diagnosing}{}%
Betancourt, M. (2016). Diagnosing suboptimal cotangent disintegrations in Hamiltonian Monte Carlo. \emph{arXiv Preprint arXiv:1604.00695}.

\leavevmode\hypertarget{ref-betancourt2017conceptual}{}%
Betancourt, M. (2017). A conceptual introduction to Hamiltonian Monte Carlo. \emph{arXiv Preprint arXiv:1701.02434}.

\leavevmode\hypertarget{ref-betancourt2018calibrating}{}%
Betancourt, M. (2018). Calibrating model-based inferences and decisions. \emph{arXiv Preprint arXiv:1803.08393}.

\leavevmode\hypertarget{ref-Betancourt2019calibration}{}%
Betancourt, M. (2019). Probabilistic modeling and statistical inference. \emph{GitHub repository}. \url{https://github.com/betanalpha/knitr_case_studies/tree/master/modeling_and_inference}; commit: b474ec1a5a79347f7c9634376c866fe3294d657a.

\leavevmode\hypertarget{ref-Betancourt2020mcmc}{}%
Betancourt, M. (2020a). Markov chain Monte Carlo. \emph{GitHub repository}. \url{https://github.com/betanalpha/knitr_case_studies/tree/master/markov_chain_monte_carlo}; commit: b474ec1a5a79347f7c9634376c866fe3294d657a.

\leavevmode\hypertarget{ref-Betancourt2020workflow}{}%
Betancourt, M. (2020b). Towards a principled Bayesian workflow (RStan). \emph{GitHub repository}. \url{https://github.com/betanalpha/knitr_case_studies/tree/master/principled_bayesian_workflow}; commit: aeab31509b8e37ff05b0828f87a3018b1799b401.

\leavevmode\hypertarget{ref-bishop2006pattern}{}%
Bishop, C. M. (2006). \emph{Pattern recognition and machine learning}. New York: Springer.

\leavevmode\hypertarget{ref-Buerkner2017brms}{}%
Bürkner, P.-C. (2017). brms: An R package for Bayesian multilevel models using Stan. \emph{Journal of Statistical Software}, \emph{80}(1), 1--28. \url{https://doi.org/10.18637/jss.v080.i01}

\leavevmode\hypertarget{ref-Buerkner2018brms}{}%
Bürkner, P.-C. (2018). Advanced Bayesian multilevel modeling with the R package brms. \emph{The R Journal}, \emph{10}(1), 395--411. \url{https://doi.org/10.32614/RJ-2018-017}

\leavevmode\hypertarget{ref-carpenter2017stan}{}%
Carpenter, B., Gelman, A., Hoffman, M. D., Lee, D., Goodrich, B., Betancourt, M., \ldots{} Riddell, A. (2017). Stan: A probabilistic programming language. \emph{Journal of Statistical Software}, \emph{76}(1).

\leavevmode\hypertarget{ref-chow2017bayesian}{}%
Chow, S.-M., \& Hoijtink, H. (2017). Bayesian estimation and modeling: Editorial to the second special issue on Bayesian data analysis. \emph{Psychological Methods}, \emph{22}(4), 609--615.

\leavevmode\hypertarget{ref-cumming2014new}{}%
Cumming, G. (2014). The new statistics: Why and how. \emph{Psychological Science}, \emph{25}(1), 7--29.

\leavevmode\hypertarget{ref-DickeyLientz1970}{}%
Dickey, J. M., Lientz, B., \& others. (1970). The weighted likelihood ratio, sharp hypotheses about chances, the order of a Markov chain. \emph{The Annals of Mathematical Statistics}, \emph{41}(1), 214--226.

\leavevmode\hypertarget{ref-dillon2013contrasting}{}%
Dillon, B., Mishler, A., Sloggett, S., \& Phillips, C. (2013). Contrasting intrusion profiles for agreement and anaphora: Experimental and modeling evidence. \emph{Journal of Memory and Language}, \emph{69}(2), 85--103.

\leavevmode\hypertarget{ref-dillon2011structured}{}%
Dillon, B. W. (2011). \emph{Structured access in sentence comprehension} (PhD thesis).

\leavevmode\hypertarget{ref-van2021bayes}{}%
Doorn, J. van, Aust, F., Haaf, J. M., Stefan, A., \& Wagenmakers, E.-J. (2021). Bayes factors for mixed models. \emph{PsyArXiv Preprint PsyArXiv:Y65h8}.

\leavevmode\hypertarget{ref-engelmann2019effect}{}%
Engelmann, F., Jäger, L. A., \& Vasishth, S. (2019). The effect of prominence and cue association on retrieval processes: A computational account. \emph{Cognitive Science}, \emph{43}(12), e12800.

\leavevmode\hypertarget{ref-etz2018how}{}%
Etz, A., Gronau, Q. F., Dablander, F., Edelsbrunner, P. A., \& Baribault, B. (2018). How to become a Bayesian in eight easy steps: An annotated reading list. \emph{Psychonomic Bulletin \& Review}, \emph{25}(1), 219--234.

\leavevmode\hypertarget{ref-etz2018introduction}{}%
Etz, A., \& Vandekerckhove, J. (2018). Introduction to Bayesian inference for psychology. \emph{Psychonomic Bulletin \& Review}, \emph{25}(1), 5--34.

\leavevmode\hypertarget{ref-Freedman1984}{}%
Freedman, L. S., Lowe, D., \& Macaskill, P. (1984). Stopping rules for clinical trials incorporating clinical opinion. \emph{Biometrics}, \emph{40}(3), 575--586.

\leavevmode\hypertarget{ref-gabry2019visualization}{}%
Gabry, J., Simpson, D., Vehtari, A., Betancourt, M., \& Gelman, A. (2019). Visualization in Bayesian workflow. \emph{Journal of the Royal Statistical Society: Series A (Statistics in Society)}, \emph{182}(2), 389--402.

\leavevmode\hypertarget{ref-ge2018turing}{}%
Ge, H., Xu, K., \& Ghahramani, Z. (2018). Turing: A language for flexible probabilistic inference. In \emph{International conference on artificial intelligence and statistics} (pp. 1682--1690). PMLR.

\leavevmode\hypertarget{ref-gelman2013bayesian}{}%
Gelman, A., Carlin, J. B., Stern, H. S., Dunson, D. B., Vehtari, A., \& Rubin, D. B. (2013). \emph{Bayesian data analysis}. New York: CRC press.

\leavevmode\hypertarget{ref-Gelman14}{}%
Gelman, A., Carlin, J. B., Stern, H. S., Dunson, D. B., Vehtari, A., \& Rubin, D. B. (2014). \emph{Bayesian data analysis} (Third). Boca Raton, FL: Chapman; Hall/CRC.

\leavevmode\hypertarget{ref-gelman2020bayesian}{}%
Gelman, A., Vehtari, A., Simpson, D., Margossian, C. C., Carpenter, B., Yao, Y., \ldots{} Modrák, M. (2020). Bayesian workflow. \emph{arXiv Preprint arXiv:2011.01808}.

\leavevmode\hypertarget{ref-Good:1950aa}{}%
Good, I. J. (1950). \emph{Probability and the weighing of evidence}. New York: Hafners.

\leavevmode\hypertarget{ref-gronau2017tutorial}{}%
Gronau, Q. F., Sarafoglou, A., Matzke, D., Ly, A., Boehm, U., Marsman, M., \ldots{} Steingroever, H. (2017a). A tutorial on bridge sampling. \emph{Journal of Mathematical Psychology}, \emph{81}, 80--97.

\leavevmode\hypertarget{ref-gronauTutorialBridgeSampling2017}{}%
Gronau, Q. F., Sarafoglou, A., Matzke, D., Ly, A., Boehm, U., Marsman, M., \ldots{} Steingroever, H. (2017b). A tutorial on bridge sampling. \emph{Journal of Mathematical Psychology}, \emph{81}, 80--97. \url{https://doi.org/10.1016/j.jmp.2017.09.005}

\leavevmode\hypertarget{ref-Gronau2020bridgesampling}{}%
Gronau, Q. F., Singmann, H., \& Wagenmakers, E.-J. (2020). bridgesampling: An R package for estimating normalizing constants. \emph{Journal of Statistical Software}, \emph{92}(10), 1--29. \url{https://doi.org/10.18637/jss.v092.i10}

\leavevmode\hypertarget{ref-grunwald2000model}{}%
Grünwald, P. (2000). Model selection based on minimum description length. \emph{Journal of Mathematical Psychology}, \emph{44}(1), 133--152.

\leavevmode\hypertarget{ref-hammerly2019grammaticality}{}%
Hammerly, C., Staub, A., \& Dillon, B. (2019). The grammaticality asymmetry in agreement attraction reflects response bias: Experimental and modeling evidence. \emph{Cognitive Psychology}, \emph{110}, 70--104.

\leavevmode\hypertarget{ref-heck2020review}{}%
Heck, D. W., Boehm, U., Böing-Messing, F., Bürkner, P.-C., Derks, K., Dienes, Z., \ldots{} others. (2020). A review of applications of the Bayes factor in psychological research.

\leavevmode\hypertarget{ref-hoijtink2017bayesian}{}%
Hoijtink, H., \& Chow, S.-M. (2017). Bayesian hypothesis testing: Editorial to the special issue on Bayesian data analysis. \emph{Psychological Methods}, \emph{22}(2), 211--216.

\leavevmode\hypertarget{ref-jager2017similarity}{}%
Jäger, L. A., Engelmann, F., \& Vasishth, S. (2017). Similarity-based interference in sentence comprehension: Literature review and Bayesian meta-analysis. \emph{Journal of Memory and Language}, \emph{94}, 316--339.

\leavevmode\hypertarget{ref-jager2020interference}{}%
Jäger, L. A., Mertzen, D., Van Dyke, J. A., \& Vasishth, S. (2020). Interference patterns in subject-verb agreement and reflexives revisited: A large-sample study. \emph{Journal of Memory and Language}, \emph{111}, 104063.

\leavevmode\hypertarget{ref-jeffreys1939theory}{}%
Jeffreys, H. (1939). \emph{Theory of probability}. Oxford: Clarendon Press.

\leavevmode\hypertarget{ref-kass1995bayes}{}%
Kass, R. E., \& Raftery, A. E. (1995). Bayes factors. \emph{Journal of the American Statistical Association}, \emph{90}(430), 773--795.

\leavevmode\hypertarget{ref-kruschke2011bayesian}{}%
Kruschke, J. K. (2011). Bayesian assessment of null values via parameter estimation and model comparison. \emph{Perspectives on Psychological Science}, \emph{6}(3), 299--312.

\leavevmode\hypertarget{ref-lago2015agreement}{}%
Lago, S., Shalom, D., Sigman, M., Lau, E. F., \& Phillips, C. (2015). Agreement processes in spanish comprehension. \emph{Journal of Memory and Language}, \emph{82}, 133--149.

\leavevmode\hypertarget{ref-lee2011cognitive}{}%
Lee, M. D. (2011). How cognitive modeling can benefit from hierarchical Bayesian models. \emph{Journal of Mathematical Psychology}, \emph{55}(1), 1--7.

\leavevmode\hypertarget{ref-Lewandowski:2009aa}{}%
Lewandowski, D., Kurowicka, D., \& Joe, H. (2009). Generating random correlation matrices based on vines and extended onion method. \emph{Journal of Multivariate Analysis}, \emph{100}(9), 1989--2001.

\leavevmode\hypertarget{ref-liu2008bayes}{}%
Liu, C. C., \& Aitkin, M. (2008). Bayes factors: Prior sensitivity and model generalizability. \emph{Journal of Mathematical Psychology}, \emph{52}(6), 362--375.

\leavevmode\hypertarget{ref-lunn2000winbugs}{}%
Lunn, D. J., Thomas, A., Best, N., \& Spiegelhalter, D. (2000). WinBUGS-a bayesian modelling framework: Concepts, structure, and extensibility. \emph{Statistics and Computing}, \emph{10}(4), 325--337.

\leavevmode\hypertarget{ref-matuschek2017balancing}{}%
Matuschek, H., Kliegl, R., Vasishth, S., Baayen, H., \& Bates, D. (2017). Balancing Type I error and power in linear mixed models. \emph{Journal of Memory and Language}, \emph{94}, 305--315.

\leavevmode\hypertarget{ref-mengSimulatingRatiosNormalizing1996}{}%
Meng, X.-l., \& Wong, W. H. (1996). Simulating ratios of normalizing constants via a simple identity: A theoretical exploration. \emph{Statistica Sinica}, 831--860.

\leavevmode\hypertarget{ref-Moreyetal2011}{}%
Morey, R., \& Rouder, J. (2011). Bayes factor approaches for testing interval null hypotheses. \emph{Psychological Methods}, \emph{16}, 406--419. \url{https://doi.org/10.1037/a0024377}

\leavevmode\hypertarget{ref-mulder2016editors}{}%
Mulder, J., \& Wagenmakers, E.-J. (2016). Editors' introduction to the special issue ``Bayes factors for testing hypotheses in psychological research: Practical relevance and new developments''. \emph{Journal of Mathematical Psychology}, \emph{72}, 1--5.

\leavevmode\hypertarget{ref-myung1997applying}{}%
Myung, I. J., \& Pitt, M. A. (1997). Applying Occam's razor in modeling cognition: A Bayesian approach. \emph{Psychonomic Bulletin \& Review}, \emph{4}(1), 79--95.

\leavevmode\hypertarget{ref-navarro2015learning}{}%
Navarro, D. (2015). \emph{Learning statistics with R}. https://learningstatisticswithr.com.

\leavevmode\hypertarget{ref-navarro2019between}{}%
Navarro, D. J. (2019). Between the devil and the deep blue sea: Tensions between scientific judgement and statistical model selection. \emph{Computational Brain \& Behavior}, \emph{2}(1), 28--34.

\leavevmode\hypertarget{ref-NicenboimVasishth2016}{}%
Nicenboim, B., \& Vasishth, S. (2016). Statistical methods for linguistic research: Foundational Ideas - Part II. \emph{Language and Linguistics Compass}, \emph{10}(11), 591--613. \url{https://doi.org/10.1111/lnc3.12207}

\leavevmode\hypertarget{ref-nicenboim2020words}{}%
Nicenboim, B., Vasishth, S., \& Rösler, F. (2020). Are words pre-activated probabilistically during sentence comprehension? Evidence from new data and a bayesian random-effects meta-analysis using publicly available data. \emph{Neuropsychologia}, 107427.

\leavevmode\hypertarget{ref-oelrich2020bayesian}{}%
Oelrich, O., Ding, S., Magnusson, M., Vehtari, A., \& Villani, M. (2020). When are Bayesian model probabilities overconfident? \emph{arXiv Preprint arXiv:2003.04026}.

\leavevmode\hypertarget{ref-ohagan2006uncertain}{}%
O'Hagan, A., Buck, C. E., Daneshkhah, A., Eiser, J. R., Garthwaite, P. H., Jenkinson, D. J., \ldots{} Rakow, T. (2006). \emph{Uncertain judgements: Eliciting experts' probabilities}. John Wiley \& Sons.

\leavevmode\hypertarget{ref-phillips2011grammatical}{}%
Phillips, C., Wagers, M. W., \& Lau, E. F. (2011). Grammatical illusions and selective fallibility in real-time language comprehension. In \emph{Experiments at the Interfaces} (Vol. 37, pp. 147--180). Emerald Bingley, UK.

\leavevmode\hypertarget{ref-plummer2003jags}{}%
Plummer, M., \& others. (2003). JAGS: A program for analysis of Bayesian graphical models using Gibbs sampling. In \emph{Proceedings of the 3rd international workshop on distributed statistical computing} (Vol. 124, pp. 1--10). Vienna, Austria.

\leavevmode\hypertarget{ref-Rabe2021designr}{}%
Rabe, M. M., Kliegl, R., \& Schad, D. J. (2021). \emph{Designr: Balanced factorial designs}. Retrieved from \url{https://maxrabe.com/designr}

\leavevmode\hypertarget{ref-robert2007bayesian}{}%
Robert, C. (2007). The Bayesian choice. Springer-Verlag.

\leavevmode\hypertarget{ref-rouder2018bayesian}{}%
Rouder, J. N., Haaf, J. M., \& Vandekerckhove, J. (2018). Bayesian inference for psychology, part IV: Parameter estimation and Bayes factors. \emph{Psychonomic Bulletin \& Review}, \emph{25}(1), 102--113.

\leavevmode\hypertarget{ref-rouder2009bayesian}{}%
Rouder, J. N., Speckman, P. L., Sun, D., Morey, R. D., \& Iverson, G. (2009). Bayesian t tests for accepting and rejecting the null hypothesis. \emph{Psychonomic Bulletin \& Review}, \emph{16}(2), 225--237.

\leavevmode\hypertarget{ref-salvatier2016probabilistic}{}%
Salvatier, J., Wiecki, T. V., \& Fonnesbeck, C. (2016). Probabilistic programming in python using pymc3. \emph{PeerJ Computer Science}, \emph{2}, e55.

\leavevmode\hypertarget{ref-schad2020toward}{}%
Schad, D. J., Betancourt, M., \& Vasishth, S. (2021). Toward a principled Bayesian workflow in cognitive science. \emph{Psychological Methods}, \emph{26}(1), 103--126. \url{https://doi.org/10.1037/met0000275}

\leavevmode\hypertarget{ref-schad2018posterior}{}%
Schad, D. J., \& Vasishth, S. (2019). The posterior probability of a null hypothesis given a statistically significant result. \emph{arXiv Preprint arXiv:1901.06889}.

\leavevmode\hypertarget{ref-schad2020capitalize}{}%
Schad, D. J., Vasishth, S., Hohenstein, S., \& Kliegl, R. (2020). How to capitalize on a priori contrasts in linear (mixed) models: A tutorial. \emph{Journal of Memory and Language}, \emph{110}, 104038. \url{https://doi.org/https://doi.org/10.1016/j.jml.2019.104038}

\leavevmode\hypertarget{ref-schonbrodt2018bayes}{}%
Schönbrodt, F. D., \& Wagenmakers, E.-J. (2018). Bayes factor design analysis: Planning for compelling evidence. \emph{Psychonomic Bulletin \& Review}, \emph{25}(1), 128--142.

\leavevmode\hypertarget{ref-TQMP12-3-175}{}%
Sorensen, T., Hohenstein, S., \& Vasishth, S. (2016). Bayesian linear mixed models using Stan: A tutorial for psychologists, linguists, and cognitive scientists. \emph{Quantitative Methods for Psychology}, \emph{12}(3), 175--200. Retrieved from \url{http://www.ling.uni-potsdam.de/~vasishth/statistics/BayesLMMs.html}

\leavevmode\hypertarget{ref-spiegelhalter1994bayesian}{}%
Spiegelhalter, D. J., Freedman, L. S., \& Parmar, M. K. (1994). Bayesian approaches to randomized trials. \emph{Journal of the Royal Statistical Society. Series A (Statistics in Society)}, \emph{157}(3), 357--416.

\leavevmode\hypertarget{ref-taatgen2006modeling}{}%
Taatgen, N. A., Lebiere, C., \& Anderson, J. R. (2006). Modeling paradigms in ACT-R. \emph{Cognition and Multi-Agent Interaction: From Cognitive Modeling to Social Simulation}, 29--52.

\leavevmode\hypertarget{ref-Talts:2018aa}{}%
Talts, S., Betancourt, M., Simpson, D., Vehtari, A., \& Gelman, A. (2018). Validating Bayesian inference algorithms with simulation-based calibration. \emph{arXiv Preprint arXiv:1804.06788}.

\leavevmode\hypertarget{ref-vandekerckhove2018bayesian}{}%
Vandekerckhove, J., Rouder, J. N., \& Kruschke, J. K. (2018). Bayesian methods for advancing psychological science. \emph{Psychonomic Bulletin \& Review}, \emph{25}(1), 1--4.

\leavevmode\hypertarget{ref-vanpaemel2010prior}{}%
Vanpaemel, W. (2010). Prior sensitivity in theory testing: An apologia for the Bayes factor. \emph{Journal of Mathematical Psychology}, \emph{54}(6), 491--498.

\leavevmode\hypertarget{ref-VasishthEtAl2017EDAPS}{}%
Vasishth, S., Nicenboim, B., Beckman, M. E., Li, F., \& Kong, E. (2018). Bayesian data analysis in the phonetic sciences: A tutorial introduction. \emph{Journal of Phonetics}, \emph{71}, 147--161. \url{https://doi.org/10.1016/j.wocn.2018.07.008}

\leavevmode\hypertarget{ref-vehtari2020rank}{}%
Vehtari, A., Gelman, A., Simpson, D., Carpenter, B., Bürkner, P.-C., \& others. (2020). Rank-normalization, folding, and localization: An improved \(\widehat R\) for assessing convergence of MCMC. \emph{Bayesian Analysis}.

\leavevmode\hypertarget{ref-wagenmakersPrinciplePredictiveIrrelevance2019}{}%
Wagenmakers, E.-J., Lee, M. D., Rouder, J. N., \& Morey, R. D. (2019). \emph{The Principle of Predictive Irrelevance, or Why Intervals Should Not be Used for Model Comparison Featuring a Point Null Hypothesis} (preprint). PsyArXiv. \url{https://doi.org/10.31234/osf.io/rqnu5}

\leavevmode\hypertarget{ref-wagenmakersBayesianHypothesisTesting2010}{}%
Wagenmakers, E.-J., Lodewyckx, T., Kuriyal, H., \& Grasman, R. (2010). Bayesian hypothesis testing for psychologists: A tutorial on the Savage-Dickey method. \emph{Cognitive Psychology}, \emph{60}(3), 158--189. \url{https://doi.org/10.1016/j.cogpsych.2009.12.001}

\leavevmode\hypertarget{ref-wagers2009agreement}{}%
Wagers, M. W., Lau, E. F., \& Phillips, C. (2009). Agreement attraction in comprehension: Representations and processes. \emph{Journal of Memory and Language}, \emph{61}(2), 206--237.

\leavevmode\hypertarget{ref-Zellner1980PosteriorOR}{}%
Zellner, A., \& Siow, A. (1980). Posterior odds ratios for selected regression hypotheses. \emph{Trabajos de Estadistica Y de Investigacion Operativa}, \emph{31}, 585--603.
\end{cslreferences}

\endgroup

\end{document}